\documentclass[prb,twocolumn,
nobalancelastpage,amsmath,amssymb,
nofootinbib,showpacs]{revtex4-1}

\usepackage{graphics}      
\usepackage{graphicx}      
\usepackage{longtable}     
\usepackage{url}           
\usepackage{bm}            
\usepackage{braket}
\usepackage{color}
\usepackage{comment}

\begin{document}
\title{Glide-symmetric magnetic topological crystalline insulators with inversion symmetry}
\author{Heejae Kim$^1$, Ken Shiozaki$^{2}$, and Shuichi Murakami$^{1,3}$}
\affiliation{$^1$Department of Physics, Tokyo Institute of Technology, Meguro-ku, Tokyo 152-8551, Japan \\
$^2$Condensed Matter Theory Laboratory, RIKEN, Wako, Saitama 351-0198, Japan \\
$^3$
TIES, Tokyo Institute of Technology, Meguro-ku, Tokyo 152-8551, Japan}
\date{\today}

\begin{abstract}
It is known that three-dimensional magnetic systems with glide symmetry 
can be characterized by a $Z_2$ topological invariant together with the Chern number associated with the normal vector of the glide plane, and they are expressed in terms of integrals
of the Berry curvature. In the present paper,
we study the fate of this topological invariant when inversion symmetry is added
while time-reversal symmetry is not enforced.
There are two ways to add inversion symmetry, leading to space groups No.~13 and No.~14. 
In space group No.~13, we find that the glide-$Z_2$ invariant is expressed solely from the irreducible 
representations at high-symmetry points in $k$-space. It constitutes the $\mathbb{Z}_2\times \mathbb{Z}_2$ 
symmetry-based indicator for this space group, together with another $\mathbb{Z}_2$ representing the
Chern number modulo 2. In space group No.~14, 
we find that the symmetry-based indicator $\mathbb{Z}_2$ is given by a combination of the 
glide-$Z_2$ invariant and the Chern number. Thus, in space group No.~14, from the irreducible representations
at high-symmetry points we can only know possible combinations of the 
glide-$Z_2$ invariant and the Chern number, but in order to know each value of these 
topological numbers, we should calculate integrals of the Berry curvature. 
Finally, we show that in both cases, the symmetry-based indicator $\mathbb{Z}_4$ for inversion symmetric 
systems leading to the higher-order topological insulators is directly related with the glide-$Z_2$ invariant and the Chern number.
As an independent approach to these results, we also construct all invariants from the layer construction for these space groups,
and we show complete agreement with the above results for the topological invariants constructed from 
$k$-space topology. 
\end{abstract}

\maketitle

\section{Introduction}

Over the decades, important roles of the relations between topology and symmetry in modern condensed matter physics have been recognized by the tour de force works.
The trigger to encourage those researches is the discovery of a topological insulator (TI) \cite{HasanKane2010RevModPhys, QiZhang2011RevModPhys}.
TIs which appear in the presence of time-reversal symmetry (TRS) have attracted tremendous attention because of their unique and robust surface properties, and they are characterized by the $Z_2$ topological invariant, 
where anti-unitariness of TRS plays a key role.
Physicists have focused on the relations between topology and symmetry, and they have classified various topological phases ensured by combinations of internal symmetries, such as TRS, particle-hole symmetry, and chiral symmetry \cite{Schnyder2008prb78, Kitaev2009aip1134, Ryu2010njp12}.

Beyond internal symmetries, crystal symmetries of crystalline materials, such as spatial inversion, mirror, and rotational symmetries, play a crucial role to understand such topological phases.
An introduction of crystal symmetries has enriched our knowledge for topological invariants \cite{Fu2007prb76, Teo2008prb78}.
For instance, the $Z_2$ topological invariant for the TI has a complicated expression when inversion symmetry is absent, while it is expressed as a simple formula as a product of parity eigenvalues  at time-reversal invariant momenta (TRIMs) when inversion symmetry is preserved \cite{Fu2007prb76}.
In addition, since the seminal proposal of a topological crystalline insulator (TCI) \cite{Fu2011prl106}, whose topological phase is ensured by crystal symmetries, various combinations of internal symmetry and crystal symmetries have turned out to give an immense list of new topological phases \cite{Mong2010prb81, Hughes2011prb83, Fang2012prb86, Kargarian2013prl110, Zhang2013prl111, Ueno2013prl111, Chiu2013prb88, Morimoto2013prb88, Slager2013nphys9, Jadaun2013prb88, Alexandradinata2014prl113, Fulga2014prb89, Benalcazar2014prb89, Shiozaki2014prb90}. 
In particular, the materials of the SnTe class are revealed to be mirror-symmetric TCIs, and they are 
the first material realization for the TCIs protected by crystal symmetries.
Besides symmorphic symmetries, topological phases with nonsymmorphic symmetries have been vigorously discussed \cite{Parameswaran2013nphys9, Liu2014prb90, Young2015prl115, Fang2015prb91, Shiozaki2015prb91, Varjas2015prb92, Watanabe2015proc112, Po2016sciadv2, Lu2016nphys12, Dong2016prb93, Chen2016prb93, Kim2016prb93, Shiozaki2016prb93, Wieder2016prb94, Zhao2016prb94, Wang2016nature532, Bzduvsek2016nature538, Yang2017prb95, Takahashi2017prb96, Furusaki2017scibul62, Chen2018nphys14}.
Furthermore,
a new class of TCIs hosting gapless boundary states whose dimension is less than $d-1$ for a $d$-dimensional ($d$D) insulating bulk has also been established, and they are dubbed a higher-order TI
\cite{Benalcazar2017sci357, Benalcazar2017prb96, Langbehn2017prl119, Song2017prl119, Schindler2018sciadv4, Khalaf2018prb97, Schindler2018nphys14, Song2018prx8, Khalaf2018prx8}.

For comprehensive classifications of topological phases protected by a space group (SG), an approach based on the $K$-theory, which is a mathematical approach for gapped systems, has been adopted
\cite{Freed2013springer, Read2017prb95, Shiozaki2017prb95, Shiozaki2018arXiv180206694, Shiozaki2018arXiv181000801}.
Other approaches, such as topological quantum chemistry and symmetry-based indicators, 
have been attracting research interest \cite{Kruthoff2017prx7, Po2017ncommun, Bradlyn2017nature547, Watanabe2018sciadv4}; in these
approaches, topology of the bands and their compatibility relations are studied 
in the context of topological phases protected by SG symmetries. 
Particularly, symmetry-based indicators are useful in diagnosing topologically distinct band structures from the combinations of irreducible representations (irreps) at high-symmetry momenta \cite{Po2017ncommun, Watanabe2018sciadv4}.
These approaches focus on different aspects of topological phases. In the $K$-theory approach, one can comprehensively classify nontrivial phases, whereas 
an explicit formula for the topological invariant does not follow immediately from the theory. On the 
other hand, the symmetry-based indicator can reveal only the topological phases characterized by combinations of irreps. Thus, this theory cannot capture topological phases which cannot be known only from the irreps. 
Thus even with these powerful tools, one cannot reach a full understanding of nontrivial topological phases protected by SG symmetries, and there is much room for further
investigation.

In this paper, 
we study topological phases when inversion symmetry is introduced in magnetic systems with glide symmetry, to see 
an intriguing interplay between topology and glide symmetry.
We focus on topological phases in 3D spinless insulating systems protected by glide symmetry  \cite{Fang2015prb91, Shiozaki2015prb91}, without any internal symmetry, such as TRS, particle-hole symmetry, and chiral symmetry. This symmetry class is known as class A, one of the 10 Altland-Zirnbauer symmetry classes \cite{AZ1997prb55}.
A glide operation is a product of a reflection and a fractional translational operation, and it is nonsymmorphic. 
Such topological phases
are characterized by the $Z_2$ topological invariant \cite{Fang2015prb91, Shiozaki2015prb91} and the Chern number 
associated with the normal vector of the glide plane. The nature of a $Z_2$ topology for the magnetic glide-symmetric systems is very different from that for the time-reversal symmetric TIs, ensured by the antiunitary property of the time-reversal operation, 
because of an absence of antiunitary symmetry in magnetic glide-symmetric systems.
Since the glide symmetry is 
contained in many SGs, this glide-symmetric TCI phase, ensured by the $Z_2$ topological invariant, can exist in many SGs as well. 
Nonetheless, the formula of the glide-$Z_2$ invariant is quite complicated 
because it involves various terms of 2D integrals in momentum space \cite{Fang2015prb91, Shiozaki2015prb91}, 
and its behavior in the presence of additional symmetries has not been addressed.
Therefore, it should be useful to make clear the fate of the glide-$Z_2$ invariant in the presence of inversion symmetry in order to search for candidate materials of the glide-symmetric TCIs.

We consider the type-I magnetic SGs (MSGs), whose symbols are the same as those for the corresponding crystallographic SGs, because they do not contain time-reversal. For example, SG \#7 in the text refers to the type-I magnetic SG G=$Pc$ (MSG7.24 in the notation of Bilbao Crystallographic Server \cite{Aroyo2006bilbao}) which consists of the same unitary operations as that of SG \#7. 
In this paper we study SGs \#7, \#13, and \#14, which are the type-I MSGs 
7.24, 13.65, and 14.75, respectively.
If we add inversion symmetry to SG \#7, it becomes either SG \#13 or SG \#14.
In SG \#13, thanks to inversion symmetry, we show that the $Z_2$ topological invariant characterizing glide-symmetric systems can be written in terms of the irreps at high-symmetry points in the momentum space.
In particular, we find that the glide-symmetric $Z_2$ TCI is related to a higher-order TI ensured by inversion symmetry.
This glide-$Z_2$ invariant constitutes the $\mathbb{Z}_2\times \mathbb{Z}_2$ 
symmetry-based indicator for this SG, together with another $\mathbb{Z}_2$ representing the
Chern number modulo 2. In SG \#14, 
we find that the symmetry-based indicator $\mathbb{Z}_2$ is given by a combination of the 
glide-$Z_2$ invariant and the Chern number.

We also construct all invariants which characterize topology of layer constructions (LCs)
for the systems with these SGs, in a similar way as in 
Ref.~\onlinecite{Song2018ncommun9}.
This construction is solely based on geometries of layers. We then show that these invariants 
completely agree with the set of topological invariants discussed so far in this paper. This means that they
exhaust all the topological invariants in these SGs. To show this agreement, we find that it is convenient to modify the definition of glide-$Z_2$ invariant, 
and this modification does not affect the results in the previous works on this topological invariant. 

The organization in this paper is the following.
In Sec.~\ref{sec:preliminaries}, we briefly review general properties of glide-symmetric TCIs, together with the results of classifications based on the $K$-theory and those from the symmetry-based indicators.
Then we alter the definition of  the glide-$Z_2$ invariant from the previous works
in Sec.~\ref{sec:redef}, in order to have complete agreement with the LC. 
We then derive 
new Fu-Kane-type simplified formulas of the $Z_2$ topological invariant
and establish representative models for the glide-symmetric TCI, for SG \#13 in Sec.~\ref{sec:SG13} and for SG \#14 in Sec.~\ref{sec:SG14}, respectively. Throughout the paper, we restrict ourselves to bulk-insulating systems.
In particular, we exclude gapless phases with topological band-crossing points in bulk, such as Weyl semimetals \cite{Murakami2007njp9, Wan2011prb83}.

\section{Preliminaries}
\label{sec:preliminaries}

In the present section, we will review basic properties of the glide-symmetric $Z_2$ TCI \cite{Fang2015prb91, Shiozaki2015prb91}.
We start by introducing the $Z_2$ topological invariant characterizing the glide-symmetric system. We also introduce the previous works on 
topological phases for SGs \#7, \#13, and \#14.

\subsection{$Z_2$ topological invariant for glide-symmetric magnetic systems}

Here we briefly review the TCI ensured by glide symmetry proposed in Refs.~\onlinecite{Fang2015prb91, Shiozaki2015prb91}.
To be concrete, let us begin with a 3D system invariant under a glide operation 
\begin{equation}
\hat{G}_y = \left\{ m_y \bigg | \frac{1}{2} \hat{\bm z} \right\} ,
\label{eq:Gy}
\end{equation}
where $m_y$ represents the mirror reflection 
with respect to the $xz$ plane, and $\hat{\bm z}$ is a unit vector along the $z$ axis. 
Henceforth, the lengths of the primitive vectors are set to be unity.
Then the Bloch Hamiltonian $H(\bm{k})$ for the glide-symmetric system satisfies the equation
\begin{equation}
G_y (k_z) H(k_x ,k_y ,k_z) G_y (k_z)^{-1} = H(k_x, -k_y, k_z) ,
\end{equation}
where $G_y (k_z)$ is the $k$-dependent glide operator representing $\hat{G}_y$.
Since the Hamiltonian commutes with the glide operator on glide-invariant planes $k_y = 0$ and $k_y = \pi$, the Hamiltonian can be block-diagonalized into two blocks which are featured by the eigenvalues of $G_y(k_z)$. As we have $\hat{G}_y^2 = \{ E | \hat{\bm z} \}$, the eigenvalues of $G_y (k_z)$ are given by
\begin{equation}
g_\pm (k_z) = \pm e^{- ik_z /2} ,
\label{eq:glide_eigenval}
\end{equation}
in spinless systems.
Therefore, the two branches for those eigenvalues are related to each other, $g_\pm (k_z \pm 2\pi) = g_\mp (k_z)$, and are interchanged when an eigenstate goes across the {\it branch cut}.
This is a remarkable property of systems with nonsymmorphic symmetries with a fractional translation.
Several previous works have addressed topological properties from the interplay between nonsymmorphic natures and other crystalline symmetries, such as 2D Dirac semimeatls \cite{Young2015prl115} and semimetallic structures in the layer groups \cite{Wieder2016prb94} in systems with TRS.
Nonetheless, in this paper, we do not assume TRS.

\begin{figure}
\centering
\includegraphics[scale=0.4]{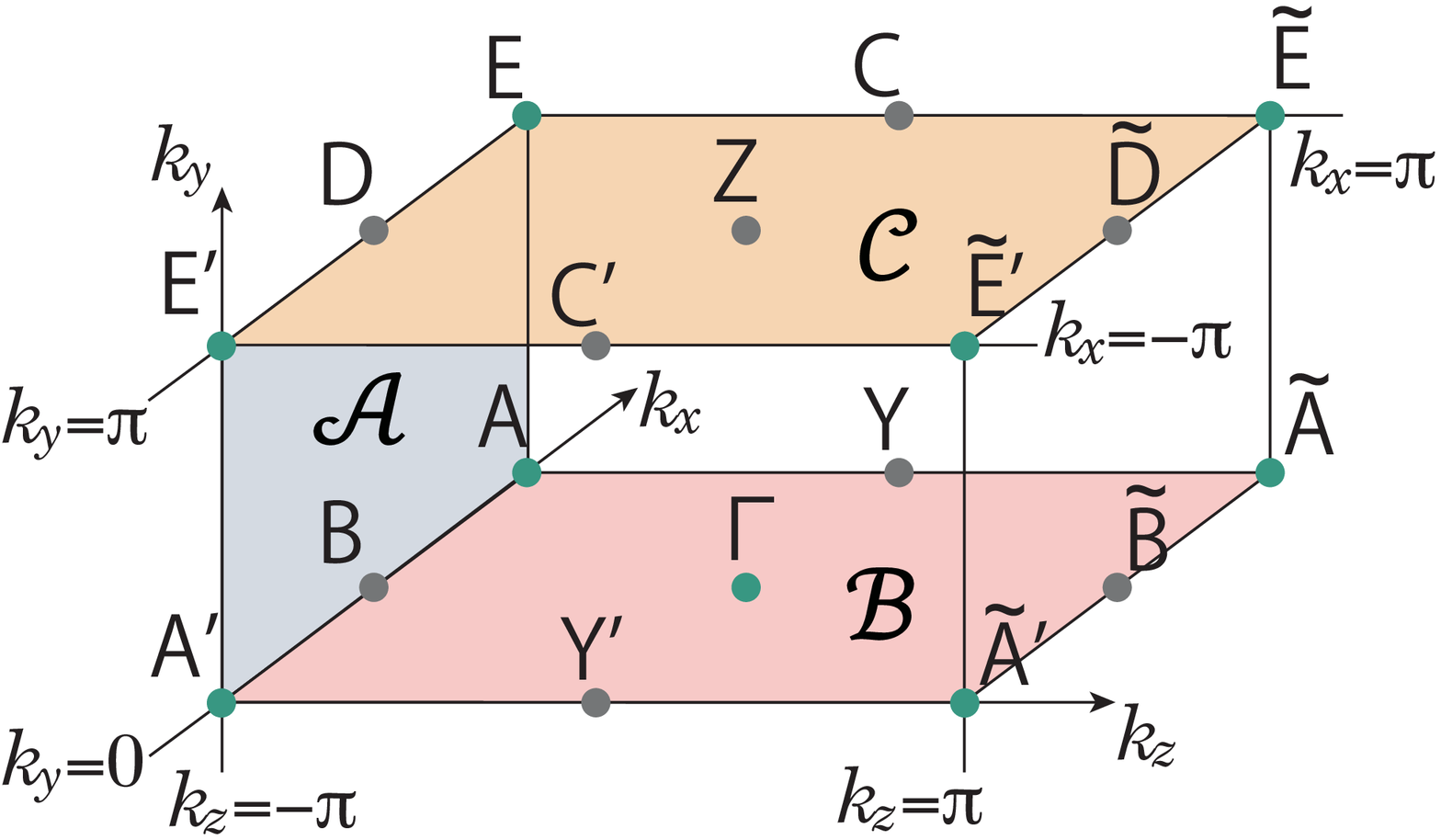}
\caption{(color online) Upper half of the Brillouin zone in SGs \#7, \#13 and \#14. $\Gamma$, A, E, Y, Z, B, C and D denote the high-symmetry points in SG \#13 and SG \#14.}
\label{fig:bz13}
\end{figure}

One can define the $Z_2$ topological invariant for such glide-symmetric systems.
Although the integral of Berry curvature for each glide sector on the glide-invariant planes can be defined, this quantity is not a quantized topological invariant due to the existence of the branch cut.
Instead, a 3D gapped system with the glide symmetry is characterized by the $Z_2$ topological invariant \cite{Fang2015prb91, Shiozaki2015prb91} defined as
\begin{align}
\nu &= \frac{1}{2\pi} \left[ \int_{\mathcal A} F_{xy} dk_x dk_y + \int_{\mathcal{B} - \mathcal{C}} F^-_{zx} dk_z dk_x \right] \nonumber \\
& \ \ \ - \frac{1}{\pi} \left( \gamma^+_{\mathrm{A}^\prime\mathrm{BA}} + \gamma^+_{\mathrm{ED}\mathrm{E}^\prime} \right) \pmod{2} .
\label{eq:z2glide}
\end{align}
where $\mathcal{A}$, $\mathcal{B}$ and $\mathcal{C}$ are the integral regions shown in Fig.~\ref{fig:bz13}.
We have defined the Berry connections
\begin{align}
& \bm{A}(\bm k) \equiv \sum_{n\in \mathrm{occ}} i \bra{u_{n\bm k}} \nabla_{\bm k} \ket{u_{n\bm k}} , \\
& \bm{A}^\pm(\bm k) \equiv \sum_{n\in \mathrm{occ}} i \bra{u^\pm_{n\bm k}} \nabla_{\bm k} \ket{u^\pm_{n\bm k}} ,
\end{align}
and the corresponding Berry curvatures 
\begin{align}
F_{ij} (\bm{k}) = \partial_{k_i} A_j (\bm{k}) - \partial_{k_j} A_i (\bm{k}), \\
F^\pm_{ij} (\bm{k}) = \partial_{k_i} A^\pm_j (\bm{k}) - \partial_{k_j} A^\pm_i (\bm{k}) ,
\end{align}
where $n$ is the band index, the summation $\sum_{n\in \mathrm{occ}}$ is over the occupied states, and $\ket{u^\pm_{n\bm k}}$ is the Bloch wavefunction belonging to the glide sector with the glide eigenvalue $g_\pm (k_z) = \pm e^{-ik_z/2}$.
The Berry phase $\gamma^{\pm} (\bm{k})$ along a closed path $\lambda$ is hence defined as
\begin{equation}
\gamma^{\pm}_\lambda (\bm{k}) = \oint_\lambda \bm{A}^{\pm} (\bm{k}) \cdot d\bm{k}.
\end{equation}
In Eq.~(\ref{eq:z2glide}),  $\gamma^+_{\mathrm{A}^\prime\mathrm{BA}}$ and $\gamma^+_{\mathrm{E}\mathrm{DE}^\prime}$ are Berry phases with  the paths $\lambda$ 
taken as straight paths $\mathrm{A}'\rightarrow\mathrm{B}\rightarrow\mathrm{A}$ and $\mathrm{E}\rightarrow\mathrm{D}\rightarrow\mathrm{E}'$, respectively, 
where the high-symmetry points are shown in Fig.~\ref{fig:bz13}. We take the gauge of the wavefunctions to be periodic along the $k_x$ direction. Namely, we take the wavefunction to be identical between $k_x=\pi$ and $k_x=-\pi$ throughout the paper.

\subsection{Previous works on topological phases for glide-symmetric systems with and without inversion symmetry in class A}
\label{subsec:glide_inversion_systems}

Here we consider what happens to the glide-$Z_2$ invariant, when inversion symmetry is added.
SG \#7 ($Pc$), i.e., SG with the glide symmetry only, becomes either SG \#13 ($P2/c$) or SG \#14 ($P2_1/c$) by adding inversion symmetry.
The difference between SG \#13 and SG \#14 comes from the position of inversion center; namely, while inversion center in SG \#13 is within the glide plane, that in SG \#14 is not.
The resulting SGs have twofold ($C_2$) rotational symmetry in SG \#13 and screw symmetry
in SG \#14.

\begin{table}
$$
\begin{array}{c | c | c | c}
 & \mathrm{SG} \ \#7 \ (Pc) & \mathrm{SG} \ \#13 \ (P2/c) & \mathrm{SG} \ \#14 \ (P2_1/c) \\ \hline
\mathrm{Generators} & \{ m_y | 00\frac{1}{2} \} & \begin{matrix} \{ I | 000 \} , \\ \{ C_{2y} | 00\frac{1}{2} \} \end{matrix} & \begin{matrix} \{ I | 000 \} , \\ \{ C_{2y} | 0\frac{1}{2}\frac{1}{2} \} \end{matrix} \\ \hline
K \mbox{-} \mathrm{group} & \mathbb{Z}^2 \times \mathbb{Z}_2 & \mathbb{Z}^8 & \mathbb{Z}^6 \\ \hline
\begin{matrix}
\mathrm{Symmetry} \\ \mathrm{indicators}
\end{matrix} 
 & \mathrm{N/A} & \mathbb{Z}_2 \times \mathbb{Z}_2 & \mathbb{Z}_2 \\ \hline 
E^{2,0}_\infty & \mathbb{Z} \times \mathbb{Z}_2 & \mathbb{Z} & \mathbb{Z} \\ 
\end{array}
$$
\caption{Summary of the generators, the $K$-groups, the symmetry-based indicators and the $E_{\infty}^{2,0}$ terms in the Atiyah-Hirzebruch spectral sequence for SGs \#7, \#13 and \#14.
The $E_\infty^{2,0}$ terms represent the existence of topological invariants whose definitions need the integral of Berry curvature on a 2D subspace in the Brillouin zone. Here, $\mathbb{Z}^2\equiv \mathbb{Z}\times \mathbb{Z}$ and $\mathbb{Z}_2\equiv \mathbb{Z}/2\mathbb{Z}$. }
\label{table:sum_table}
\end{table}

For these three SGs, the $K$-groups, the classifications of insulators including atomic ones, are known to be $\mathbb{Z}^2 \times \mathbb{Z}_2$ (SG \#7), $\mathbb{Z}^8$ (SG \#13) and $\mathbb{Z}^6$ (SG \#14) \cite{Shiozaki2018arXiv180206694} (Table~\ref{table:sum_table}). 
The symmetry-based indicators, which are defined as the quotients of the $K$-groups by the abelian groups generated by atomic insulators, are trivial (SG \#7), $\mathbb{Z}_2 \times \mathbb{Z}_2$ (SG \#13) and $\mathbb{Z}_2$ (SG \#14) \cite{Po2017ncommun, Watanabe2018sciadv4} (Table~\ref{table:sum_table}). 
On the other hand, the topological invariants including the integral of Berry curvature on a 2D subspace of the Brillouin zone, which are classified by the $E_{\infty}^{2,0}$ terms in the Atiyah-Hirzebruch spectral sequence, are classified by $\mathbb{Z} \times \mathbb{Z}_2$ (SG \#7), $\mathbb{Z}$ (SG \#13) and $\mathbb{Z}$ (SG \#14) \cite{Shiozaki2018arXiv180206694} (Table~\ref{table:sum_table}). 
However, physical meaning of each $\mathbb{Z}^n$ or $\mathbb{Z}_n$ factor in Table~\ref{table:sum_table} is not
obvious.
The recent work in Ref.~\onlinecite{Ono2018prb98} gives physical interpretations of symmetry-based indicators in class A by analyzing those explicit formula in several key SGs.
Nonetheless, from the symmetry-based indicators, one can neither study properties of 
topological invariants which cannot be expressed by symmetry-based indicators nor
know how those topological invariants evolve when additional symmetry is added and the
SG becomes its supergroup.


\section{Redefinition of the glide-$Z_2$ invariant}
\label{sec:redef}
In this section we revisit the definition of the glide-$Z_2$ invariant $\nu$. 
First we show how this topological invariant $\nu$ depends on the choice of the glide plane.
We then introduce 
another topological invariant $\delta_{\rm g}$ for glide-symmetric systems from the LC similarly to Ref.~\onlinecite{Song2018ncommun9}.
To achieve agreement between the two glide topological invariants, we
show that it is better to change the definition of the glide-$Z_2$ invariant 
from $\nu$ to $\tilde{\nu}\equiv \nu+n_{\rm Ch}$ (mod 2), where $n_{\rm Ch}$ is 
the Chern number along the glide-invariant plane $k_y=0$. 
Henceforth, $n_{\mathrm{Ch}}$ refers the Chern number along the normal vector of the glide-invariant planes.
We will also explain that this redefinition does not affect physical properties of the
glide-$Z_2$ invariant discussed in previous works \cite{Fang2015prb91, Shiozaki2015prb91}.

\subsection{Gauge dependence of the glide-$Z_2$ invariant}
\label{sec:gauge_dep_z2}
In systems with glide symmetry, we have two glide planes which are inequivalent under 
the lattice translation. For the glide operation given by Eq.~(\ref{eq:Gy}), 
the glide planes are $y=0$ and $y=\frac{1}{2}$. In such systems, when we
take another glide operation given by 
\begin{equation}
\hat{G}'_y \equiv \left\{ m'_y \bigg | \frac{1}{2} \hat{\bm z} \right\} ,
\label{eq:G'y}
\end{equation}
where $m^\prime_y$ is a mirror reflection with respect to the plane $y=\frac{1}{2}$. 
Because $\hat{G}'_y=T_y\hat{G}_y$ where $T_y$ is a unit translation along the $y$ direction, 
this change of the glide operation switches the $g_+$ and $g_-$ sectors on the $k_y=\pi$ plane,
while the glide sectors on the $k_y=0$ plane remain intact. Therefore, the value of the glide $Z_2$ invariant 
changes from $\nu$ to 
\begin{align}
\nu' &= \frac{1}{2\pi} \left[ \int_{\mathcal A} F_{xy} dk_x dk_y + \int_{\mathcal{B} } F^-_{zx} dk_z dk_x \right. \nonumber \\
 &\ \ \ \ \ - \left. \int_{ \mathcal{C}} F^+_{zx} dk_z dk_x \right] \nonumber \\
& \ \ \ - \frac{1}{\pi} \left( \gamma^+_{\mathrm{A}^\prime\mathrm{BA}} + \gamma^-_{\mathrm{ED}\mathrm{E}^\prime} \right) \pmod{2} .
\label{eq:z2glide'}
\end{align}
Its difference from the original value $\nu$ is given by 
\begin{align}
\nu'-\nu&= -\frac{1}{2\pi}  \int_{\mathcal{C}} (F^+_{zx}-F^-_{zx}) dk_z dk_x 
\nonumber \\
& \ \ \ - \frac{1}{\pi} \left(
 \gamma^-_{\mathrm{ED}\mathrm{E}^\prime}
- \gamma^+_{\mathrm{ED}\mathrm{E}^\prime}
 \right) \pmod{2}.
\end{align}
Because, the second term can be rewritten as
$\gamma^-_{\mathrm{ED}\mathrm{E}^\prime}
- \gamma^+_{\mathrm{ED}\mathrm{E}^\prime}
=\gamma^+_{\mathrm{\tilde{E}\tilde{D}}\mathrm{\tilde{E}}^\prime}-\gamma^+_{\mathrm{ED}\mathrm{E}^\prime}
=-\int_{\mathcal{C}} F^+_{zx}dk_z dk_x $, 
points $\mathrm{\tilde{D}}$, $\mathrm{\tilde{E}}$, and $\mathrm{\tilde{E}}^\prime$ are specified in
Fig.~\ref{fig:bz13}, we get
\begin{align}
\nu'-\nu&= \frac{1}{2\pi}  \int_{\mathcal{C}} F_{zx} dk_z dk_x =n_{\rm Ch} \pmod{2},
\label{eq:nu'nu}\end{align}
where $n_{\rm Ch}$ is the Chern number along the $k_z$-$k_x$ plane. 
Therefore, the value of the glide-$Z_2$ invariant changes with the change of the 
glide plane, when $n_{\rm Ch}$ is an odd integer.

\subsection{Redefinition of the glide-$Z_2$ invariant}
In systems with glide symmetry, one can introduce a topological invariant associated with 
the glide symmetry, from the viewpoint of the real-space LC. This is a different approach from 
the approach from the $k$-space topology, which led us to the glide-$Z_2$ invariant $\nu$.  
This approach of LC has been formulated in Ref.~\onlinecite{Song2018ncommun9} for systems with TRS. 
We here extend this theory to systems with glide symmetry without TRS. 
We introduce a glide-invariant $\delta_{\rm g}$ based on the LC by a straightforward
calculation based on the geometry of the layers. The details of the definitions and
calculations of $\delta_{\rm g}$ are straightforward but lengthy, and are given in Appendix \ref{sec:LC}. 

By comparing this glide-invariant $\delta_{\rm g}$ with the glide-$Z_2$ invariant $\nu$ in 
Eq.~(\ref{eq:z2glide}), we find that they are not equal. Nonetheless, 
if we redefine the glide-$Z_2$ invariant 
to be 
\begin{equation}
\tilde{\nu}\equiv \nu+n_{\rm Ch} \pmod{2} ,
\label{eq:redef}
\end{equation}
it is equal to the glide invariant $\delta_{\rm g}$:
\begin{equation}
\tilde{\nu}\equiv\delta_{\rm g}.
\end{equation}

One may wonder whether this redefinition may invalidate the bulk-boundary correspondence 
of the glide-$Z_2$ invariant $\nu$, meaning the number of helical surface
states in the gap given
 in the previous works
\cite{Fang2015prb91, Shiozaki2015prb91}. Nevertheless, it is not the case, 
because in the arguments on the bulk-boundary correspondence in Refs.~\onlinecite{Fang2015prb91, Shiozaki2015prb91}, the Chern number $n_{\rm Ch}$ is assumed to be equal to zero, in which case
this redefinition does not alter the value of the glide-$Z_2$ invariant: $\nu=\tilde{\nu}$. 
Furthermore, the glide-$Z_2$ invariant does not depend on the choice of the glide plane,
as follows from Eq.~(\ref{eq:nu'nu}). 
On the other hand, we remark that when $n_{\rm Ch}\neq 0$, the glide-$Z_2$ invariant may lose its meaning as the
number of helical surface states, because its value is gauge dependent.

Thus the properties of the glide-$Z_2$ invariant  in  Refs.~\onlinecite{Fang2015prb91, Shiozaki2015prb91} remain valid even with this redefinition, and henceforth we will adopt the redefined glide-$Z_2$ invariant (\ref{eq:redef}).
This redefinition is convenient when we compare the results with the LC, 
and is physically meaningful as we discuss later.  

Here we give an expression for the new glide-$Z_2$ invariant $\tilde{\nu}$. From 
Eqs.~(\ref{eq:z2glide'}) and (\ref{eq:nu'nu}), we conclude that $\tilde{\nu}$ is equal to $\nu'$. Thus, we get a formula for the new glide-$Z_2$ invariant
\begin{align}
\tilde{\nu} &= \frac{1}{2\pi} \left[ \int_{\mathcal A} F_{xy} dk_x dk_y + \int_{\mathcal{B} } F^-_{zx} dk_z dk_x \right. \nonumber \\
 &\ \ \ \ \ - \left. \int_{ \mathcal{C}} F^+_{zx} dk_z dk_x \right] \nonumber \\
& \ \ \ - \frac{1}{\pi} \left( \gamma^+_{\mathrm{A}^\prime\mathrm{BA}} + \gamma^-_{\mathrm{ED}\mathrm{E}^\prime} \right) \pmod{2}.
\label{eq:nu_tilde}
\end{align}

\section{Glide-symmetric magnetic topological crystalline insulators for space group \#13}
\label{sec:SG13}

In the present section, we consider SG \#13, which is realized by adding inversion symmetry to SG \#7.
As explained in Sec.~\ref{subsec:glide_inversion_systems}, SG \#13 has $C_2$ rotational symmetry, unlike SG \#14.
Here we derive a new formula of the glide-$Z_2$ invariant (Eq.~(\ref{eq:nu_tilde})) for SG \#13. 
We then discuss how our new formula is related to symmetry-based indicators \cite{Po2017ncommun, Watanabe2018sciadv4} and the topological invariants based on the $K$-theory \cite{Shiozaki2018arXiv180206694}.
To check our results, we establish models realizing the topological phases in SG \#13 based on the LC \cite{Song2018ncommun9}.

\subsection{Topological invariants for SG \#13}

We will show that in SG \#13, the $Z_2$ topological invariant for the glide-symmetry system ${\tilde{\nu}}$ (mod 2) is given in terms of the irreps at high-symmetry points.
For simplicity, here we consider the spinless case.
Then, later we obtain
\begin{equation}
(-1)^{\tilde{\nu}} = \prod_{i \in \mathrm{occ}} \frac{\zeta_i^- (\Gamma) \zeta_i^+(\mathrm{C})}{\zeta_i^- (\mathrm{Y}) \zeta_i^+ (\mathrm{Z})} ,
\end{equation}
where $\zeta_i^\pm(=\pm 1)$ is an eigenvalue of the $C_2$ rotation for the eigenstates in the $g_\pm$ sector at high-symmetry points $\Gamma$, Y, Z, and C.
In the following we show this formula by calculating $(-1)^{\tilde{\nu}}(=e^{i\pi{\tilde{\nu}}})$ from Eq.~(\ref{eq:nu_tilde}).

By adding an inversion $\hat{I} = \{ I | \bm{0} \}$ around the origin to SG \#7, we obtain SG \#13. Then 
$C_2$ rotational symmetry 
around the axis $x=0$, $z=1/4$
\begin{equation}
\hat{C}_{2}=   \hat{G}_y \hat{I}= \left\{C_{2y} \bigg | \frac{1}{2} \hat{\bm z} \right\} ,
\label{eq:C213}
\end{equation}
is also added to the symmetry operations.
The key commutation relation between the glide operator and the $C_2$ rotational operator is given by
\begin{equation}
\hat{C}_2 \hat{G}_y = \hat{G}_y \hat{C}_2 \{ E | \hat{\bm z} \} .
\label{eq:CR13}
\end{equation}
This will be used in the following discussion.

We now rewrite the formula of the glide-$Z_2$ invariant (\ref{eq:nu_tilde}) with the help of additional symmetries.
First, the Berry curvature on the $k_z=-\pi$ plane in Eq.~(\ref{eq:nu_tilde}) satisfies
\begin{equation}
F_{xy} (k_x, k_y,-\pi) = -F_{xy} (-k_x , k_y,\pi)= -F_{xy} (-k_x , k_y,-\pi) ,
\end{equation}
owing to the $C_2$ symmetry.
This immediately leads to
\begin{equation}
\int_{\mathcal A} F_{xy} dk_x dk_y = 0 .
\label{eq:FA13}
\end{equation}

Next, we address the Berry curvature on the glide-invariant planes $k_y = 0$ and $k_y = \pi$.
Equation~(\ref{eq:CR13}) indicates that the glide sector for wavefunctions within the glide-invariant planes is unchanged under the $C_2$ rotation. Therefore, 
the Berry curvatures $F^\pm_{zx} (\bm{k})$ on those planes have to be an even function of $k_x$ and $k_z$,
\begin{equation}
F^\pm_{zx} (k_x, k_y, k_z) = F^\pm_{zx} (-k_x, k_y, -k_z) ,
\label{eq:F13-1}
\end{equation}
where $k_y$ is either $0$ or $\pi$.

Let us first consider the integral on the glide-invariant plane $k_y = 0$.
From Eq.~(\ref{eq:F13-1}), we get
\begin{equation}
\int_{\mathcal B} F^-_{zx} dk_z dk_x = 2 \int_{\mathcal{B}^\prime} F^-_{zx} dk_z dk_x ,
\label{eq:FB13-1}
\end{equation}
where $\mathcal{B}^\prime$ is a half of the region $\mathcal{B}$ given by $-\pi \le k_x < \pi$, $0 \le k_z < \pi$. Note 
that the $\hat{C}_2$ does not change the glide sector of the eigenstates within the $k_y = 0$ plane and we have taken the gauge to be periodic along the $k_x$ direction.
Therefore, by using the Stokes' theorem on the $k_y = 0$ plane, we have
\begin{align}
& \exp \left[ \frac{i}{2} \int_{\mathcal B} F^-_{zx} dk_z dk_x - i \gamma^+_{\mathrm{A}^\prime \mathrm{BA}} \right]  \nonumber \\
&= \exp \left[ i \int_{\mathcal{B}^\prime} F^-_{zx} dk_z dk_x - i \gamma^+_{\mathrm{A}^\prime \mathrm{BA}} \right] = \exp \left( - i \gamma^-_{\mathrm{Y}^\prime \Gamma \mathrm{Y}} \right) .
\label{eq:FB13}
\end{align}
We have used the relation of the branch cut $\gamma^+_{\tilde{\mathrm{A}}^\prime \tilde{\mathrm{B}} \tilde{\mathrm{A}}} = \gamma^-_{\mathrm{A}^\prime\mathrm{BA}}$ (mod $2\pi$),
points $\tilde{\mathrm{A}}$, $\tilde{\mathrm{A}}^\prime$, and $\tilde{\mathrm{B}}$ are specified in Fig.~\ref{fig:bz13}.

Similarly, the terms on the other glide-invariant plane $k_y = \pi$ are given by
\begin{equation}
\exp \left[ -\frac{i}{2} \int_{\mathcal C} F^+_{zx} dk_z dk_x - i \gamma^-_{\mathrm{EDE}^\prime} \right] = \exp \left( i \gamma^+_{\mathrm{C}^\prime\mathrm{ZC}} \right).
\label{eq:FC13}
\end{equation}
We note that Eqs.~(\ref{eq:FB13}) and (\ref{eq:FC13}) are gauge invariant.

Consequently, from Eqs.~(\ref{eq:FA13}), (\ref{eq:FB13}), and (\ref{eq:FC13}), we obtain a formula for the glide-$Z_2$ invariant (\ref{eq:nu_tilde}) as
\begin{equation}
(-1)^{\tilde{\nu}} = e^{- i \gamma^-_{\mathrm{Y}^\prime\Gamma\mathrm{Y}} + i \gamma^+_{\mathrm{C}^\prime\mathrm{ZC}}} = \prod_{i \in \mathrm{occ}}
\frac{\zeta^-_i (\Gamma) \zeta^+_i (\mathrm{C}) }{\zeta^-_i (\mathrm{Y}) \zeta^+_i (\mathrm{Z})} ,
\label{eq:13rewrite_z2}
\end{equation}
where $\zeta^+_i$ and  $\zeta^-_i$ are the $C_2$ eigenvalues ($=\pm 1$) of the $i$-th occupied state in the $g_+$ and $g_-$ sectors respectively, and $\Gamma, \mathrm{Y}, \mathrm{Z},$ and $\mathrm{C}$ are the high-symmetry points on the plane $k_z = 0$ shown in Fig.~\ref{fig:bz13}.
To show this formula we used the properties of the sewing matrix as introduced in Ref.~\onlinecite{Fang2012prb86}
and its details are explained in Appendix \ref{app:sewing}.

At the four high-symmetry points $\Gamma$, Y, Z, and C, all the irreps are 1D, and their characters are shown in Table~\ref{table:13irreps}.
Therefore, an alternative expression for ${\tilde{\nu}}$ is
\begin{equation}
{\tilde{\nu}} = N_{B_g} (\Gamma)+N_{B_g} (\mathrm{Y})+N_{B_u} (\mathrm{C})+N_{B_u} (\mathrm{Z})
\pmod 2 ,
\label{eq:13z2}
\end{equation}
where $N_{R}(P)$ is the number of occupied states at the high-symmetry point $P$ with an irrep $R$ for the $k$-group of SG \#13.

Equations (\ref{eq:13rewrite_z2}) and (\ref{eq:13z2}) explicitly depend on glide eigenvalues at high-symmetry points, which looks convention-dependent 
due to $4\pi$ periodicity of the glide eigenvalues. Nevertheless, it can be rewritten in terms of the
parity eigenvalues and thus is independent of conventions. 
By using compatibility relations for the irreps of SG \#13, we can rewrite this formula for the glide-$Z_2$ invariant 
$\tilde{\nu}$ as follows. We introduce the $z_4$ indicator for systems with inversion symmetry
\begin{align}
&z_4=\sum_{\bm{K}\in\text{TRIM}}\frac{n_{\bm{K}}^+ -n_{\bm{K}}^-}{2} \pmod 4 ,
\label{eq:z4}
\end{align}
where $n_{\bm{K}}^+$ and $n_{\bm{K}}^-$ are the number of occupied even-parity and odd-parity states at a TRIM $\bm{K}$, and the sum is taken over the eight TRIMs.

As mentioned earlier, we are assuming insulating systems, and in particular 
we exclude the Weyl semimetal phase, which means that 
this $z_4$ indicator is always an even integer \cite{Hughes2011prb83,Ono2018prb98}.
Here we briefly explain why the $z_4$ indicator should be even. 
When the $z_4$ indicator is odd, 
which means that the parity of the Chern number on the $k_y = 0$ plane and that on the $k_y = \pi$ plane are not equal, the gap should close somewhere between these planes, and the system is no longer an insulator \cite{Hughes2011prb83}.
In such case, the gap closes at Weyl nodes, which are apices of the Dirac cones
and carry topological charges characterized by the monopole density $\rho(\bm{k}) \equiv \frac{1}{2\pi} \bm{\nabla}_{\bm k} \cdot \bm{F} (\bm{k})$ \cite{Murakami2007njp9, Wan2011prb83}.
Therefore, the system is no longer an insulator, which is to be excluded here.
Thus, the $z_4$ indicator is always an even integer 
and when it takes a value $z_4\equiv 2 \pmod 2$ the 
system is a higher-order topological insulator with gapless hinge states. 
Then from the compatibility relations for
SG \#13, summarized in Appendix \ref{sec:irrep}, we obtain
\begin{align}
\tilde{\nu}\equiv \frac{z_4}{2} \pmod 2 .
\label{eq:nuz4-13}
\end{align}
Thus, in SG \#13, the glide-$Z_2$ invariant is equivalent to the $z_4$ indicator for inversion symmetry.

Since TRS is not assumed here, the Chern number $n_{\mathrm{Ch}}$ on the $k_x$-$k_z$ plane, which is parallel to the glide plane, can be nonzero. 
On the other hand, the Chern numbers on the $k_x$-$k_y$ plane and on the $k_y$-$k_z$ plane identically vanish due to the $C_2$ symmetry.
As argued in Refs.~\onlinecite{Fang2012prb86, Hughes2011prb83}, the Chern number on the $k_x$-$k_z$ plane, 
which is perpendicular to the $\hat{C}_2$ axis, is calculated in terms of $C_2$ eigenvalues up to modulo 2. On the $k_y=0$ plane, 
it is expressed as 
\begin{equation}
(-1)^{n_{\mathrm{Ch}}} = \prod_{i\in \mathrm{occ}}  \zeta_i (\Gamma)\zeta_i (\mathrm{Y}) \zeta_i (\mathrm{B}) \zeta_i (\mathrm{A}) ,
\label{eq:Chern_num_13}
\end{equation}
where $\zeta_i$ is the $C_2$ eigenvalue of the $i$-th occupied state at the high-symmetry points, $\Gamma$, Y, B, and A on the $k_y = 0$ plane shown in Fig.~\ref{fig:bz13}.
In the same manner, the Chern number on the $k_y = \pi$ plane is calculated from the product of $C_2$ eigenvalues at Z, C, D, and E, and, 
it is equal to $n_{\mathrm{Ch}}$ calculated on the $k_y=0$ plane, since we have assumed a 
presence of the bulk gap. 
Otherwise, if the Chern numbers calculated at $k_y = 0$ and at $k_y = \pi$ are not equal, 
the band gap vanishes somewhere between these planes \cite{Hughes2011prb83}, which occurs when $z_4$ is odd. On the other hand,
this cannot occur if $z_4$ is even, i.e., $z_4 = 0 \pmod 2$.

\begin{table}
$$
\begin{array}{c|cccc}
{C}_{2h} & E & C_{2z} & G_y & I \\ \hline
A_g & +1 & +1 & +1 & +1 \\
A_u & +1 & +1 & -1 & -1 \\
B_g & +1 & -1 & -1 & +1 \\
B_u & +1 & -1 & +1 & -1 \\
\end{array}
$$
\caption{Characters of the irreps of $C_{2h}$. It is also a character table for the irreps of the $k$-group of SG \#13 at $\Gamma$, Y, Z, and C. }
\label{table:13irreps}
\end{table}

\subsection{Symmetry-based indicators and $K$-theory}
Let us compare our results with the results on the symmetry-based indicator theory in the previous work~\cite{Po2017ncommun, Watanabe2018sciadv4}.
The symmetry-based indicator for SG \#13 is $\mathbb{Z}_2 \times \mathbb{Z}_2$ \cite{Po2017ncommun, Watanabe2018sciadv4}.  Following Ref.~\onlinecite{Po2017ncommun, Watanabe2018sciadv4},  we can show that one $\mathbb{Z}_2$ means the $Z_2$ topological invariant ${\tilde{\nu}}$ for the glide-symmetric systems in Eq.~(\ref{eq:13rewrite_z2}), and that 
the other $\mathbb{Z}_2$ is the Chern number $n_{\mathrm{Ch}}$ modulo 2 in Eq.~(\ref{eq:Chern_num_13}). 
Thus, when inversion symmetry is present, ${\tilde{\nu}}$ and $n_{\mathrm{Ch}}$ (mod 2) are both symmetry-based indicators, 
expressed in terms of irreps at high-symmetry points. 

Let us discuss the relationship between the higher-order TI and these topological invariants. 
In a system with only inversion symmetry, obeying SG \#2 ($P\bar{1}$), symmetry-based indicators are given by $(\mathbb{Z}_2)^3 \times \mathbb{Z}_4$ \cite{Po2017ncommun, Watanabe2018sciadv4}, where the three factors of $\mathbb{Z}_2$ are weak indices for Chern insulators, while the factor of $\mathbb{Z}_4$ corresponds to a strong index $z_4$ in Eq.~(\ref{eq:z4}), counting the number of eigenstates below the Fermi energy having an 
odd parity.
When $z_4 = 1$ or $z_4 = 3$, it means that Weyl nodes appear in the momentum space.
The case with $z_4 = 2$ corresponds to a higher-order TI, leading to an existence of 1D chiral hinge modes on the surface.
In the present case of SG \#13, 
Eq.~(\ref{eq:nuz4-13}) shows that the $z_4=2$ phase is equivalent to ${\tilde{\nu}}=1 \pmod{2}$.
Thus, the system with nontrivial glide-$Z_2$ invariant is in the phase with $z_4 = 2$, 
and it is a higher-order TI ensured by inversion symmetry. Thus they exhibit 1D chiral hinge modes.

On the other hand, in the classification based on the $K$-theory \cite{Shiozaki2018arXiv180206694}, the topological invariant given by 
 the 
$E_\infty^{2,0}$ term expressing a topological invariant in terms of integrals of wavefunctions is the Chern number $\mathbb{Z}$ in SG \#13 (Table~\ref{table:sum_table}).
We conclude that the topological phase characterized by the glide-$Z_2$ invariant is related to the higher-order TI with trivial weak indices $(\mathbb{Z}_2, \mathbb{Z}_2, \mathbb{Z}_2; \mathbb{Z}_4) = (0, 0, 0; 2)$, while that by an odd value of the Chern number is $(\mathbb{Z}_2, \mathbb{Z}_2, \mathbb{Z}_2; \mathbb{Z}_4) = (0, 1, 0; 0)$, where three $\mathbb{Z}_2$'s
refer to the Chern numbers modulo 2 along the $k_x=0$, $k_y=0$ and $k_z=0$ planes, and 
$\mathbb{Z}_4$ refers to the $z_4$ indicator in Eq.~(\ref{eq:z4}).
This is among principle results of the present paper, and we check this scenario by constructing simple tight-binding models exhibiting 
topological phases in the following section.

\subsection{Layer constructions for glide- and inversion-symmetric systems}
\label{subsec:layer_construction13}
One can introduce various real-space topological invariants by the LC, similarly to 
Ref.~\onlinecite{Song2018ncommun9}. In this LC for a SG, we introduce a
set of planes, which are located periodically to be compatible with lattice translation symmetry. 
Here we consider each plane to be a 2D Chern insulator with a Chern number equal to $+1$.
Then for a certain SG operation, such as inversion, rotation, and glide, we can 
show whether the configurations of the layers can be continuously trivialized or not, when the symmetry elements of SG other than symmetry for defining the invariant are ignored.
Consequently, to indicate nontrivial configurations of planes we introduce an invariant for each SG operation. 

To be more specific, we introduce a layer consisting of planes
\begin{align}
(mnl;d) &= \left \{ \bm{r} | \bm{r} \cdot (m\bm{b}_1 + n\bm{b}_2 + l\bm{b}_3 ) = 2\pi (d+q), \ q \in \mathbb{Z}\right \} \nonumber \\
&= \left \{ \bm{r} | \frac{mx}{a} + \frac{ny}{b} + \frac{lz}{c} = d+q, \ q \in \mathbb{Z} \right \} ,
\label{eq:mnl}
\end{align}
where $\bm{b}_i$'s are reciprocal vectors corresponding to the primitive vectors 
$\bm{a}_1 = (a,0,0), \bm{a}_2 = (0,b,0), \bm{a}_3 = (0,0,c),$ and $0 \le d < 1$.
The integer $q$ is introduced because of the translation symmetry, and each layer consists of an infinite number of planes.
Each layer is decorated with a 2D Chern insulator with a Chern number $+1$, 
with its orientation is given by a reciprocal lattice vector given by $\bm{G}=m\bm{b}_1 + n\bm{b}_2 + l\bm{b}_3$.
Depending on SGs, we minimally introduce other layers to make the set of layers to be compatible with
the given SG. Then we evaluate invariants for individual SG operations. 
The details are shown in Appendix \ref{sec:LC}.

In SG \#13, we can define four kinds of invariants, the Chern invariants $\delta_{n_\mathrm{Ch}, i=1,2,3}$, the glide invariant
$\delta_{\text{g}}$, the inversion invariant $\delta_{\text{i}}$, and the $C_2$ invariant $\delta_{\text{r}}$,
only from geometric configuration of the layer
as discussed in detail in Appendix \ref{sec:LC}. 
The Chern invariant $\delta_{n_\mathrm{Ch}, i}$ takes an integer value, 
while the other invariants
$\delta_{\text{g}}$ and $\delta_{\text{i}}$ are defined modulo 2, and $\delta_{\text{r}}$ turns out to be always trivial.
Then for every LC, one can evaluate these invariants, from which we establish relations between these
invariants:
\begin{align}
&\delta_{\text{i}}\equiv\delta_{\text{g}} \pmod 2 , \\
& \delta_{\text{r}}\equiv 0 ,
\end{align}
as shown in Appendix \ref{sec:LC}.

Now we can compare these invariants with the glide-$Z_2$ invariant $\tilde{\nu}$ and the 
Chern number $n_{\text{Ch}}$. One can calculate these two topological invariants for general 
layers and compare them with the 
invariants $\delta_{n_\mathrm{Ch}, i}$, $\delta_{\text{g}}$, $\delta_{\text{i}}$, and $\delta_{\text{r}}$ from the LC. Then we get
\begin{align}
n_{\text{Ch}}&=\delta_{n_\mathrm{Ch}, 2},\label{eq:SG13Ch}\\
\tilde{\nu}&\equiv\delta_{\text{i}}\equiv\delta_{\text{g}}.\label{eq:SG13nu}
\end{align}
Thus the construction of the topological invariants based on the LC completely agrees with 
the known topological invariants, the glide-$Z_2$ invariant $\tilde{\nu}$ and the 
Chern number $n_{\text{Ch}}$. 

One can calculate  the glide-$Z_2$ invariant $\tilde{\nu}$ and the 
Chern number $n_{\text{Ch}}$ for general LCs, in order to show Eqs.~(\ref{eq:SG13Ch}) and (\ref{eq:SG13nu}) after a straightforward but lengthy calculation.
From these results we can demonstrate generators of LCs for nontrivial values of topological invariants,
i.e., minimal layer configurations for nontrivial 
combinations of the topological invariants. They are listed in Table
\ref{table:eLCs} in Appendix \ref{sec:LC}.
In this section we study the
two minimal LCs $(001;0)$ and $(010;\frac{1}{2})$ in Table \ref{table:eLCs} for SG \#13. These two LCs correspond
to the glide-symmetric $Z_2$ TCI and the Chern insulator phases in SG \#13, respectively, and 
in this subsection 
we construct simple tight-binding models for them in order to examine our scenario.
We show the outline of the calculations, and the details are presented in Appendix \ref{app:layer13}.

\subsubsection{Glide-symmetric $Z_2$ TCI phase: $(001;0)$}

\begin{figure}
\centering
\includegraphics[scale=0.35]{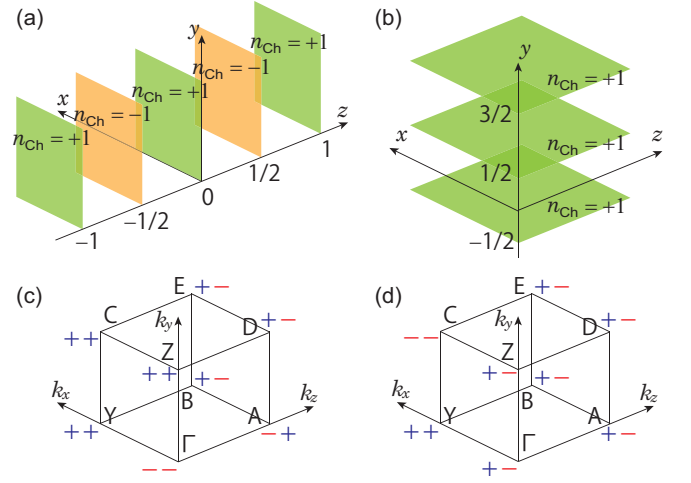}
\caption{(color online) Layer constructions  (a) $(001;0)$, showing a glide-$Z_2$ topological phase (${\tilde{\nu}}=1$, $n_{\mathrm{Ch}}=0$) and (b) $(010;\frac{1}{2})$, showing a Chern insulator (${\tilde{\nu}}=0$, $n_{\mathrm{Ch}}=1$) with SG \#13. The parities at high-symmetry points for the models (a) and (b) are shown in (c) and (d), respectively.}
\label{fig:lc13}
\end{figure}

We begin with the elementary LC (eLC) $(001;0)$ showing the glide-symmetric $Z_2$ TCI phase (Fig.~\ref{fig:lc13}(a)).
In the absence of inversion symmetry, i.e., for SG \#7, a minimal configuration for a glide-symmetric $Z_2$ TCI can be
realized as a LC $(001;d)$, representing 2D Chern insulators with $n_{\mathrm{Ch}} = +1$ along the $xy$ plane placed at $z = z_0+n$ where $n$ is an integer. 
Then, the glide symmetry requires presence of other layers of a 2D Chern insulator with $n_{\mathrm{Ch}} = -1$ at $z = z_0 + n+1/2$, because the glide operation changes the sign of the Chern number
along the $xy$ plane.
This model gives a nontrivial value of the $Z_2$ topological invariant ${\tilde{\nu}}=1$ for glide-symmetric systems by a direct calculation. 

We now consider the case when inversion symmetry is added, and SG becomes SG \#13. 
Inversion symmetry fixes the value of $z_0$ to be $z_0=0$, and the configuration allowed by symmetry  is
the one with the 2D Chern insulators on $z = n$ with $n_{\mathrm{Ch}} = +1$ and on $z = n + 1/2$ with $n_{\mathrm{Ch}} = -1$, where $n_{\mathrm{Ch}}$ is the Chern number along the $xy$ plane and $n$ is an integer (Fig.~\ref{fig:lc13}(a)). This is the LC with $(001;0)$.
Therefore, a representative Hamiltonian for the glide-symmetric $Z_2$ TCI phase with additional inversion symmetry is obtained as
\begin{equation}
H^{\mathrm{LC}}_{{\tilde{\nu}}} (\bm{k}) = (m + \cos k_x + \cos k_y) \sigma_z + \sin k_x \sigma_x + \sin k_y \sigma_y \tau_z ,
\label{eq:13z2_layer_construction}
\end{equation}
where $\bm{\sigma}$ and $\bm{\tau}$ are Pauli matrices denoting the orbital and lattice degrees of freedom.
We here set the Fermi energy to be $E_F=0$.
Hence, the corresponding $k$-dependent glide operator and the $k$-dependent inversion operator are
\begin{align}
G_y (k_z) &= e^{-i k_z/2} \left( \cos\frac{k_z}{2} \tau_x + \sin \frac{k_z}{2} \tau_y \right) , \label{eq:glide_operator} \\
I (k_z) &= \sigma_z
\begin{pmatrix}
1 & \\ & e^{-ik_z}
\end{pmatrix}
_{\tau} .
\label{eq:inversion_operator}
\end{align}
Here, the subscript $\tau$ meant the matrix for the lattice degrees of freedom.
The Hamiltonian (\ref{eq:13z2_layer_construction}) satisfies
\begin{align}
G_y(k_z) H^{\mathrm{LC}}_{{\tilde{\nu}}}(k_x, k_y, k_z) G_y (k_z)^{-1} &= H^{\mathrm{LC}}_{{\tilde{\nu}}}(k_x, -k_y, k_z) , \label{eq:relation_ham_glide} \\
I(k_z)H^{\mathrm{LC}}_{{\tilde{\nu}}}(\bm{k}) I(k_z)^{-1} &= H^{\mathrm{LC}}_{{\tilde{\nu}}}(-\bm{k}) ,
\label{eq:relation_ham_inversion}
\end{align}
and these operators satisfy the commutation relation
\begin{equation}
G_y(-k_z) I(k_z) = e^{ik_z} I(k_z) G_y(k_z) ,
\end{equation}
which is just a Bloch form for $\hat{G}_y \hat{I} = \hat{T}_z \hat{I} \hat{G}_y$ 
given by the combination of Eqs.~(\ref{eq:C213}) and (\ref{eq:CR13}).

The Hamiltonian $H^{\mathrm{LC}}_{{\tilde{\nu}}}$ is gapped unless $m = \pm 2, 0$.
The gap closes at $\Gamma$ when $m = -2$, at Y, Z when $m = 0$, and at C when $m=2$. The glide-$Z_2$ invariant is nontrivial (trivial) if $ 0 < |m| < 2$ ($|m| > 2$).

The effective Hamiltonian at the four high-symmetry points $P=(k_x,k_y,0)$ ($\Gamma$, Y, Z, and C) in Eq.~(\ref{eq:13rewrite_z2}) can be written as
\begin{equation}
H^{\mathrm{LC}}_{{\tilde{\nu}}} (P) = (m + \cos k_x + \cos k_y) \sigma_z .
\end{equation}
The irreps of the occupied states are summarized in Table \ref{table:irreps13}.
The sum of the numbers of the $B_g$ irreps at $\Gamma$ and Y and those of the $B_u$ irreps at Z and C in the occupied bands is odd at $m = -1.8$ and the system is $Z_2$ nontrivial, whereas it is even at $m=-2.2$ and the system is $Z_2$ trivial.

The parities at high-symmetry points 
at $m=-1.8$ are shown in Fig.~\ref{fig:lc13}(c).
Therefore, it is a higher-order TI with $(\mathbb{Z}_2, \mathbb{Z}_2, \mathbb{Z}_2; \mathbb{Z}_4) = (0, 0, 0; 2)$ in SG \#2 shown in Ref.~\onlinecite{Ono2018prb98}. Thus, the glide-symmetric topological phase with 
the nontrivial $Z_2$ invariant  at $m=-1.8$ is a higher-order TI in accordance with Sec.~\ref{sec:gauge_dep_z2}.

\begin{center}
\begin{table}
\begin{tabular}{c | c | c | c | c}
 & $\Gamma$ & Y & Z & C \\ \hline
$H_{{\tilde{\nu}}}^{\mathrm{LC}} (m=-2.2)$ & $A_g + B_g$ & $A_g + B_g$ & $A_g + B_g$ & $A_g + B_g$ \\ \hline
$H_{{\tilde{\nu}}}^{\mathrm{LC}} (m=-1.8)$ & $A_u + B_u$ & $A_g + B_g$ & $A_g + B_g$ & $A_g + B_g$ \\ \hline
$H_{\mathrm{Ch}}^{\mathrm{LC}} (m=-2.2)$ & $A_g + B_g$ & $A_g + B_g$ & $A_u + B_u$ & $A_u + B_u$ \\ \hline
$H_{\mathrm{Ch}}^{\mathrm{LC}} (m=-1.8)$ & $B_g + B_u$ & $A_g + B_g$ & $B_g + B_u$ & $A_u + B_u$
\end{tabular}
\caption{Irreps for the two tight-binding models for the topological phases in SG \#13 from the layer construction.}
\label{table:irreps13}
\end{table}
\end{center}

\subsubsection{Chern insulator phase: $(010;\frac{1}{2})$}

Next, we consider the eLC $(010;\frac{1}{2})$ showing a weak 3D Chern insulator with a nontrivial Chern number along the $k_x$-$k_z$ direction (Fig.~\ref{fig:lc13}(b)).
In the absence of inversion symmetry, such a weak 3D Chern insulator phase can be realized as a
stacking of 2D Chern insulators with $n_{\mathrm{Ch}} = +1$ parallel to the $xz$ plane, located at $y = y_0+n$ with an integer $n$.
When we add inversion symmetry with inversion center at the origin, $y_0$ is set to be 0 or $\frac{1}{2}$ (Fig.~\ref{fig:lc13}(b)). To make $\tilde{\nu}$ to vanish, we choose $y_0=\frac{1}{2}$. This layer is described as $(010;\frac{1}{2})$. 
Therefore, we have a representative Hamiltonian
\begin{align}
H^{\mathrm{LC}}_{\mathrm{Ch}} (\bm{k}) &= \left(m + \cos k_x + \cos\frac{k_z}{2}\right) \sigma_z + \sin k_x \sigma_y + \sin\frac{k_z}{2} \sigma_x .
\label{eq:LC-Z-SG13}
\end{align} 
Here, in order to make the model consistent with the glide symmetry, the primitive 
translation vector along the $z$ axis is taken as $(0,0,1/2)$. Nonetheless, 
for the calculation of topological invariants from Eqs.~(\ref{eq:13rewrite_z2}) and (\ref{eq:Chern_num_13}) we should instead regard the primitive translation vector to be $(0,0,1)$; namely we double the unit cell along the $z$ direction. In this doubled unit cell, the Hamiltonian is rewritten as
\begin{align}
\tilde{H}^{\mathrm{LC}}_{\mathrm{Ch}}  (\bm{k}) &= [ (m + \cos k_x ) \sigma_z + \sin k_x \sigma_y] \nonumber \\&+
\left(\cos\frac{k_z}{2}\sigma_z+\sin\frac{k_z}{2}\sigma_x\right) \left(\cos\frac{k_z}{2}\tau_x+\sin\frac{k_z}{2}\tau_y\right) .
\label{eq:LC-Z-SG13-2}
\end{align}
The Hamiltonian $\tilde{H}^{\mathrm{LC}}_{\mathrm{Ch}}$ indeed satisfies the relations in Eqs.~(\ref{eq:relation_ham_glide}) and (\ref{eq:relation_ham_inversion}) under the corresponding operators given by Eqs.~(\ref{eq:glide_operator}) and 
(\ref{eq:inversion_operator}).

This model shows phase transitions at $m=0, \pm 2$. For example, we focus on the transition at $m=-2$. It is a trivial insulator at $m = -2.2$ and a weak 3D Chern insulator with $n_{\mathrm{Ch}} = +1$ at $m = -1.8$, as can be seen from Eq.~(\ref{eq:LC-Z-SG13}). Meanwhile, both $m = -2.2$ and $m = -1.8$, the glide-$Z_2$ invariant (\ref{eq:13rewrite_z2}) is trivial.
Therefore, the Hamiltonian $\tilde{H}_{\mathrm{Ch}}^{\mathrm{LC}}$ realizes a topological phase with a nonzero Chern number but a trivial glide-$Z_2$ invariant at $m = -1.8$ as we intuitively expected.
The parities at high-symmetry points in this model are shown in Fig.~\ref{fig:lc13}(d).
It gives the values of the symmetry-based indicators as $(\mathbb{Z}_2, \mathbb{Z}_2, \mathbb{Z}_2; \mathbb{Z}_4) = (0, 1, 0; 0)$.

\section{Glide-symmetric magnetic topological crystalline insulators for space group \#14}
\label{sec:SG14}

In this section, we consider SG \#14 which is realized by adding $C_2$ screw symmetry to SG \#7 having glide symmetry only.
We mostly follow the argument for SG \#13 in the previous section.
Meanwhile, unlike SG \#13, the inversion center in SG \#14 is not within the glide plane, leading to the property that
the glide sector on the $k_y=\pi$ plane changes by $C_2$ screw rotation. This makes the calculation very different from that in SG \#13.

\subsection{Topological invariants for SG \#14}

To begin with, we derive the glide-$Z_2$ invariant in SG \#14 by adding inversion symmetry to SG \#7.
Here we consider the spinless case for simplicity.
In this case, we consider glide and inversion symmetries,
\begin{align}
\hat{G}_y = \left\{ m_y \bigg | \frac{1}{2} \hat{\bm y} + \frac{1}{2} \hat{\bm z} \right\}, \quad \hat{I}  = \{ I | \bm{0} \} ,
\end{align}
with its inversion center at the origin. 
It means that the glide plane is $y=1/4$, and it does not contain inversion center, unlike that in SG \#13.
Then we get $C_2$ screw symmetry
\begin{equation}
\hat{S}_y = \hat{G}_y\hat{I}=\left\{ C_{2y} \bigg | \frac{1}{2} \hat{\bm y} + \frac{1}{2} \hat{\bm z} \right\}.
\end{equation}
The commutation relation between the glide and the $C_2$ screw operators is given by
\begin{equation}
\hat{S}_y \hat{G}_y = \hat{G}_y \hat{S}_y \{ E |- \hat{\bm y} + \hat{\bm z} \} .\
\label{eq:CR14}
\end{equation}

Below, we mainly follow the same spirit of the previous section to derive a formula of the glide-$Z_2$ invariant for SG \#14.
The symmetry-based indicator is $\mathbb{Z}_2$ for SG \#14 \cite{Po2017ncommun, Watanabe2018sciadv4}. 
In the following, we show that the indicator $\mathbb{Z}_2$ gives information on the combination of the glide-$Z_2$ invariant and the Chern number.
Its formula here we find reads
\begin{equation}
n_{\mathrm{Ch}} \in 2\mathbb{Z}, \quad (-1)^{\tilde{\nu}} (-1)^{n_{\mathrm{Ch}}/2} = \prod_{i \in \mathrm{occ}} \frac{\xi_i^-(\Gamma) \xi_i^+(\mathrm{D})}{\xi_i^-(\mathrm{Y}) \xi_i^+(\mathrm{E})} ,
\end{equation}
where $\xi_i^\pm$ is an eigenvalue of the $C_2$ screw operation in the $g_\pm$ sector at the high-symmetry points $\Gamma$, Y, D, and E.
In the following we show this formula by calculating $(-1)^{\tilde{\nu}}(=e^{i\pi{\tilde{\nu}}})$ with ${\tilde{\nu}}$ given by Eq.~(\ref{eq:nu_tilde}).
In the derivation, we should pay attention to the glide sectors and the branch cut.

Let us start with the ${\mathcal A}$ term in Eq.~(\ref{eq:nu_tilde}) on the $k_z = -\pi$ plane.
The Berry curvature on the $xy$ plane with $k_z = -\pi$ should be an odd function of $k_x$ because of the $C_2$ screw symmetry.
Therefore, we get 
\begin{equation}
\int_{\mathcal A} F_{xy} dk_x dk_y = 0 ,
\label{eq:FA14}
\end{equation}
as similar to Eq.~(\ref{eq:FA13}) in SG \#13.

On the $k_y = 0$ plane, because $C_2$ screw merely behaves as $C_2$ rotation, and it does not alter the
glide sector, we have
\begin{equation}
\exp \left[ \frac{i}{2} \int_{\mathcal B} F_{zx}^- dk_z dk_x - i \gamma^+_{\mathrm{A}^\prime \mathrm{BA}} \right] = \exp \left( -i \gamma^-_{\mathrm{Y}^\prime \Gamma \mathrm{Y}} \right) ,
\label{eq:FB14}
\end{equation}
which is the same as Eq.~(\ref{eq:FB13}) in SG \#13.

On the other hand, we confront a difference from the case of SG \#13, when we evaluate the integral over the other glide-invariant plane $k_y = \pi$.
One cannot use the same trick as in SG \#13 because $C_2$ screw changes the glide sectors, leading to the relation $F^+_{zx} (k_x, \pi,k_z) = F^-_{zx} ( -k_x, \pi, -k_z)$.
Due to this relation, we find that the integral of the Berry curvature for the $g_+$ sector is a half of the integral of the total Berry curvature on the $k_y = \pi$ plane:
\begin{align}
&\exp \left[ -\frac{i}{2} \int_{\mathcal C} F^+_{zx} dk_z dk_x - i \gamma^-_{\mathrm{EDE}^\prime} \right] \nonumber \\
& \ \ \ = \exp \left[ -\frac{i}{4} \int_{\mathcal C} F_{zx} dk_z dk_x - i \gamma^-_{\mathrm{EDE}^\prime} \right] \nonumber \\
& \ \ \ = (-i)^{n_{\mathrm{Ch}}} \exp \left( - i \gamma^-_{\mathrm{EDE}^\prime} \right) .
\end{align}
In the similar manner, noticing that
\begin{align}
\exp \left( - i \gamma^-_{\mathrm{EDE}^\prime} \right) &= \exp \left( - i \gamma^+_{\tilde{\mathrm{E}} \tilde{\mathrm{D}} \tilde{\mathrm{E}}^\prime} \right) \nonumber\\
&=\exp \left[ i \int_{\mathcal C} F^+_{zx} dk_z dk_x \right]\exp \left( -i \gamma^+_{\mathrm{EDE}^\prime} \right) \nonumber\\
&=
(-1)^{n_{\mathrm{Ch}}} \exp \left( -i \gamma^+_{\mathrm{EDE}^\prime} \right) ,
\end{align}
we can rewrite the terms on the $k_y = \pi$ plane as
\begin{equation}
\exp \left[ - \frac{i}{2} \int_{\mathcal C} F^+_{zx} dk_z dk_x - i \gamma^-_{\mathrm{EDE}^\prime} \right] = i^{n_{\mathrm{Ch}}} \exp \left( -i \gamma^+_{\mathrm{EDE}^\prime} \right) .
\label{eq:FC14}
\end{equation}
As a consequence, by combining Eqs.~(\ref{eq:FA14}), (\ref{eq:FB14}), and (\ref{eq:FC14}), we recast the formula for the glide-$Z_2$ invariant into
\begin{align}
(-1)^{\tilde{\nu}} &= i^{n_{\mathrm{Ch}}} \times e^{-i \gamma^-_{\mathrm{Y}^\prime \Gamma \mathrm{Y}}} \times e^{-i \gamma^+_{\mathrm{EDE}^\prime}} \nonumber \\
&= i^{n_\mathrm{Ch}} \prod_{i \in \mathrm{occ}} \frac{\xi^-_i (\Gamma) \xi^+_i (\mathrm{D})}{\xi^-_i (\mathrm{Y}) \xi^+_i (\mathrm{E})} ,
\end{align}
where $\prod_{i \in \mathrm{occ}} \xi^\pm_i (P)$ is the product of the $C_2$ screw eigenvalues over the occupied states for the $g_\pm$ sector at the high-symmetry point $P$.
An alternative expression is
\begin{equation}
(-1)^{\tilde{\nu}} (-i)^{n_{\mathrm{Ch}}}= \prod_{i \in \mathrm{occ}} \frac{\xi^-_i (\Gamma) \xi^+_i (\mathrm{D})}{\xi^-_i (\mathrm{Y}) \xi^+_i (\mathrm{E})}.
\label{eq:indicator14}
\end{equation}
Since the product of $\xi^\pm_i$ in Eq.~(\ref{eq:indicator14}) is equal to $\pm 1$, the Chern number is found to be an even integer.
Therefore, we have eventually shown that
\begin{equation}
n_{\mathrm{Ch}} \in 2\mathbb{Z} , \quad (-1)^{\tilde{\nu}} (-1)^{n_{\mathrm{Ch}}/2}= \prod_{i \in \mathrm{occ}} \frac{\xi^-_i (\Gamma) \xi^+_i (\mathrm{D})}{\xi^-_i (\mathrm{Y}) \xi^+_i (\mathrm{E})} .
\label{eq:14z2}
\end{equation}
This corresponds to the symmetry-based indicator $\mathbb{Z}_2$ for SG \#14. 
Therefore, an alternative expression is
\begin{equation}
{\tilde{\nu}}+\frac{n_{\mathrm{Ch}}}{2} = N_{B_g} (\Gamma)+N_{B_g} (\mathrm{Y})+N_{B_u} (\mathrm{D})+N_{B_u} (\mathrm{E})
\pmod 2 .
\label{eq:14z2-irrep}
\end{equation}
Furthermore, from the compatibility relations, one can also show that the $z_4$ indicator for 
inversion symmetry is directly related with this symmetry-based indicator Eq.~(\ref{eq:14z2-irrep}):
\begin{equation}
{\tilde{\nu}}+\frac{n_{\mathrm{Ch}}}{2} = \frac{z_4}{2}
\pmod 2 .
\label{eq:nuz4-14}
\end{equation}

Thus the value of the symmetry-based indicator $\mathbb{Z}_2$ for SG \#14 gives possible combinations
of the 
value of  the glide-$Z_2$ invariant ${\tilde{\nu}}$ and that of the Chern number $
n_{\mathrm{Ch}}$. Meanwhile, one cannot uniquely determine the values of these topological numbers, solely from the
$\mathbb{Z}_2$ symmetry-based indicator.  
Furthermore, both 
the glide-$Z_2$ topological phase and a Chern insulator with the Chern number equal to $4n+2$ ($n$: integer) correspond to higher-order TIs with $z_4 = 2$, 
ensured by inversion symmetry. It  corresponds to $(\mathbb{Z}_2, \mathbb{Z}_2, \mathbb{Z}_2; \mathbb{Z}_4) = (0, 0, 0; 2)$ phase. 
On the other hand, the $E_\infty^{2,0}$ term of the $K$-theory classification \cite{Shiozaki2018arXiv180206694} shows the $\mathbb{Z}$ topological invariant, i.e., the Chern number (Table~\ref{table:sum_table}). 
In the following section, we confirm these conclusions on the topological invariants 
 by LCs.

\begin{table}
$$
\begin{array}{c|cccc}
 & E & S_y & G_y & I \\ \hline
A_g & +1 & +i & +i & +1 \\
A_u & +1 & +i & -i & -1 \\
B_g & +1 & -i & -i & +1 \\
B_u & +1 & -i & +i & -1 \\
\end{array}
$$
\caption{Characters of the irreps of the $k$-group of SG \#14 at D and E on the $k_y = \pi$ plane. This is obtained from the character
table of $C_{2h}$ with an additional phase factor for $S_y$ and $G_{y}$. }
\label{table:14irreps}
\end{table}

\subsection{Layer constructions for glide- and inversion-symmetric systems}
One can define topological invariants based on the LC, as has been 
done for SG \#13. 
In SG \#14, we can define four kinds of invariants, the Chern invariants $\delta_{n_{\mathrm{Ch}}, i=1,2,3}$, the glide invariant
$\delta_{\text{g}}$, the inversion invariant $\delta_{\text{i}}$, and the $C_2$ screw invariant $\delta_{\text{s}}$,
from geometric configurations of the layers 
with its details in Appendix \ref{sec:LC}. 
The Chern invariant $\delta_{n_{\mathrm{Ch}}, i}$ takes an integer value, 
while the other invariants
$\delta_{\text{g}}$, $\delta_{\text{i}}$, and $\delta_{\text{s}}$ are defined modulo 2. 
Then for every LC, one can evaluate these invariants, from which we establish relations between these
invariants:
\begin{align}
&\delta_{n_{\mathrm{Ch}}, 2}\in 2\mathbb{Z},\\
&\delta_{\text{i}}\equiv\delta_{\text{g}}+\delta_{n_{\mathrm{Ch}}, 2}/2 \pmod{2},\\ 
&\delta_{\text{s}}\equiv\delta_{n_{\mathrm{Ch}}, 2}/2 \pmod{2}.
\end{align}

We can compare these invariants with the glide-$Z_2$ invariant $\tilde{\nu}$ and the 
Chern number $n_{\text{Ch}}$, and we get
\begin{align}
&\delta_{n_{\mathrm{Ch}}, 2}=n_{\text{Ch}}\in 2\mathbb{Z},\label{eq:SG14Ch}\\
&\delta_{\text{g}}\equiv\tilde{\nu},\label{eq:SG14g}\\
&\delta_{\text{i}}\equiv\tilde{\nu}+\frac{1}{2}n_{\text{Ch}},\label{eq:SG14nu}\\
&\delta_{\text{s}}\equiv\frac{1}{2}n_{\text{Ch}}.\label{eq:SG14nCh}
\end{align}
Thus the topological invariants constructed from the LC completely agree with
the known topological invariants, the glide-$Z_2$ invariant $\tilde{\nu}$ and the 
Chern number $n_{\text{Ch}}$.

One can calculate  the glide-$Z_2$ invariant $\tilde{\nu}$ and the 
Chern number $n_{\text{Ch}}$ for general LCs, in order to show Eqs.~(\ref{eq:SG14Ch}) to (\ref{eq:SG14nCh}) after a straightforward but lengthy calculation.
From these results we can demonstrate generators of LCs for nontrivial values of topological invariants,
i.e., minimal layer configurations for nontrivial 
combinations of the topological invariants. They are listed in Table
\ref{table:eLCs} in Appendix \ref{sec:LC}.
In this section we study the
two minimal LCs $(001;0)$ and $(020;0)$ in Table \ref{table:eLCs} for SG \#14.
These two LCs correspond
to the glide-symmetric $Z_2$ TCI and the Chern insulator phases in SG \#14, respectively, and 
in this subsection 
we construct simple tight-binding models for them in order to examine our scenario.
We show the outline of the calculations, and the details are presented in Appendix \ref{app:layer14}.

\subsubsection{Glide-symmetric $Z_2$ TCI phase: $(001;0)$}
\begin{figure}
\centering
\includegraphics[scale=0.35]{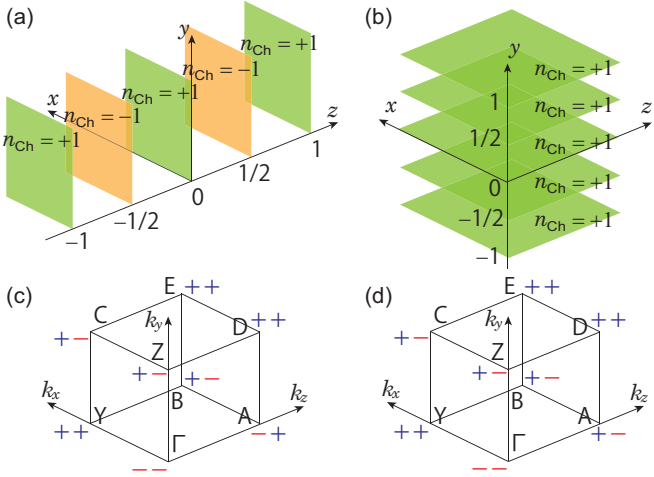}
\caption{(color online) Layer constructions for (a) a glide-$Z_2$ topological phase and (b) a Chern insulator with SG \#14. The parities at the high-symmetry points for the models (a) and (b) are shown in (c) and (d), respectively.
Unlike the case of SG \#13, the combinations of parities at high-symmetry points are equivalent between (c) and (d).}
\label{fig:lc14}
\end{figure}

First, a Hamiltonian exhibiting a glide-symmetric $Z_2$ TCI is constructed by putting 2D Chern insulator layers with $n_{\mathrm{Ch}} = +1$ on the 
$z = z_0+n$ planes and those with $n_{\mathrm{Ch}} = -1$ on the $z = z_0+n+ 1/2$ planes, where $n_{\mathrm{Ch}}$ is the Chern number along the $xy$ plane and $n$ is an integer.
Since this configuration becomes compatible with inversion symmetry by setting $z_0=0$, the layer is now
described as $(001;0)$ as shown in Fig.~\ref{fig:lc14}(a). Its representative Hamiltonian is given by
\begin{equation}
H^{\mathrm{LC}}_{{\tilde{\nu}}} (\bm{k}) = (m + \cos k_x + \cos k_y) \sigma_z + \sin k_x \sigma_x + \sin k_y \sigma_y \tau_z.
\label{eq:14z2_layer_construction}
\end{equation}
The form of the Hamiltonian is the same as Eq.~(\ref{eq:13z2_layer_construction}), 
whereas the corresponding glide and inversion operators in momentum space are taken as
\begin{align}
G_y (k_z) &= e^{-i k_z/2} \left( \cos\frac{k_z}{2} \tau_x + \sin \frac{k_z}{2} \tau_y \right) , \label{eq:glide_operator14} \\
I(k_y, k_z) &= \sigma_z
\begin{pmatrix}
1 & \\ & e^{-i(k_y + k_z)}
\end{pmatrix}_{\tau} .
\label{eq:inversion14}
\end{align}
Thus, we can see easily
\begin{equation}
G_y(-k_z) I(k_y, k_z) = e^{i(-k_y+k_z)} I(-k_y, k_z) G_y(k_z) ,
\end{equation}
which is Bloch form of $\hat{G}_y\hat{I} = \hat{T}^{-1}_y \hat{T}_z \hat{I} \hat{G}_y$.

As we mentioned in the previous section, the representative Hamiltonian $H_{{\tilde{\nu}}}^{\mathrm{LC}} (\bm{k})$ in Eq.~(\ref{eq:14z2_layer_construction}) closes the gap at $\Gamma$ when $m = -2$, at Y, Z when $m = 0$, and at C when $m = 2$.
The irreps for the phases at $m = -2.2$ and $m = -1.8$ are summarized in Table~\ref{table:LCirreps14}.
By calculating the $C_2$ screw eigenvalues at $\Gamma$, D, Y, and E, 
the value of the symmetry-based indicator in our new formula of Eq.~(\ref{eq:14z2}) is $-1$ and 
is nontrivial.
Since the Chern number should be 0 in this configuration, this nontrivial indicator is
attributed to a nontrivial glide-$Z_2$ invariant.
Since this $Z_2$ topological phase corresponds to $(\mathbb{Z}_2, \mathbb{Z}_2, \mathbb{Z}_2; \mathbb{Z}_4) = (0, 0, 0; 2)$ in SG \#2 by calculating the parities  (Fig.~\ref{fig:lc14}(c)), this model is also a higher-order TI ensured by inversion symmetry.

\subsubsection{Chern insulator phase: $(020;0)$}
Next we construct a model in the Chern insulator phase.
To make a model compatible with inversion symmetry, we must put {\it two} 2D Chern insulator layers with $n_{\mathrm{Ch}} = +1$ within the lattice constant i.e., along the $xz$ plane at $y = n$ and $y = n + 1/2$ ($n$ : integer) (Fig.~\ref{fig:lc14}(b)).
It is written as $(020;0)$.
One can write down the model as
\begin{align}
H^{\mathrm{LC}}_{\mathrm{Ch}} (\bm{k}) &= [(m + \cos k_x + \cos k_z) \sigma_z + \sin k_x \sigma_y + \sin  k_z \sigma_x ]\tau_0
\label{eq:14Ch_layer_construction}
\end{align}
with the glide operator and inversion operator given by Eqs.~(\ref{eq:glide_operator14}) and (\ref{eq:inversion14}), respectively.
The Hamiltonian $H^{\mathrm{LC}}_{\mathrm{Ch}}$ closes its gap 
at $m = 0,\pm 2$, and phase transitions occur.  We here focus on the transition at $m=-2$, when the gap closes at $\Gamma$, 
and the irreps at the both sides of the gap closing are given in Table~\ref{table:LCirreps14}.
Surprisingly, they are the same as the previous model with the Hamiltonain $H_{{\tilde{\nu}}}^{\mathrm{LC}}$ for the 
glide-symmetric TCI (${\tilde{\nu}} = 1, n_{Ch} = 0$) despite of the difference in the models.
Because the Chern number for $H^{\mathrm{LC}}_{\mathrm{Ch}}$ is obviously equal to two by construction, the glide-$Z_2$ invariant is zero in this model. 
Thus, the Chern insulator with the Chern number equal to $4n+2$ ($n$: integer) is also a higher-order TI ensured by inversion symmetry which has $(\mathbb{Z}_2, \mathbb{Z}_2, \mathbb{Z}_2; \mathbb{Z}_4) = (0, 0, 0; 2)$ in SG \#2 \cite{Ono2018prb98} (Fig.~\ref{fig:lc14}(d)).

The two models 
(\ref{eq:14z2_layer_construction}) and (\ref{eq:14Ch_layer_construction}) 
give the same values for the symmetry-based indicator.
This is natural from our result in (\ref{eq:14z2}). 
From Eq.~(\ref{eq:14z2}), the symmetry-based indicator is given by a combination of the 
glide-$Z_2$ invariant and the Chern number, 
and therefore, although the values of 
the 
glide-$Z_2$ invariant and the Chern number are different between the two models, the 
symmetry-based indicators are the same.

\begin{center}
\begin{table}
\begin{tabular}{c | c | c | c | c}
 & $\Gamma$ & Y & D & E \\ \hline
$H_{{\tilde{\nu}}}^{\mathrm{LC}} (m=-2.2)$ & $A_g + B_g$ & $A_g + B_g$ & $A_g + B_g$ & $A_g + B_g$ \\ \hline
$H_{{\tilde{\nu}}}^{\mathrm{LC}} (m=-1.8)$ & $A_u + B_u$ & $A_g + B_g$ & $A_g + B_g$ & $A_g + B_g$ \\ \hline
$H_{\mathrm{Ch}}^{\mathrm{LC}} (m=-2.2)$ & $A_g + B_g$ & $A_g + B_g$ & $A_g + B_g$ & $A_g + B_g$ \\ \hline
$H_{\mathrm{Ch}}^{\mathrm{LC}} (m=-1.8)$ & $A_u + B_u$ & $A_g + B_g$ & $A_g + B_g$ & $A_g + B_g$ 
\end{tabular}
\caption{Irreps for the two tight-binding models for the topological phases in SG \#14 from the layer construction.}
\label{table:LCirreps14}
\end{table}
\end{center}

\section{Conclusion}
In this paper we study the fate of the glide-symmetric $Z_2$ topological invariant when inversion symmetry is added while TRS is not enforced.
There are two ways to add inversion symmetry, leading to SGs 13 and 14.
In SG \#13, we derive the formula for the  glide-symmetric $Z_2$ topological invariant, and found that 
it is expressed in terms of the irreps at high-symmetry points. The symmetry-based indicator 
is $\mathbb{Z}_2\times \mathbb{Z}_2$, and one $\mathbb{Z}_2$ is this glide-symmetric $Z_2$ topological invariant, while the 
other $\mathbb{Z}_2$ is the Chern number modulo 2 along the normal vector of the glide plane. On the other hand, 
in SG \#14, the symmetry-based indicator 
is $\mathbb{Z}_2$, and we show that this $\mathbb{Z}_2$ is equal to the sum of the glide-symmetric $Z_2$ topological
invariant and a half of the Chern number. Here we note that in this SG \#14 the Chern number is always an even integer. It is interesting that the symmetry-based indicator gives a combination of
two different topological invariants, while it does not uniquely determine the values of the individual topological 
invariants. 
We also construct invariants for LCs in these SGs. The list of invariants for 
LC turns out to agree with the topological invariants discussed in the present paper. 
We found the list of eLCs for SGs \#13 and \#14, and construct tight-binding models for these eLCs. By direct calculations, we can show that they 
have nontrivial values either for the glide-$Z_2$ topological 
invariant and for the Chern number, 
which confirms the above conclusions. 

Furthermore, we show that both the glide-symmetric $Z_2$ TCI in SG \#13  is a higher-order TI. In SG \#14, the glide-symmetric $Z_2$ TCI  with the Chern number equal to
an integer multiple of four is a higher-order TI. In SG \#14, the Chern insulator with the Chern number equal to $4n+2$ ($n$: integer) and the trivial glide-$Z_2$ invariant is also a higher-order TI. 
Nevertheless, in the Chern insulators the surface is gapless, and one cannot call the latter case in SG \#14 a higher-order TI in the strict sense because the gapless hinge states are always hidden 
behind the gapless surface states due to the nonzero Chern number.

Our results will be useful for gaining insights for relating various topological phases. 
So far, various topological phases have been discovered by means of powerful methods, while their mutual relationships
are not obvious in general.  
For example, 
we have shown that 
the glide-symmetric $Z_2$ TCI becomes the higher-order TI in the presence of 
inversion symmetry, but such an equivalence is far from obvious from the known formula
(\ref{eq:z2glide}). Because the glide symmetry is one of the fundamental symmetries in 
crystals, and it is contained in many SGs, 
the present study will provoke studies on relating various topological invariants
in some SGs and those in their supergroups.

It must be an intriguing and promising topic to search real materials of the glide-symmetric spinless TCI and the higher-order TI.
Since our work have mainly discussed spinless systems, it does not suffer from the constraint of strong spin-orbit coupling when one does first-principle calculations.
Nonetheless, we have not found any candidate materials so far, since it requires breaking of TRS.
For realizing systems breaking of TRS but preserving glide symmetry or inversion symmetry, one candidate is magnetic insulators, such as insulators in magnetic field or localized spin systems.
Bosonic systems, such as magnonic systems or photonic crystal systems in magnetic field, are one of the best candidates to find specific materials realizing these physics \cite{Lu2016nphys12}.
From the results in this paper, it is also promising to discover new TCI materials or the higher-order TI materials.

\vspace{5mm}

\begin{acknowledgments}
We appreciate Tiantian Zhang for fruitful discussions.
This work was supported by Grant-in-Aid for Scientific Research
No, 26287062 and No, 18H03678.
H. K. is supported by Japan Society for the Promotion of Science (JSPS) KAKENHI Grant-in-Aid for JSPS Fellows Grant No.~JP17J10672.
K. S. is supported by PRESTO, JST (JPMJPR18L4). 
\end{acknowledgments}

\begin{widetext}
\appendix

\section{Sewing matrices with twofold rotational or screw symmetry}
\label{app:sewing}

We exploit several convenient quantities argued in Ref.~\onlinecite{Fang2012prb86} for evaluating the integral of the Berry curvature in the presence of additional $C_2$ rotational symmetry or screw symmetry.

First, we define a sewing matrix $w_{mn} (\bm{k})$.
The Bloch Hamiltonian $H(\bm{k})$ having a point group symmetry $R$ satisfies $\hat{R} H(\bm{k}) = H(R\bm{k}) \hat{R}$, where $\hat{R}$ is an operator
corresponding to the point-group operation $R$, and  $R\bm{k}$ is transformed from $\bm{k}$ by $\hat{R}$.
From the operator $\hat{R}$ which acts on the Bloch wavefunction $\ket{\psi_n(\bm{k})}= e^{i\bm{k}\cdot\bm{r}} \ket{u_n (\bm{k})}$, 
we define a corresponding operator $\tilde{R}$, which acts on the occupied cell-periodic eigenstates $\ket{u_n (\bm{k})}$.
Then, we define a unitary matrix as
\begin{equation}
w_{mn} (\bm{k}) \equiv \bra{u_m (R\bm{k})} \tilde{R} \ket{u_n (\bm{k})} , 
\end{equation}
where $m$ and $n$ run over the occupied states. 
At the wavevector $\bm{k}$ invariant under $R$, the Hamiltonian $H(\bm{k})$ commutes with $\hat{R}$, so that we can find the common eigenstates of the operators $H(\bm{k})$
and
$\tilde{R}$.
Here, the sewing matrix $w_{mn}$ should be diagonal at such a high-symmetry point $\bm{k}_i$,
\begin{equation}
w_{mn} (\bm{k}_i) = R_m (\bm{k}_i) \delta_{mn} ,
\end{equation}
where $R_m$ is the eigenvalue of $\hat{R}$ for the $m$th band.
Therefore, the determinant of $w_{mn}$ is given by the product of the eigenvalues of occupied eigenstates at $\bm{k}_i$,
\begin{equation}
\det [w (\bm{k}_i)] = \prod_{n \in \mathrm{occ}} R_n(\bm{k}_i) .
\label{eq:relation_sewmat_eigval}
\end{equation}
This quantity is gauge invariant because the determinant does not depend on the choice of basis.

Apart from the Berry phase, we also define a path-ordered exponential of the Berry connection,
\begin{equation}
\mathcal{U}_{\bm{k}_1 \bm{k}_2} = P \exp \left[ i \int^{\bm{k}_2}_{\bm{k}_1} \bm{A} (\bm{k}) \cdot d\bm{k} \right] ,
\label{eq:Wilson_loop}
\end{equation}
evaluated in the subspace of the occupied states, along the path from $\bm{k}_1$ to $\bm{k}_2$ not necessarily being a closed path.

We can relate some formulas containing the Berry phase in the main text to rotation or screw eigenvalues, following
Ref.~\onlinecite{Fang2012prb86}.
As an example, here we show the detail of Eq.~(\ref{eq:13rewrite_z2}).
The path-ordered integral along Y$^\prime \rightarrow \Gamma \rightarrow$ Y can be rewritten in terms of the $C_2$ eigenvalues.
Due to the $C_2$ symmetry,
\begin{align*}
e^{-i \gamma^-_{\mathrm{Y}^\prime \Gamma \mathrm{Y}}} &
=\det\left\{\mathcal{U}_{\mathrm{Y}\Gamma}\mathcal{U}_{\Gamma\mathrm{Y'}}\right\} \nonumber \\
&= \det \left\{ \mathcal{U}_{\mathrm{Y}\Gamma}w^-_{C_2} (\Gamma) \mathcal{U}_{\Gamma\mathrm{Y}}\left( w^-_{C_2} (\mathrm{Y}) \right)^{-1}  \right\} \nonumber \\
&= \det \left[ w^-_{C_2} (\Gamma) \left( w^-_{C_2} (\mathrm{Y}) \right)^{-1} \right] =  \prod_{i \in \mathrm{occ}} \frac{\zeta_i^- (\Gamma)}{\zeta_i^- (\mathrm{Y})} ,
\end{align*}
where $w^-_{C_2} (\Gamma)$ is the sewing matrix for the $C_2$ operator, restricted within the $g_-$ subspace at the high-symmetry point $P$, and 
$\zeta_i^-(P)$ is the $C_2$ eigenvalue for the $g_-$ sector at the high-symmetry point $P$.

\section{Layer construction}
\label{sec:LC}
In this Appendix, we consider a LC in class A for SGs \#7, \#13, and \#14.
In Ref.~\onlinecite{Song2018ncommun9}, a LC is introduced
for time-reversal-invariant systems, and in this previous work, 
each layer is decorated with a 2D TI and a 2D mirror TCI. In contrast, in the present 
paper, because we do not assume 
TRS, we decorate each layer with a 2D Chern insulator. 
We then construct all invariants based on each space-group operations, based solely on 
geometric properties of the layers. Then we show that these set of invariants are in 
complete agreement with the topological invariants based on $k$-space geometry discussed
in the main text of the paper. 
Based on these arguments, we show eLCs, which 
constitute a basis for the values of  topological invariants.

\subsection{setup}
A layer $(mnl;d)$  is a set of planes given by the Miller indices $(mnl)$ displaced from the origin by $d$ ($0 \le d < 1$). It is given by 
\begin{align}
(mnl;d) &= \left \{ \bm{r} | \bm{r} \cdot (m\bm{b}_1 + n\bm{b}_2 + l\bm{b}_3 ) = 2\pi (d+q), \ q \in \mathbb{Z}\right \} \nonumber \\
&= \left \{ \bm{r} | \frac{mx}{a} + \frac{ny}{b} + \frac{lz}{c} = (d+q), \ q \in \mathbb{Z} \right \} ,
\end{align}
where $\bm{b}_i$'s are reciprocal vectors corresponding to the primitive vectors $\bm{a}_1 = (a,0,0), \bm{a}_2 = (0,b,0), \bm{a}_3 = (0,0,c)$ and  
the integer $q$ is introduced because of the translation symmetry.

We then formulate general LCs, in the similar way as in Ref.~\onlinecite{Song2018ncommun9}.
In a given SG $\mathcal{G}$, symmetry property of a layer $L$ is described by its little group, which is defined as a subgroup of $\mathcal{G}$ that leaves $L$ invariant:
\begin{equation}
\mathcal{S} (L) = \{ s \in \mathcal{G} | sL = L \} .
\end{equation}
Then, $\mathcal{S}$ must also be a SG containing the full translation subgroup since by definition any lattice vector would translate the layer.
We can get a finite number of cosets of $\mathcal{S}$ for SG $\mathcal{G}$,
\begin{equation}
\mathcal{G} = g_0 \mathcal{S} + g_1 \mathcal{S} + \cdots .
\end{equation}
Thus, we have a set of symmetric layers
\begin{equation}
\{ g_0 L, g_1 L , \cdots \} ,
\end{equation}
by applying all the coset representatives on $L$.
An eLC is obtained by decorating the layers in $\{ g_0 L, g_1 L, \cdots \}$ with a 2D Chern insulator with a Chern number $+1$.

\subsection{topological invariants for glide-symmetric systems}
First of all, we consider how a general layer $(mnl;d)$ gives topological invariants in SG \#7.
Then, we expand our argument to SG \#13 and SG \#14 by adding inversion symmetry.
Then, inversion symmetry and either $C_2$ rotational symmetry or $C_2$ screw symmetry are added as generators.
We briefly explain how topological invariants can be calculated or added by these symmetries.
Here we construct topological invariants associated with respective symmetry operations. 
We focus on one of the symmetry operations in the given SG, and in constructing a
topological invariant associated with the symmetry operation, we neglect all the other symmetry operations in SG. We then construct possible LCs compatible with SG,
and see whether each LC can be trivialized by deforming it while keeping its invariance
under the focused symmetry operation. 
For example, in considering a glide invariant $\delta_{\mathrm{g}}$, we construct LCs satisfying the glide operation only,  and then deform the LCs while keeping the glide symmetry, to
see whether the LCs can be trivialized. This consideration leads to the definition of the 
glide topological invariant $\delta_{\mathrm{g}}$.

\subsubsection{Chern invariant}
The Chern number along a reciprocal lattice plane $\mathcal{D}_{i}$ normal to one of the primitive vectors $\bm{a}_{i=1,2,3}$ is 
written as 
defined as
\begin{align}
\delta_{n_\mathrm{Ch},i}=\frac{1}{2\pi}\int_{\mathcal{D}_{i}} d^2\bm{k}\ \bm{n}\cdot\bm{\Omega}_{\bm{k}} ,
\end{align}
where $\bm{n}_i\equiv \bm{a}_i/|\bm{a}_i|$ is a unit vector along $\bm{a}_i$ and the integral is taken over the
2D Brillouin zone on the plane $\mathcal{D}_i$.  
For a layer $L=(mnl;d)$, it is equal to the number of intersections of $L=(mnl;d)$ with 
an axis parallel to the $\bm{a}_{i}$ within the unit cell (Fig.~\ref{fig:1-LC} (a)). 
One can see easily its intersections with $\bm{a}_{1,2,3}$ as
\begin{equation}
\frac{d+q}{m}, \ \frac{d+q}{n}, \ \frac{d+q}{l} ,
\end{equation}
respectively.
Thus, the Chern numbers $\bm{n}_{\mathrm{Ch}}$ associated with three primitive lattice vector $\bm{a}_{1,2,3}$ should be given by
\begin{align}
\delta_{n_\mathrm{Ch}, 1} (E) &= \sum_{L \in E} m_L , \\
\delta_{n_\mathrm{Ch}, 2} (E) &= \sum_{L \in E} n_L , \\
\delta_{n_\mathrm{Ch}, 3} (E) &= \sum_{L \in E} l_L ,
\end{align}
where $E$ is a general LC. Namely, it can be written as
\begin{align}
\delta_{n_\mathrm{Ch}, i} (E) &= \frac{1}{2\pi}\sum_{L \in E} {\bf g}_L\cdot \bm{a}_i ,
\end{align}
where ${\bf g}_L = m\bm{b}_1 + n\bm{b}_2 + l\bm{b}_3$ for the layer $L \in E$.

\subsubsection{glide invariant $\delta_{\rm g}$}
In addition, we can find a glide invariant for a glide-symmetric systems based on the LC
in the similar way as in time-reversal-invariant systems treated in 
Ref.~\onlinecite{Song2018ncommun9}. 
In glide-invariant systems, there are two glide planes in a unit cell, which 
are displaced by a half of the primitive translation vector. In constructing the glide invariant $\delta_{\rm g}$, we should first choose one of the glide planes, and we fix this choice throughout our theory.

In our case each layer is decorated with the 2D Chern insulator because TRS
is not assumed, whereas in Ref.~\onlinecite{Song2018ncommun9}, each layer is decorated with the 2D TIs. Nonetheless, nontrivial LC configurations are similar, classified into two types.
In the first type of the nontrivial LC configuration, the glide plane is included in one of the layers in the LC (Fig.~\ref{fig:1-LC} (d)). It is characterized by the glide-occupation number (glide-ON) $N_{m,{\bf t}_{\|}}^{\rm o}(L)$ where $m$ specifies the glide plane. 
In the second type of the nontrivial LC configuration, there is a pair of layers which are connected by the glide operation (Fig.~\ref{fig:1-LC} (b)), which can be continuously deformed into the previous case (see Fig.~\ref{fig:1-LC} (c)). It is represented by a  glide-stacking number (glide-SN)
$N_{m,{\bf t}_{\|}}^{\rm s}(L)$. Similarly to the cases in 
Ref.~\onlinecite{Song2018ncommun9}, two pairs of such layers can be trivialized. Therefore, 
the resulting topological invariant is defined modulo 2, and one can write the glide invariant $\delta_{\rm g}$ as 
\begin{align}
&\delta_{\mathrm{g}} (E) = \sum_{L \in E} (N_{m,{\bf t}_{\|}}^{\rm o}(L)+N_{m,{\bf t}_{\|}}^{\rm s}(L)) \pmod 2 , \\
&N_{m,{\bf t}_{\|}}^{\rm o}(L)=\left\{\begin{array}{ll}
1& \text{if}\  m\in L\\ 0 &\text{otherwise}\end{array}\right. , \label{eq:Nmt}\\
&
N_{m,{\bf t}_{\|}}^{\rm s}(L)=\frac{1}{2\pi}|{\bf t}_{\|}\cdot{\bf g}_{L}| .
\end{align}
We again emphasize that in Eq.~(\ref{eq:Nmt}), $m$ refers to the glide plane which 
we have chosen from the two different choices of the glide plane.

\begin{figure}
\centering
\includegraphics[scale=0.25]{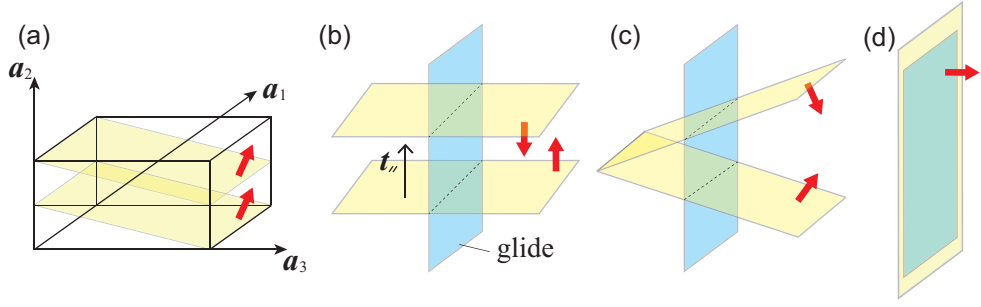}
\caption{(Color online) Chern invariant $\delta_{n_{\text{Ch}}, i=1,2,3}$ and glide invariant $\delta_{\rm g}$ in layer constructions.  The yellow planes are the planes in the layer construction, and the blue planes are
glide planes. The red arrows represent the directions of the reciprocal lattice vector $\bm{G}=m\bm{b}_1+n\bm{b}_2+l\bm{b}_3$, which specifies the 
orientation of the normal vector of the plane giving a positive Chern number $+1$. (a) Chern invariant $\delta_{n_{\text{Ch}},2}$
along the $\bm{a}_3$-$\bm{a}_1$ plane is determined as the number of intersections between the planes in the layer, and the ${\bf a}_2$ axis within the lattice constant. The figure shows the case with
$(021;0)$, which yields $\delta_{n_\mathrm{Ch}, 1}=0, \delta_{n_\text{Ch},2}=2$, and $\delta_{n_\mathrm{Ch}, 1}=1$.  (b)-(d) Glide invariant $\delta_{\text{g}}$. (b) shows one of the nontrivial configurations of the layer. Two planes are perpendicular to the glide plane, and 
they are related by the glide operation, which means that they are separated by a glide vector ${\bf t}_{\|}$. Note that 2${\bf t}_{\|}$ is a lattice translation vector. 
(c) The configuration in (b) can be deformed into that in (d). 
(d) shows another nontrivial configuration, where a plane in the layer matches with the glide plane. }
\label{fig:1-LC}
\end{figure}

\subsubsection{inversion invariant $\delta_{\rm i}$} 
Systems with inversion symmetry have eight inversion centers within a unit cell. We first choose one of the inversion centers, and we fix the choice throughout our theory. 
If a layer in the LC includes the inversion center (Fig.~\ref{fig:2-LC} (a)), this LC cannot be trivialized as
long as the inversion symmetry is preserved. Meanwhile, if the inversion center is
occupied by two layers decorated with the 2D Chern insulator, one can trivialize this
configuration up to the creation of the atomic insulator at the center of inversion. Thus the inversion invariant $\delta_\mathrm{i}$ can be 
written as a inversion-ON $N_{i}^{\rm o}(L)$ of $L$ modulo 2:
\begin{align}
&\delta_{\mathrm{i}} (E) = \sum_{L \in E} N_{i}^{\rm o}(L) \pmod 2 , \\
&N_{i}^{\rm o}(L)=\left\{\begin{array}{ll}
1& \text{if}\  i\in L\\ 0 &\text{otherwise}\end{array}\right. .
\end{align}

\subsubsection{$C_2$ rotation invariant $\delta_{\rm r}$} 
In a similar manner, the $C_2$ rotation  invariant $\delta_{\rm r}$ is obtained.
A layer of the 2D Chern insulator is $C_2$ invariant only when the plane is perpendicular to the $C_2$ axis (Fig.~\ref{fig:2-LC} (b)), and this layer  
can be trivialized, as we see from Fig.~\ref{fig:2-LC} (c).
Consequently, the $C_2$ rotation invariant is always trivial:
\begin{align}
&\delta_{\mathrm{r}} (E) = 0.
\end{align}
It is a remarkable result, because the $C_2$ rotation invariant is always trivial in the system without TRS, while it is not in the presence of TRS \cite{Song2018ncommun9}.
In the presence of TRS, 2D TI layers are used for LCs, and the $C_2$ rotation invariant becomes nontrivial when a TI layer includes the $C_2$ axis.
On the other hand, in our case where TRS is not enforced, layers of a 2D Chern insulator are used.
In this case, the configuration where a layer of a 2D Chern insulator includes the $C_2$ axis is not $C_2$ invariant, unlike the case with TRS. 
Thus, there is no topologically nontrivial configuration for the $C_2$ rotation invariant.

\begin{figure}
\centering
\includegraphics[scale=0.25]{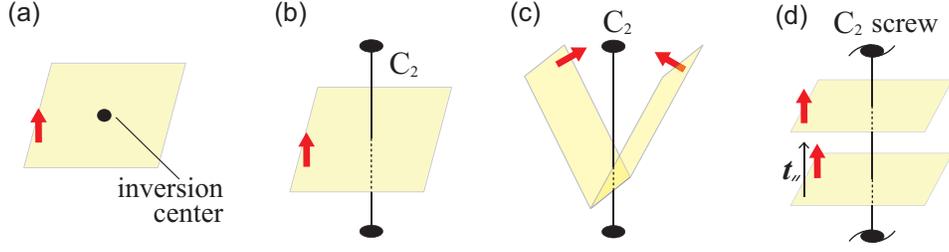}
\caption{(Color online) Inversion invariant $\delta_{\rm i}$, $C_2$ rotation invariant $\delta_{\text{r}}$ and $C_2$ screw invariant $\delta_{\text{s}}$ in layer constructions. (a) Inversion invariant $\delta_{\text{i}}$. The plane includes the inversion center, which cannot be trivialized. (b)(c) $C_2$ rotation invariant $\delta_{\text{r}}$. In (b), the plane is perpendicular to the $C_2$ rotation axis. This configuration is trivial, because as shown in (c) it can be trivialized while keeping the $C_2$ rotation symmetry. (d) $C_2$ screw invariant $\delta_{\text{s}}$. The two planes in the layer are perpendicular to the $C_2$ screw axis, and they are
related by the screw operation. ${\bf t}_{\|}$ is a screw vector, and is a half of a lattice translation vector. This configuration is nontrivial.}
\label{fig:2-LC}
\end{figure}

\subsubsection{$C_2$ screw invariant $\delta_{\rm s}$}
In systems with $C_2$ screw symmetry
the topological invariant is defined by a screw-SN. 
The nontrivial LC configuration is the LC with a pair of layers
combined by the screw operation. In such cases, one pair of layers combined by screw symmetry 
cannot be trivialized (Fig.~\ref{fig:2-LC}(d)), while two pairs can be trivialized. Thus we get
\begin{align}
&\delta_{\mathrm{s}} (E) = \sum_{L \in E} N_{s,{\bf t}_{\|}}^{\rm s}(L) \pmod 2 , \\
&
N_{s,{\bf t}_{\|}}^{\rm s}(L)=\frac{1}{2\pi}|{\bf t}_{\|}\cdot{\bf g}_{L}| ,
\end{align}
where ${\bf t}_{\|}$ is a screw vector,  and is equal to a half of a lattice translation vector.

\subsection{Convention dependence of topological invariants}

As remarked in Ref.~\onlinecite{Song2018ncommun9}, topological invariants in the previous section depend on the conventions.
For example, there are eight inversion centers in a unit cell in the presence of inversion symmetry, and we should choose one inversion center among them in defining an inversion invariant $\delta_{\mathrm i}$.
When we choose the inversion center at the origin, $(mnl;0)$ passing the origin has $\delta_{\mathrm i} = 1$ whereas $(mnl; \frac{1}{2})$ has $\delta_{\mathrm i} = 0$ because it does not pass the inversion center.
On the other hand, when we choose the inversion center at $(0, 0, \frac{1}{2})$, $(mnl;0)$ has $\delta_{\mathrm i} = 0$ whereas $(mnl; \frac{1}{2})$ has $\delta_{\mathrm i} = 1$.
Thus, the inversion invariant $\delta_{\mathrm i}$ depends on the convention on the choice of the inversion center among the eight inversion centers in the unit cell.
Meanwhile, when we consider an LC with two eLC passing all eight inversion centers in a unit cell, the inversion invariant is independent of the choice of the inversion center, and the system is considered to be topologically nontrivial. 
Similarly, the glide-ON is also dependent on the convention. There are two glide planes within the unit cell, displaced from each other by half the primitive translation vector, and 
the glide invariant $\delta_{\rm g}$ depends on  the choice.  
When the LC occupies both of these glide planes odd times each, 
the glide invariant $\delta_{\rm g}$ is nontrivial, irrespective of the choice of the glide plane, as
 is similar to the cases with time-reversal symmetry in Ref.~\onlinecite{Song2018ncommun9}

In our cases in SG \#13 as an example, the configuration in Fig.~\ref{fig:lc13}(a) has a nontrivial inversion 
invariant $\delta_{\rm i}$, irrespective of the convention of the inversion center since the layers pass all the eight inversion centers in the unit cell. Meanwhile, in the configuration in Fig.~\ref{fig:lc13}(b), the  inversion invariant $\delta_{\rm i}$ depends on the convention.
In SG \#14, both of the two configurations in Figs.~\ref{fig:lc14}(a)(b) have a convention-independent
nontrivial   inversion 
invariant $\delta_{\rm i}$,

\subsection{Invariants for layer constructions}
Based on the definitions of the invariants given above, one can 
calculate their values for LCs in SGs \#7, \#13 and \#14.
We note that in the glide symmetry and in the inversion symmetry, we choose
the glide planes and inversion centers to be the standard ones given in the main text,
which are the same as those in the Bilbao Crystallographic Server \cite{Aroyo2006bilbao}.

\subsubsection{SG \#7}
In SG \#7, only the Chern invariants $\delta_{n_{\mathrm{Ch}}, i=1,2,3}$ and the 
glide invariant $\delta_{\rm g}$ are defined. 
In the calculation
we note that the glide operator $G_y = \{ m_{010} | \bm{a}_3 / 2 \}$ transforms the layer $L=(mnl;d)$ to
\begin{align}
G_y L &= \left\{ g_y \bm{r} | \frac{m}{a}x - \frac{n}{b}y + \frac{l}{c} \left( z+\frac{c}{2} \right) = d+q, \ q \in \mathbb{Z} \right\} \nonumber \\
&= \left\{ g_y \bm{r} | \frac{(-m)}{a}x + \frac{n}{b}y + \frac{(-l)}{c} z = \frac{l}{2}-d-q, \ q \in \mathbb{Z} \right\} \nonumber \\
&= \left( \bar{m} n \bar{l} \Big| \frac{l}{2} - d \right) ,
\label{apeq:glide_layer}
\end{align}
Note that it is not written as $(m\bar{n}l|d-\frac{l}{2})$. 
Therefore, the Chern numbers for the LC $L(mnl;d)$ in SG \#7 are given as
\begin{align}
\delta_{n_\mathrm{Ch}, 2} (E) &= \sum_{L \in E} N_{\mathrm{Ch}} (L) \pmod{\mathbb{Z}} , \\
N_{\mathrm{Ch}} (L) &=
\begin{cases}
n_L &  (m_L=0,\ l_L=0, \ d_L=0,1/2)\\ 2n_L & (\text{otherwise})
\end{cases}
, \\
\delta_{n_\mathrm{Ch}, 1} (E) &= 0 , \\
\delta_{n_\mathrm{Ch}, 3} (E) &= 0 ,
\end{align}
where the classification is necessary for  distinguishing the cases $L=G_yL$ and $L\neq
G_y L$. Here $\delta_{n_\mathrm{Ch}, 1} (E) = 0$ and 
$\delta_{n_\mathrm{Ch}, 3} (E) = 0$ follows from glide symmetry. 

The glide invariant is given by 
\begin{align}
&\delta_{\rm g}\equiv \left\{
\begin{array}{ll}
1 & (m_L=0,\ l_L=0,\ d_L=0)\\
l_L & (\text{otherwise})
\end{array}\right.
\label{eq:dgsg7}\end{align}
modulo 2. We note that the glide plane is chosen to be $y=0$, and not $y=1/2$. That
is why (\ref{eq:dgsg7}) does not contain $d_L=1/2$ in the first case of $\delta_{\rm g}=1$.

\subsubsection{SG \#13}
SG \#13 is constructed from SG \#7 by adding the inversion symmetry. 
The inversion transforms the layer $(mnl;d)$ into $(mnl;-d)$. 
This additional inversion symmetry doubles the number of layers, 
if the additional layers are not identical with the original ones, 
For the layer $L(mnl;d)$, the values of the Chern invariants are obtained as follows:
\begin{align}
&\delta_{n_\mathrm{Ch}, 1}=0,\\
&\delta_{n_\mathrm{Ch}, 3}=0,\\
&\delta_{n_\mathrm{Ch}, 2}=\left\{
\begin{array}{ll}
n_L & (m_L=0,\ l_L=0,\ d_L=0,1/2)\\
2n_L & (m_L=0,\ l_L=0,\ d_L\neq0,1/2)\\
2n_L & ((m_L,l_L)\neq (0,0),\ d_L=0,1/2)\\
4n_L & ((m_L,l_L)\neq (0,0),\ d_L\neq 0,1/2)\\
\end{array}\right. .
\end{align}

The glide invariant and the inversion invariant are calculated, 
and they turn out to be exactly equal;
\begin{align}
&\delta_{\rm g}=\delta_{\rm i}\equiv \left\{
\begin{array}{ll}
1 & (m_L=0,\ l_L=0,\ d_L=0)\\
l_L & ((m_L,l_L)\neq(0,0),\ d_L=0,1/2)\\
0 & (\text{otherwise})
\end{array}\right.
\end{align}
modulo 2.

\subsubsection{SG \#14}
SG \#14 is constructed from SG \#7 by adding inversion symmetry. 
For the layer $L(mnl;d)$, the values of the Chern invariants are obtained as follows:
\begin{align}
&\delta_{n_\mathrm{Ch}, 1}=0,\\
&\delta_{n_\mathrm{Ch}, 3}=0,\\
&\delta_{n_\mathrm{Ch}, 2}=\left\{
\begin{array}{ll}
n_L & (m_L=0,\ n_L=\text{even},\ l_L=0,\ d_L=0,1/2)\\
2n_L& (m_L=0,\ n_L=\text{odd},\ l_L=0,\ d_L=1/4,3/4)\\
2n_L& ((m_L,l_L)\neq(0,0),\ n_L=\text{odd},\ d=0,1/2)\\
4n_L & (\text{otherwise})\\
\end{array}\right. .
\end{align}

The glide invariant is given by
\begin{align}
&\delta_{\rm g}\equiv \left\{
\begin{array}{ll}
1 & (m_L=0,\ l_L=0,\ d_L\equiv \pm\frac{n_L}{4}\ \pmod 1)\\
1 & (l_L=\text{odd},\ d_L=0,1/2)\\
0 & (\text{otherwise})
\end{array}\right.
\end{align}
modulo 2. 

The inversion invariant is given by
\begin{align}
&\delta_{\rm i}\equiv \left\{
\begin{array}{ll}
1 & (m_L=0,\ n_L=\text{even},\ l_L=0,\ d_L=0)\\
1 & (n_L+l_L=\text{odd},\ d_L=0,1/2)\\
0 & (\text{otherwise})
\end{array}\right.
\end{align}
modulo 2. 

The $C_2$ screw invariant is given by
\begin{align}
&\delta_{\rm s}\equiv \left\{
\begin{array}{ll}
1 & (n_L=\text{odd},\ d_L=0,1/2)\\
1 & (m_L=0,\ n_L=\text{odd},\ l_L=0,\ d_L=0,1/4,1/2,3/4)\\
1 & (m_L=0,\ n_L=2\times\text{odd},\ l_L=0,\ l_L=0, 1/2)\\
0 & (\text{otherwise})
\end{array}\right.
\end{align}
modulo 2. 

They are related by the following equations:
\begin{align}
&\delta_{\rm i}\equiv\frac{1}{2}\delta_{n_\mathrm{Ch}, 2}+\delta_{\rm g} \pmod{2},\\
&\delta_{\rm s}\equiv
\frac{1}{2}\delta_{n_\mathrm{Ch}, 2} \pmod{2} .
\end{align}

\subsection{Elementary layer constructions}
Following the argument in Ref.~\onlinecite{Song2018ncommun9}, we can find eLCs in SGs \#7, \#13, and \#14 summarized in Table~\ref{table:eLCs}. 
All SGs listed here have two independent topological invariants 
as we have seen in this paper, Therefore, each SG has two eLCs.
\begin{table}[h] 
$$
\begin{array}{c | c | c | c c c c}
\mathrm{SG} & \mathrm{eLC} & Z_{2,2,2,4} & \multicolumn{4}{c}{\mathrm{Invariants}} \\ \hline
 & (mnl;d) &  & \delta_{n_\mathrm{Ch}, i} & g^{010}_{00\frac{1}{2}} & & \\ \hline
\#7 \ \ (Pc) & 001; d_0 & \mathrm{N} \slash \mathrm{A} & 000 & 1 & & \\ 
 & 010; 0 & & 010 & 0 & & \\ \hline
 & (mnl;d) & & \delta_{n_\mathrm{Ch}, i} & g^{010}_{00\frac{1}{2}} & i & 2^{010} \\ \hline
\#13 \ (P2/c) & 001; 0 & 0002 & 000 & 1 & 1 & 0 \\
 & 010; \frac{1}{2} & 0102 & 010 & 0 & 0 & 0 \\ \hline
 & (mnl;d) & & \delta_{n_\mathrm{Ch}, i} & g^{010}_{0\frac{1}{2}\frac{1}{2}} & i & 2^{010}_{0\frac{1}{2}\frac{1}{2}} \\ \hline
\#14 \ (P2_1/c)  & 001; 0 & 0002 & 000 & 1 & 1 & 0 \\
 & 020; 0 & 0002 & 020 & 0 & 1 & 1 \\ \hline
\end{array}
$$
\caption{Summary of elementary layer construction, symmetry-based indicators, and topological invariants for SGs \#7, \#13, and \#14.}
\label{table:eLCs}
\end{table}

\section{Layer construction for SG \#13}
\label{app:layer13}
In this section we show the detailed construction of the models for the glide-symmetric TCI phase and the 
Chern insulator phase in SG \#13. 
The symmetry generators are given by 
\begin{align}
\hat{G}_y = \left\{ m_y \bigg | \frac{1}{2} \hat{\bm z} \right\}, \quad \hat{I}  = \{ I | \bm{0} \} .
\end{align}

\subsection{Glide-symmetric TCI phase in the layer construction $(001;0)$ (${\tilde{\nu}} = 1, n_{\mathrm{Ch}}=0$)}
We first calculate the topological invariants for the LC $(001;0)$.
The model Hamiltonian is given by putting a Chern insulator with $n_{\mathrm{Ch}}=+1$ on the $xy$ plane at $z=0$ with inversion symmetry, which is written as 
\begin{align*}
&\sum_{x,y \in \mathbb{Z}} \left[ 
\psi^{\dag}(x+1,y) \frac{\sigma_z+i \sigma_x}{2} \psi(x,y)
+\psi^{\dag}(x,y+1) \frac{\sigma_z+i \sigma_y}{2} \psi(x,y)
+ \mathrm{h.c.} 
\right]
+ \sum_{x,y \in \mathbb{Z}} m \psi^{\dag}(x,y) \sigma_z \psi(x,y),
\end{align*}
\begin{align*}
\hat I \psi^{\dag}(x,y) \hat I^{-1} = \psi^{\dag}(-x,-y) \sigma_z, 
\end{align*}
and making copies by the glide transformation 
\begin{align*}
\hat G_y \psi^{\dag}(x,y,z) \hat G_y^{-1}
= 
\psi^{\dag}\left(x,-y,z+\frac{1}{2}\right).
\end{align*}
The Hamiltonian reads 
\begin{align*}
\hat H
&=\sum_{x,y,z \in \mathbb{Z}} \left[ 
\psi^{\dag}(x+1,y,z) \frac{\sigma_z+i \sigma_x}{2} \psi(x,y,z)
+\psi^{\dag}(x,y+1,z) \frac{\sigma_z+i \sigma_y}{2} \psi(x,y,z)
+ \mathrm{h.c.} 
\right]
+ \sum_{x,y,z \in \mathbb{Z}} m \psi^{\dag}(x,y,z) \sigma_z \psi(x,y,z) \\
&+\sum_{x,y,z \in \mathbb{Z}} \left[ 
\psi^{\dag}(x+1,-y,z+\frac{1}{2}) \frac{\sigma_z+i \sigma_x}{2} \psi(x,-y,z+\frac{1}{2})
+\psi^{\dag}(x,-y-1,z+\frac{1}{2}) \frac{\sigma_z+i \sigma_y}{2} \psi(x,-y,z+\frac{1}{2})
+ \mathrm{h.c.} 
\right] \\
&+ \sum_{x,y,z \in \mathbb{Z}} m \psi^{\dag}(x,-y,z+\frac{1}{2}) \sigma_z \psi(x,-y,z+\frac{1}{2}) \\
&=\sum_{x,y,z \in \mathbb{Z}} \left[ 
\psi^{\dag}(x+1,y,z) \frac{\sigma_z+i \sigma_x}{2} \psi(x,y,z)
+\psi^{\dag}(x,y+1,z) \frac{\sigma_z+i \sigma_y}{2} \psi(x,y,z)
+ \mathrm{h.c.} 
\right]
+ \sum_{x,y,z \in \mathbb{Z}} m \psi^{\dag}(x,y,z) \sigma_z \psi(x,y,z) \\
&+\sum_{x,y,z \in \mathbb{Z}} \left[ 
\psi^{\dag}(x+1,y,z+\frac{1}{2}) \frac{\sigma_z+i \sigma_x}{2} \psi(x,y,z+\frac{1}{2})
+\psi^{\dag}(x,y+1,z+\frac{1}{2}) \frac{\sigma_z-i \sigma_y}{2} \psi(x,y,z+\frac{1}{2})
+ \mathrm{h.c.} 
\right] \\
&+ \sum_{x,y,z \in \mathbb{Z}} m \psi^{\dag}(x,y,z+\frac{1}{2}) \sigma_z \psi(x,y,z+\frac{1}{2}) .
\end{align*}
Let us introduce the $k$-space basis 
\begin{align}
\Psi^{\dag}(k_x,k_y,k_z)
\equiv 
\sum_{x,y,z \in \mathbb{Z}} (\psi^{\dag}(x,y,z),\psi^{\dag}(x,y,z+\frac{1}{2})) e^{i (k_x x + k_y y + k_z z)}. 
\end{align}
The glide and inversion operations are represented as 
\begin{align*}
\hat G_y \Psi^{\dag}(k_x,k_y,k_z) \hat G_y^{-1}
= 
\sum_{x,y,z \in \mathbb{Z}} (\psi^{\dag}(x,-y,z+\frac{1}{2}),\psi^{\dag}(x,-y,z+1)) e^{i (k_x x + k_y y + k_z z)} 
= \Psi^{\dag}(k_x,-k_y,k_z) \begin{pmatrix}
0 & e^{-i k_z} \\
1 & 0
\end{pmatrix}_{\tau}, \\
\hat I \Psi^{\dag}(k_x,k_y,k_z) \hat I^{-1}
= 
\sum_{x,y,z \in \mathbb{Z}} (\psi^{\dag}(-x,-y,-z),\psi^{\dag}(-x,-y,-z-\frac{1}{2})) \sigma_z e^{i (k_x x + k_y y + k_z z)} 
= \Psi^{\dag}(-k_x,-k_y,-k_z) \sigma_z \begin{pmatrix}
1 & 0 \\
0 & e^{-i k_z}
\end{pmatrix}_{\tau}.
\end{align*}
Thus, the $k$-space Hamiltonian and symmetry operators are given by
\begin{align*}
&H(k_x,k_y,k_z)
= \sin k_x \sigma_x + \sin k_y \sigma_y \tau_z + (m + \cos k_x + \cos k_y) \sigma_z, \\
&G_y(k_x,k_y,k_z) = \begin{pmatrix}
0 & e^{-i k_z} \\
1 & 0 \\
\end{pmatrix}_{\tau}, \qquad 
I(k_x,k_y,k_z) = \sigma_z \begin{pmatrix}
1 & 0 \\
0 & e^{-i k_z}
\end{pmatrix}_{\tau}.
\end{align*}
The $C_2$ rotation is given by 
\begin{align*}
C_2(k_x,k_y,k_z)
\equiv 
G_y(-k_x,-k_y,-k_z) I(k_x,k_y,k_z) = \sigma_z \begin{pmatrix}
0 & 1 \\
1 & 0
\end{pmatrix}_\tau.
\end{align*}
Let us compute the indicator for $-2 < m < 0$.
At the high-symmetry points, the Hamiltonians and the symmetry operators within the occupied states are 
shown in the following table. Here an overall positive coefficients are omitted for simplicity.
\begin{align*}
&\begin{array}{c|ccccccc}
P & H(P) & G_y(P) & C_2(P)|_{\rm occ} \\
\hline
\Gamma & \sigma_z & \tau_x & -\tau_x \\
\mathrm{Y} & -\sigma_z & \tau_x & \tau_x \\
\mathrm{Z} & -\sigma_z & \tau_x & \tau_x \\
\mathrm{C} & -\sigma_z & \tau_x & \tau_x \\
\end{array} .
\end{align*}
Therefore, the glide-$Z_2$ invariant becomes 
\begin{align*}
(-1)^{\tilde{\nu}}
=
\prod_{i \in {\rm occ}} \frac{\zeta_i^-(\Gamma) \zeta^+_i(\mathrm{C})}{\zeta_i^-(\mathrm{Y}) \zeta^+_i(\mathrm{Z})}
= \frac{(-1) \times 1}{1 \times 1}
= -1,
\end{align*}
i.e., $\tilde{\nu}=1$. As a result, $\tilde{\nu}$ is equal to 1 for $|m| < 2$, and $\tilde{\nu}=0$ for other values of $m$. 

\subsection{Chern insulator phase in the layer construction $(010;\frac{1}{2})$ (${\tilde{\nu}} = 0, n_{\mathrm{Ch}}=1$)}
We here calculate the topological invariants for the LC $(010;\frac{1}{2})$.
The model Hamiltonian is given by putting a Chern insulator with $n_{\mathrm{Ch}}=1$ on the $xz$ plane at $y=\frac{1}{2}$ with inversion symmetry, which is written as 
\begin{align*}
&\sum_{x,2z \in \mathbb{Z}} \left[ 
\psi^{\dag}(x,z+\frac{1}{2}) \frac{\sigma_z+i \sigma_x}{2} \psi(x,z)
+\psi^{\dag}(x+1,z) \frac{\sigma_z+i \sigma_y}{2} \psi(x,z)
+ \mathrm{h.c.} 
\right]
+ \sum_{x,z \in \mathbb{Z}} m \psi^{\dag}(x,z) \sigma_z \psi(x,z),\\
&\hat I \psi^{\dag}(x,z) \hat I^{-1} = \psi^{\dag}(-x,-z) \sigma_z.
\end{align*}
As noted in Sec. IV C, we first take the length of the unit cell along the $z$ direction to be $1/2$, in order
to
make the model compatible with the glide symmetry. We then 
 make copies by the the translation $T_y$, 
\begin{align*}
\hat T_y \psi^{\dag}(x,y,z) \hat T_y^{-1}
= \psi^{\dag}(x,y+1,z),
\end{align*}
and the glide operation is given by 
\begin{align*}
\hat G_y \psi^{\dag}(x,y,z) \hat G_y^{-1}
= 
\psi^{\dag}(x,-y,z+\frac{1}{2}).
\end{align*}
In the momentum space, the Hamiltonian and symmetry operators are represented 
\begin{align*}
&H(k_x,k_y,k_z)
=\sin \frac{k_z}{2}  \sigma_x + \sin k_x\sigma_y + (m + \cos k_x + \cos \frac{k_z}{2}) \sigma_z, \\
&G_y(k_x,k_y,k_z) =e^{-ik_y}e^{-ik_z/2}, \qquad 
I(k_x,k_y,k_z) = e^{-ik_y}\sigma_z .
\end{align*}
This is a two-band model, when the primitive vector along the $z$ axis to be $(0,0,1/2)$. 
One can formally regard the primitive vector along the $z$ axis to be $(0,0,1)$, 
which renders 
this model to a four-band model. 
Let us introduce the $k$-space basis 
\begin{align}
\Psi^{\dag}(k_x,k_y,k_z)
\equiv 
\sum_{x,y+\frac{1}{2},z \in \mathbb{Z}} (\psi^{\dag}(x,y,z),\psi^{\dag}(x,y,z+\frac{1}{2})) e^{i (k_x x + k_y y + k_z z)}. 
\end{align}
Then the Hamiltonian is given by 
\begin{align*}
&H(k_x,k_y,k_z)
=\left[\sin k_x \sigma_y  + (m + \cos k_x ) \sigma_z\right]+
( \cos \frac{k_z}{2} \sigma_z+\sin \frac{k_z}{2} \sigma_x)
( \cos \frac{k_z}{2} \tau_x+\sin \frac{k_z}{2} \tau_y)
, \\
&G_y(k_x,k_y,k_z) = e^{-ik_y}\begin{pmatrix}
0 & e^{-i k_z} \\
1 & 0 \\
\end{pmatrix}_{\tau}, \qquad 
I(k_x,k_y,k_z) = e^{-ik_y}\sigma_z \begin{pmatrix}
1 & 0 \\
0 & e^{-i k_z}
\end{pmatrix}_{\tau}.
\end{align*}
The $C_2$ rotation is 
\begin{align*}
C_2(k_x,k_y,k_z)
\equiv 
G_y(-k_x,-k_y,-k_z) I(k_x,k_y,k_z) = \sigma_z \begin{pmatrix}
0 & 1 \\
1 & 0
\end{pmatrix}_\tau.
\end{align*}
Let us compute the indicator for $-2 < m < 0$. 
At the high-symmetry points, the Hamiltonians and the symmetry operators within the occupied states are 
$$
\begin{array}{c|ccccccc}
P & H(P) & G_y(P) & C_2(P)\\
\hline
\Gamma & (m+1)\sigma_z+\tau_x\sigma_z & \tau_x & \sigma_z\tau_x \\
\mathrm{Y} &  (m-1)\sigma_z+\tau_x\sigma_z & \tau_x & \sigma_z\tau_x \\
\mathrm{Z} &  (m+1)\sigma_z+\tau_x\sigma_z& -\tau_x & \sigma_z\tau_x  \\
\mathrm{C} &  (m-1)\sigma_z+\tau_x\sigma_z& -\tau_x &\sigma_z\tau_x  \\
\end{array} .
$$
Therefore, the glide-$Z_2$ invariant is calculated as 
\begin{align*}
(-1)^{\tilde{\nu}}
=\prod_{i \in {\rm occ}} \frac{\zeta_i^-(\Gamma) \zeta^+_i(\mathrm{C})}{\zeta_i^-(\mathrm{Y}) \zeta^+_i(\mathrm{Z})}
= 1, 
\end{align*}
for all the values of $m$
and it turned out to be trivial. On the other hand, this model has $n_{\rm Ch} = +1$  for $-2<m<0$ and $n_{\rm Ch} = -1$  for $0<m<2$ by construction.

\section{Layer construction for SG \#14}
\label{app:layer14}
In this section we show the detailed construction of the models for the glide-symmetric TCI phase and the 
Chern insulator phase in SG \#14. The symmetry generators are given by 
\begin{align}
\hat{G}_y = \left\{ m_y \bigg | \frac{1}{2} \hat{\bm y} + \frac{1}{2} \hat{\bm z} \right\}, \quad \hat{I}  = \{ I | \bm{0} \}.
\end{align}

\subsection{Glide-symmetric TCI phase in the layer construction $(001;0)$ (${\tilde{\nu}} = 1, 
n_{\mathrm{Ch}}=0$)}
We first calculate the topological invariants for the LC $(001;0)$.
The model Hamiltonian is given by putting a Chern insulator with $n_{\mathrm{Ch}}=1$ on the $xy$ plane at $z=0$ with inversion symmetry, which is written as 
\begin{align*}
&\sum_{x,y \in \mathbb{Z}} \left[ 
\psi^{\dag}(x+1,y) \frac{\sigma_z+i \sigma_x}{2} \psi(x,y)
+\psi^{\dag}(x,y+1) \frac{\sigma_z+i \sigma_y}{2} \psi(x,y)
+ \mathrm{h.c.} 
\right]
+ \sum_{x,y \in \mathbb{Z}} m \psi^{\dag}(x,y) \sigma_z \psi(x,y),
\end{align*}
\begin{align*}
\hat I \psi^{\dag}(x,y) \hat I^{-1} = \psi^{\dag}(-x,-y) \sigma_z, 
\end{align*}
and making copies by the glide transformation 
\begin{align*}
\hat G_y \psi^{\dag}(x,y,z) \hat G_y^{-1}
= 
\psi^{\dag}\left(x,-y+\frac{1}{2},z+\frac{1}{2}\right).
\end{align*}
The Hamiltonian reads 
\begin{align*}
\hat H
&=\sum_{x,y,z \in \mathbb{Z}} \left[ 
\psi^{\dag}(x+1,y,z) \frac{\sigma_z+i \sigma_x}{2} \psi(x,y,z)
+\psi^{\dag}(x,y+1,z) \frac{\sigma_z+i \sigma_y}{2} \psi(x,y,z)
+ \mathrm{h.c.} 
\right] \\
&+ \sum_{x,y,z \in \mathbb{Z}} m \psi^{\dag}(x,y,z) \sigma_z \psi(x,y,z) \\
&+\sum_{x,y,z \in \mathbb{Z}} \left[ 
\psi^{\dag}(x+1,-y+\frac{1}{2},z+\frac{1}{2}) \frac{\sigma_z+i \sigma_x}{2} \psi(x,-y+\frac{1}{2},z+\frac{1}{2}) \right. \\
&\qquad \qquad \left. +\psi^{\dag}(x,-y-\frac{1}{2},z+\frac{1}{2}) \frac{\sigma_z+i \sigma_y}{2} \psi(x,-y+\frac{1}{2},z+\frac{1}{2})
+ \mathrm{h.c.} 
\right] \\
&+ \sum_{x,y,z \in \mathbb{Z}} m \psi^{\dag}(x,-y+\frac{1}{2},z+\frac{1}{2}) \sigma_z \psi(x,-y+\frac{1}{2},z+\frac{1}{2}) \\
&=\sum_{x,y,z \in \mathbb{Z}} \left[ 
\psi^{\dag}(x+1,y,z) \frac{\sigma_z+i \sigma_x}{2} \psi(x,y,z)
+\psi^{\dag}(x,y+1,z) \frac{\sigma_z+i \sigma_y}{2} \psi(x,y,z)
+ \mathrm{h.c.} 
\right] \\
&+ \sum_{x,y,z \in \mathbb{Z}} m \psi^{\dag}(x,y,z) \sigma_z \psi(x,y,z) \\
&+\sum_{x,y,z \in \mathbb{Z}} \left[ 
\psi^{\dag}(x+1,y+\frac{1}{2},z+\frac{1}{2}) \frac{\sigma_z+i \sigma_x}{2} \psi(x,y+\frac{1}{2},z+\frac{1}{2}) \right. \\
& \qquad \qquad \left. +\psi^{\dag}(x,y+\frac{3}{2},z+\frac{1}{2}) \frac{\sigma_z-i \sigma_y}{2} \psi(x,y+\frac{1}{2},z+\frac{1}{2})
+ \mathrm{h.c.} 
\right] \\
&+ \sum_{x,y,z \in \mathbb{Z}} m \psi^{\dag}(x,y+\frac{1}{2},z+\frac{1}{2}) \sigma_z \psi(x,y+\frac{1}{2},z+\frac{1}{2}) .
\end{align*}
Let us introduce the $k$-space basis 
\begin{align}
\Psi^{\dag}(k_x,k_y,k_z)
\equiv
\sum_{x,y,z \in \mathbb{Z}} (\psi^{\dag}(x,y,z),\psi^{\dag}(x,y+\frac{1}{2},z+\frac{1}{2})) e^{i (k_x x + k_y y + k_z z)}. 
\end{align}
The glide and inversion operations are represented as 
\begin{align*}
\hat G_y \Psi^{\dag}(k_x,k_y,k_z) \hat G_y^{-1}
&= 
\sum_{x,y,z \in \mathbb{Z}} (\psi^{\dag}(x,-y+\frac{1}{2},z+\frac{1}{2}),\psi^{\dag}(x,-y,z+1)) e^{i (k_x x + k_y y + k_z z)} \\
&= \Psi^{\dag}(k_x,-k_y,k_z) \begin{pmatrix}
0 & e^{-i k_z} \\
1 & 0
\end{pmatrix}_{\tau}, \\
\hat I \Psi^{\dag}(k_x,k_y,k_z) \hat I^{-1}
&= 
\sum_{x,y,z \in \mathbb{Z}} (\psi^{\dag}(-x,-y,-z),\psi^{\dag}(-x,-y-\frac{1}{2},-z-\frac{1}{2})) \sigma_z e^{i (k_x x + k_y y + k_z z)} \\
&= \Psi^{\dag}(-k_x,-k_y,-k_z) \sigma_z \begin{pmatrix}
1 & 0 \\
0 & e^{-i k_y-i k_z}
\end{pmatrix}_{\tau}.
\end{align*}
Thus, the $k$-space Hamiltonian and the symmetry operators are given by
\begin{align*}
&H(k_x,k_y,k_z)
= \sin k_x \sigma_x + \sin k_y \sigma_y \tau_z + (m + \cos k_x + \cos k_y) \sigma_z, \\
&G_y(k_x,k_y,k_z) = \begin{pmatrix}
0 & e^{-i k_z} \\
1 & 0 \\
\end{pmatrix}_{\tau}, \qquad 
I(k_x,k_y,k_z) = \sigma_z \begin{pmatrix}
1 & 0 \\
0 & e^{-i k_y-i k_z}
\end{pmatrix}_{\tau}.
\end{align*}
The $C_2$ screw operation $C_2^{\rm s}$ is 
\begin{align*}
C_2^{\rm s}(k_x,k_y,k_z)
\equiv 
G_y(-k_x,-k_y,-k_z) I(k_x,k_y,k_z) = \sigma_z \begin{pmatrix}
0 & e^{-i k_y} \\
1 & 0
\end{pmatrix}_\tau.
\end{align*}
Let us compute the indicator for $-2 < m < 0$. At the high-symmetry points, the Hamiltonians and the symmetry operators within the occupied states are 
\begin{align*}\begin{array}{c|ccccccc}
P & H(P) & G_y(P)|_{\rm occ} & C_2^{\rm s}(P)|_{\rm occ} \\
\hline
\Gamma & \sigma_z & \tau_x & -\tau_x \\
\mathrm{Y} & -\sigma_z & \tau_x & \tau_x \\
\mathrm{D} & -\sigma_z & -i\tau_y & -i\tau_y \\
\mathrm{E} & -\sigma_z & -i\tau_y & -i\tau_y \\
\end{array} .
\end{align*}
Therefore, the symmetry-based indicator becomes 
\begin{align*}
(-1)^{\tilde{\nu}+n_{\rm Ch}/2}
=\prod_{i \in {\rm occ}} \frac{\xi_i^-(\Gamma) \xi^+_i(\mathrm{D})}{\xi_i^-(\mathrm{Y}) \xi^+_i(\mathrm{E})}
= \frac{(-1) \times i}{1 \times i}
= -1.
\end{align*}
One can calculate the value of $\tilde{\nu}$ for other values of $m$. As
a result, $\tilde{\nu}+n_{\rm Ch}/2$ is equal to $1$ for $|m|<2$ and zero otherwise. 
By construction, we know $n_{\rm Ch}=0$ for this model, and therefore we
get $\tilde{\nu}=1$ for $|m|<2$ and $\tilde{\nu}=0$ otherwise.

\subsection{Chern insulator phase in the layer construction $(020;0)$ (${\tilde{\nu}} = 0, n_{\mathrm{Ch}}=2$)}
We here calculate the topological invariants for the LC $(020;0)$.
The model Hamiltonian is given by putting a 2D Chern insulator with $n_{\mathrm{Ch}}=1$ on the $xz$ plane at $y=0$ with inversion symmetry, which is written as 
\begin{align*}
&\sum_{x,z \in \mathbb{Z}} \left[ 
\psi^{\dag}(x,z+1) \frac{\sigma_z+i \sigma_x}{2} \psi(x,z)
+\psi^{\dag}(x+1,z) \frac{\sigma_z+i \sigma_y}{2} \psi(x,z)
+ \mathrm{h.c.} 
\right]
+ \sum_{x,z \in \mathbb{Z}} m \psi^{\dag}(x,z) \sigma_z \psi(x,z),\\
&\hat I \psi^{\dag}(x,z) \hat I^{-1} = \psi^{\dag}(-x,-z) \sigma_z, 
\end{align*}
and making copies by the glide transformation and the translation $T_y$, 
\begin{align*}
\hat G_y \psi^{\dag}(x,y,z) \hat G_y^{-1}
= 
\psi^{\dag}(x,-y+\frac{1}{2},z+\frac{1}{2}), \qquad 
\hat T_y \psi^{\dag}(x,y,z) \hat T_y^{-1}
= \psi^{\dag}(x,y+1,z).
\end{align*}
In the momentum space, the Hamiltonian and symmetry operators are represented by
\begin{align*}
&H(k_x,k_y,k_z)
= \left[ \sin k_z \sigma_x + \sin k_x \sigma_y + (m + \cos k_x + \cos k_z) \sigma_z \right] \tau_0, \\
&G_y(k_x,k_y,k_z) = \begin{pmatrix}
0 & e^{-i k_z} \\
1 & 0 \\
\end{pmatrix}_{\tau}, \qquad 
I(k_x,k_y,k_z) = \sigma_z \begin{pmatrix}
1 & 0 \\
0 & e^{-i k_y-i k_z}
\end{pmatrix}_{\tau}.
\end{align*}
Let us compute the indicator for $-2 < m < 0$. 
The screw operation is 
\begin{align*}
C_2^{\rm s}(k_x,k_y,k_z)\equiv G_y(-k_x,-k_y,-k_z) I(k_x,k_y,k_z) = \sigma_z \begin{pmatrix}
0 & e^{-i k_y} \\
1 & 0
\end{pmatrix}_\tau.
\end{align*}
At the high-symmetry points, the Hamiltonians and the symmetry operators within the occupied states are 
$$
\begin{array}{c|ccccccc}
P & H(P) & G_y(P)|_{\rm occ} & C_2^{\rm s}(P)|_{\rm occ} \\
\hline
\Gamma & \sigma_z & \tau_x & -\tau_x \\
\mathrm{Y} & -\sigma_z & \tau_x & \tau_x \\
\mathrm{D} & -\sigma_z & -i\tau_y & -i\tau_y \\
\mathrm{E} & -\sigma_z & -i\tau_y & -i\tau_y \\
\end{array} .
$$
Note that the Hamiltonians at the high-symmetry points are the same as the case of ${\tilde{\nu}}=1,n_{\mathrm{Ch}}=0$. 
Therefore, the symmetry-based indicator is identical with the previous one:
\begin{align*}
(-1)^{\tilde{\nu}+n_{\rm Ch}/2}
=\prod_{i \in {\rm occ}} \frac{\xi_i^-(\Gamma) \xi^+_i(\mathrm{D})}{\xi_i^-(\mathrm{Y}) \xi^+_i(\mathrm{E})}
= -1, 
\end{align*}
i.e. $\tilde{\nu}+n_{\rm Ch}/2=1$ for $|m|<2$, and $\tilde{\nu}+n_{\rm Ch}/2=0$ otherwise.
By construction, we in fact have 
$n_{\rm Ch}=2$ for $|m|<2$, and $n_{\rm Ch}=0$ otherwise.

\section{Irreducible representations}
\label{sec:irrep}
In this section, we summarize the lists of irreps and compatibility relations for SGs \#13 and \#14. 
They are shown in Tables~\ref{table:total_irreps13} for SG \#13 and \ref{table:total_irreps14} for SG \#14.
We used the database, the Bilbao Crystallographic Server \cite{Aroyo2006bilbao}.
We show the numbers of irreps $R$ at high-symmetry points in the rightmost column of each 
table, so as to satisfy the compatibility relations. 
Namely, within the Abelian group generated by the irreps at high-symmetry points, 
the integer parameters $a,b, \cdots$ in these numbers corresponds to the generators 
of its subgroup  satisfying the compatibility relations. 
Therefore, the number of the parameters  
is equal to the dimension $d_{\mathrm{BS}}$ in $\mathbb{Z}^{d_{\mathrm{BS}}}$ in Ref.~\onlinecite{Po2017ncommun}, and we have $d_{\mathrm{BS}}=7$ for SG \#13 and $d_{\mathrm{BS}}=5$ for SG \#14.
From these numbers, one can show various relations 
for the numbers of irreps.
As an example, here we show Eqs.~(\ref{eq:nuz4-13}) for SG \#13 and 
(\ref{eq:nuz4-14}) for SG \#14. 

In SG \#13, from the Table~\ref{table:total_irreps13}, the glide-$Z_2$ invariant $\tilde{\nu}$ is equal to $m+x+y+z+l \pmod{2}$.
Similarly, the Chern number $n_{\mathrm{Ch}}$ modulo 2 is given by $m+x+y \pmod 2$.
On the other hand, the $\mathbb{Z}_4$ index $z_4$ is calculated by counting the number of odd-parity states at all the TRIMs, and we get $z_4\equiv 2m - 2x +2y - 2z - 2l = 2(m+x+y+z+l) = 2 \tilde{\nu} \pmod 4$.
Therefore, we verify that $\tilde{\nu} \pmod 2$ is one half of the $\mathbb{Z}_4$ index $z_4$, from the compatibility relation for irreps.

In SG \#14, the combinations of the value of the glide-$Z_2$ invariant $\tilde{\nu} $ and that of the half of the Chern number $n_{\mathrm{Ch}}$ is given by the number of the irreps at $\Gamma$, Y, D, and E.
It can be calculated as $\tilde{\nu}+\frac{1}{2}n_{\mathrm{Ch}}\equiv x+y+z \mod2$.
On the other hand, the $\mathbb{Z}_4$ index for inversion symmetric systems is equal to  $z_4=-2(x+y+z) = 2(x+y+z) = 2(\tilde{\nu} + n_{\mathrm{Ch}}/2) \pmod 4$.
Therefore, we conclude that both the glide-$Z_2$ topological phase and a Chern insulator with the Chern number equal to $4n+2$ ($n$: integer) correspond to the topological phase with $z_4 = 2$.


\begin{table}[htb]
$$
\begin{array}{c | c c c c || c | c}
\mathrm{Seitz} & \{ 1 | t_1, t_2, t_3 \} & \{ 2_{010} | 0, 0, 1/2 \} & \{ \bar{1} | 0, 0, 0 \} & \{ m_{010} | 0, 0, 1/2 \} & \mathrm{Mulliken}
  & \begin{matrix}
\mathrm{number} \\ \mathrm{of} \ \mathrm{irreps}
\end{matrix} \\ \hline
\begin{matrix}
\mathrm{Matrix} \\ \mathrm{presentation}
\end{matrix}  &
\begin{pmatrix}
1 & 0 & 0 & t_1 \\ 0 & 1 & 0 & t_2 \\ 0 & 0 & 1 & t_3
\end{pmatrix} & 
\begin{pmatrix}
-1 & 0 & 0 & 0 \\ 0 & 1 & 0 & 0 \\ 0 & 0 & -1 & 1/2
\end{pmatrix} &
\begin{pmatrix}
-1 & 0 & 0 & 0 \\ 0 & -1 & 0 & 0 \\ 0 & 0 & -1 & 0
\end{pmatrix} &
\begin{pmatrix}
1 & 0 & 0 & 0 \\ 0 & -1 & 0 & 0 \\ 0 & 0 & 1 & 1/2
\end{pmatrix} & \\ \hline
\Gamma^+_1 & 1 & 1 & 1 & 1 & A_g & a \\
\Gamma^-_1 & 1 & 1 & -1 & -1 & A_u & b+m\\
\Gamma^+_2 & 1 & -1 & 1 & -1 & B_g & a-m \\
\Gamma^-_2 & 1 & -1 & -1 & 1 & B_u & b \\ \hline
\mathrm{Y}^+_1 & e^{i\pi t_1} & 1 & 1 & 1 & A_g & a+x \\
\mathrm{Y}^-_1 & e^{i\pi t_1} & 1 & -1 & -1 & A_u & b+y \\
\mathrm{Y}^+_2 & e^{i\pi t_1} & -1 & 1 & -1 & B_g & a-y \\
\mathrm{Y}^-_2 & e^{i\pi t_1} & -1 & -1 & 1 & B_u & b-x \\ \hline
\mathrm{Z}^+_1 & e^{i\pi t_2} & 1 & 1 & 1 & A_g & a+z \\
\mathrm{Z}^-_1 & e^{i\pi t_2} & 1 & -1 & -1 & A_u & b+m-z \\
\mathrm{Z}^+_2 & e^{i\pi t_2} & -1 & 1 & -1 & B_g & a-m+z \\
\mathrm{Z}^-_2 & e^{i\pi t_2} & -1 & -1 & 1 & B_u & b-z \\ \hline
\mathrm{C}^+_1 & e^{i\pi (t_1+t_2)} & 1 & 1 & 1 & A_g & a+x+l \\
\mathrm{C}^-_1 & e^{i\pi (t_1+t_2)} & 1 & -1 & -1 & A_u & b+y-l \\
\mathrm{C}^+_2 & e^{i\pi (t_1+t_2)} & -1 & 1 & -1 & B_g & a-y+l \\
\mathrm{C}^-_2 & e^{i\pi (t_1+t_2)} & -1 & -1 & 1 & B_u & b-x-l \\ \hline
\mathrm{B}_1 & e^{i\pi t_3} \sigma_0 & \sigma_x & \sigma_z & -i \sigma_y & E & a+b \\ \hline
\mathrm{A}_1 & e^{i\pi (t_1+t_3)} \sigma_0 & \sigma_x & \sigma_z & -i \sigma_y & E & a+b \\ \hline
\mathrm{D}_1 & e^{i\pi (t_2+t_3)} \sigma_0 & \sigma_x & \sigma_z & -i \sigma_y & E & a+b \\ \hline
\mathrm{E}_1 & e^{i\pi (t_1+t_2+t_3)} \sigma_0 & \sigma_x & \sigma_z & -i \sigma_y & E & a+b
\end{array}
$$
\caption{Summary of irreducible representations, correspondences to characters denoted in the main text and the numbers of irreps  for SG \#13 satisfying the compatibility relations, where $a, b, m, x, y, z$ and $l$ are integers.}
\label{table:total_irreps13}
\end{table}

\clearpage

\begin{table}[htb]
$$
\begin{array}{c | c c c c || c | c}
\mathrm{Seitz} & \{ 1 | t_1, t_2, t_3 \} & \{ 2_{010} | 0, 1/2, 1/2 \} & \{ \bar{1} | 0, 0, 0 \} & \{ m_{010} | 0, 1/2, 1/2 \} & \mathrm{Mulliken}
  & \begin{matrix}
\mathrm{number} \\ \mathrm{of} \ \mathrm{irreps}
\end{matrix} \\ \hline
\begin{matrix}
\mathrm{Matrix} \\ \mathrm{presentation}
\end{matrix}  &
\begin{pmatrix}
1 & 0 & 0 & t_1 \\ 0 & 1 & 0 & t_2 \\ 0 & 0 & 1 & t_3
\end{pmatrix} & 
\begin{pmatrix}
-1 & 0 & 0 & 0 \\ 0 & 1 & 0 & 1/2 \\ 0 & 0 & -1 & 1/2
\end{pmatrix} &
\begin{pmatrix}
-1 & 0 & 0 & 0 \\ 0 & -1 & 0 & 0 \\ 0 & 0 & -1 & 0
\end{pmatrix} &
\begin{pmatrix}
1 & 0 & 0 & 0 \\ 0 & -1 & 0 & 1/2 \\ 0 & 0 & 1 & 1/2
\end{pmatrix} & \\ \hline
\Gamma^+_1 & 1 & 1 & 1 & 1 & A_g & a \\
\Gamma^-_1 & 1 & 1 & -1 & -1 & A_u & b \\
\Gamma^+_2 & 1 & -1 & 1 & -1 & B_g & a \\
\Gamma^-_2 & 1 & -1 & -1 & 1 & B_u & b \\ \hline
\mathrm{Y}^+_1 & e^{i\pi t_1} & 1 & 1 & 1 & A_g & a+x \\
\mathrm{Y}^-_1 & e^{i\pi t_1} & 1 & -1 & -1 & A_u & b-x \\
\mathrm{Y}^+_2 & e^{i\pi t_1} & -1 & 1 & -1 & B_g & a+x \\
\mathrm{Y}^-_2 & e^{i\pi t_1} & -1 & -1 & 1 & B_u & b-x \\ \hline
\mathrm{Z}_1 & e^{i\pi t_2} \sigma_0 & -i \sigma_y & \sigma_z & \sigma_x & E & a+b \\ \hline
\mathrm{C}_1 & e^{i\pi (t_1+t_2)} \sigma_0 & -i \sigma_y & \sigma_z & \sigma_x & E & a+b \\ \hline
\mathrm{B}_1 & e^{i\pi t_3} \sigma_0 & \sigma_x & \sigma_z & -i \sigma_y & E & a+b \\ \hline
\mathrm{A}_1 & e^{i\pi (t_1+t_3)} \sigma_0 & \sigma_x & \sigma_z & -i \sigma_y & E & a+b \\ \hline
\mathrm{D}^+_1 & e^{i\pi (t_2+t_3)} & i & 1 & i & A_g & a+y \\
\mathrm{D}^-_1 & e^{i\pi (t_2+t_3)} & i & -1 & -i & A_u & b-y \\
\mathrm{D}^+_2 & e^{i\pi (t_2+t_3)} & -i & 1 & -i & B_g & a+y \\
\mathrm{D}^-_2 & e^{i\pi (t_2+t_3)} & -i & -1 & i & B_u & b-y \\ \hline
\mathrm{E}^+_1 & e^{i\pi (t_1+t_2+t_3)} & i & 1 & i & A_g & a+z \\
\mathrm{E}^-_1 & e^{i\pi (t_1+t_2+t_3)} & i & -1 & -i & A_u & b-z \\
\mathrm{E}^+_2 & e^{i\pi (t_1+t_2+t_3)} & -i & 1 & -i & B_g & a+z \\
\mathrm{E}^-_2 & e^{i\pi (t_1+t_2+t_3)} & -i & -1 & i & B_u & b-z
\end{array}
$$
\caption{Summary of irreducible representations, correspondences to characters denoted in the main text, and the numbers of irreps for SG \#14 satisfying the compatibility relations, where $a, b, x, y$ and $z$ are integers.}
\label{table:total_irreps14}
\end{table}


\section{Spinful systems}
\label{sec:spinful}

In this section, we will consider the spinful systems of SG \#13 and SG \#14.
The formulas for SG \#13 in Eq.~(\ref{eq:13rewrite_z2}) and SG \#14 in Eq.~(\ref{eq:14z2}) can be also applied in the same manner to the spinful systems, because the commutation relations between the glide and $C_2$ rotation/screw operations in Eqs.~(\ref{eq:CR13}) and (\ref{eq:CR14}) are unchanged.

From the database \cite{Aroyo2006bilbao}, we summarize the lists of irreps for spinful SGs \#13 and \#14 in Tables~\ref{table:total_irreps13_dsp} and \ref{table:total_irreps14_dsp}.
The double-valued representation $\{ ^d p | a_1 a_2 a_3 \}$ of $\{ p | a_1 a_2 a_3 \}$ is given by  $-1$ times to $\{ p | a_1 a_2 a_3 \}$.
For simplicity, we have omitted some information;
matrix representations are same in Tables~\ref{table:total_irreps13} and \ref{table:total_irreps14} except for
\begin{equation}
\begin{pmatrix}
1 & 0 & 0 & 0 \\ 0 & 1 & 0 & 0 \\ 0 & 0 & 1 & 0
\end{pmatrix}
\end{equation}
for $\{ ^\mathrm{d} 1 | 0, 0, 0 \}$ and numbers of irreps are twice those in Tables~\ref{table:total_irreps13} and \ref{table:total_irreps14}, respectively.

The formulas for spinful systems of SG	\#13 and SG \#14 are given by
\begin{align}
(-1)^{\tilde{\nu}} &= \prod_{i \in \mathrm{occ}} \frac{\zeta_i^- (\Gamma) \zeta_i^+(\mathrm{C})}{\zeta_i^- (\mathrm{Y}) \zeta_i^+ (\mathrm{Z})} \quad (\mathrm{SG} \ \#13) , \\
n_{\mathrm{Ch}} \in 2\mathbb{Z}, \quad (-1)^{\tilde{\nu}}& (-1)^{n_{\mathrm{Ch}}/2} = \prod_{i \in \mathrm{occ}} \frac{\xi_i^-(\Gamma) \xi_i^+(\mathrm{D})}{\xi_i^-(\mathrm{Y}) \xi_i^+(\mathrm{E})} \quad (\mathrm{SG} \ \#14) .
\end{align}
where $\zeta_i^\pm =\pm i$ and $\xi_i^\pm = \pm i e^{-ik_y/2}$ are an eigenvalue of the $C_2$ rotation/screw for the eigenstates in the $g_\pm$ sector, 
\begin{equation}
g_\pm (k_z) = \pm i e^{-ik_z/2} ,
\end{equation}
at high-symmetry points $\Gamma$, Y, Z, and C in SG \#13 and $\Gamma$, Y, D, and E in SG \#14, respectively.

\begin{table}[htb]
$$
\begin{array}{c | cccc}
\mathrm{Seitz} & 
\begin{matrix}
\{ 1 | t_1, t_2, t_3 \} \\ \left( \{ ^\mathrm{d} 1 | 0, 0, 0 \} \right)
\end{matrix} & 
\begin{matrix} 
\{ 2_{010} | 0, 0, 1/2 \} \\ \left( \{ ^\mathrm{d} 2_{010} | 0, 0, 1/2 \} \right)
\end{matrix} & 
\begin{matrix}
\{ \bar{1} | 0, 0, 0 \} \\ \left( \{ ^\mathrm{d} \bar{1} | 0, 0, 0 \} \right)
\end{matrix} & 
\begin{matrix}
\{ m_{010} | 0, 0, 1/2 \} \\ \left( \{ ^\mathrm{d} m_{010} | 0, 0, 1/2 \} \right)
\end{matrix} \\ \hline
\mathrm{spin \ half \ integer \ rotation} & \begin{matrix} +\sigma_0 & (-\sigma_0) \end{matrix} & \begin{matrix} -i \sigma_y & (i\sigma_y) \end{matrix} & \begin{matrix} \sigma_0 & (-\sigma_0) \end{matrix} & \begin{matrix} -i \sigma_y & (i\sigma_y) \end{matrix} \\ \hline
\overline{\Gamma}_3 & \begin{matrix} +1 & (-1) \end{matrix} & \begin{matrix} -i & (+i) \end{matrix} & \begin{matrix} +1 & (-1) \end{matrix} & \begin{matrix} -i & (+i) \end{matrix} \\
\overline{\Gamma}_4 & \begin{matrix} +1 & (-1) \end{matrix} & \begin{matrix} +i & (-i) \end{matrix} & \begin{matrix} +1 & (-1) \end{matrix} & \begin{matrix} +i & (-i) \end{matrix} \\
\overline{\Gamma}_5 & \begin{matrix} +1 & (-1) \end{matrix} & \begin{matrix} -i & (+i) \end{matrix} & \begin{matrix} -1 & (+1) \end{matrix} & \begin{matrix} +i & (-i) \end{matrix} \\
\overline{\Gamma}_6 & \begin{matrix} +1 & (-1) \end{matrix} & \begin{matrix} +i & (-i) \end{matrix} & \begin{matrix} -1 & (+1) \end{matrix} & \begin{matrix} -i & (+i) \end{matrix} \\ \hline
\overline{\mathrm{Y}}_3 & \begin{matrix} e^{i\pi t_1} & (-1) \end{matrix} & \begin{matrix} -i & (+i) \end{matrix} & \begin{matrix} +1 & (-1) \end{matrix} & \begin{matrix} -i & (+i) \end{matrix} \\
\overline{\mathrm{Y}}_4 & \begin{matrix} e^{i\pi t_1} & (-1) \end{matrix} & \begin{matrix} +i & (-i) \end{matrix} & \begin{matrix} +1 & (-1) \end{matrix} & \begin{matrix} +i & (-i) \end{matrix} \\
\overline{\mathrm{Y}}_5 & \begin{matrix} e^{i\pi t_1} & (-1) \end{matrix} & \begin{matrix} -i & (+i) \end{matrix} & \begin{matrix} -1 & (+1) \end{matrix} & \begin{matrix} +i & (-i) \end{matrix} \\
\overline{\mathrm{Y}}_6 & \begin{matrix} e^{i\pi t_1} & (-1) \end{matrix} & \begin{matrix} +i & (-i) \end{matrix} & \begin{matrix} -1 & (+1) \end{matrix} & \begin{matrix} -i & (+i) \end{matrix} \\ \hline
\overline{\mathrm{Z}}_3 & \begin{matrix} e^{i\pi t_2} & (-1) \end{matrix} & \begin{matrix} -i & (+i) \end{matrix} & \begin{matrix} +1 & (-1) \end{matrix} & \begin{matrix} -i & (+i) \end{matrix} \\
\overline{\mathrm{Z}}_4 & \begin{matrix} e^{i\pi t_2} & (-1) \end{matrix} & \begin{matrix} +i & (-i) \end{matrix} & \begin{matrix} +1 & (-1) \end{matrix} & \begin{matrix} +i & (-i) \end{matrix} \\
\overline{\mathrm{Z}}_5 & \begin{matrix} e^{i\pi t_2} & (-1) \end{matrix} & \begin{matrix} -i & (+i) \end{matrix} & \begin{matrix} -1 & (+1) \end{matrix} & \begin{matrix} +i & (-i) \end{matrix} \\
\overline{\mathrm{Z}}_6 & \begin{matrix} e^{i\pi t_2} & (-1) \end{matrix} & \begin{matrix} +i & (-i) \end{matrix} & \begin{matrix} -1 & (+1) \end{matrix} & \begin{matrix} -i & (+i) \end{matrix} \\ \hline
\overline{\mathrm{C}}_3 & \begin{matrix} e^{i\pi (t_1+t_2)} & (-1) \end{matrix} & \begin{matrix} -i & (+i) \end{matrix} & \begin{matrix} +1 & (-1) \end{matrix} & \begin{matrix} -i & (+i) \end{matrix} \\
\overline{\mathrm{C}}_4 & \begin{matrix} e^{i\pi (t_1+t_2)} & (-1) \end{matrix} & \begin{matrix} +i & (-i) \end{matrix} & \begin{matrix} +1 & (-1) \end{matrix} & \begin{matrix} +i & (-i) \end{matrix} \\
\overline{\mathrm{C}}_5 & \begin{matrix} e^{i\pi (t_1+t_2)} & (-1) \end{matrix} & \begin{matrix} -i & (+i) \end{matrix} & \begin{matrix} -1 & (+1) \end{matrix} & \begin{matrix} +i & (-i) \end{matrix} \\
\overline{\mathrm{C}}_6 & \begin{matrix} e^{i\pi (t_1+t_2)} & (-1) \end{matrix} & \begin{matrix} +i & (-i) \end{matrix} & \begin{matrix} -1 & (+1) \end{matrix} & \begin{matrix} -i & (+i) \end{matrix} \\ \hline
\overline{\mathrm{B}}_2 & \begin{matrix} e^{i\pi t_3} \sigma_0 & (-\sigma_0) \end{matrix} & \begin{matrix} -i \sigma_y & (i \sigma_y) \end{matrix} & \begin{matrix} \sigma_z & (-\sigma_z) \end{matrix} & \begin{matrix} \sigma_x & (-\sigma_x) \end{matrix} \\ \hline
\overline{\mathrm{A}}_2 &\begin{matrix}  e^{i\pi (t_1+t_3)} \sigma_0 & (-\sigma_0) \end{matrix} & \begin{matrix} -i \sigma_y & (i \sigma_y) \end{matrix} & \begin{matrix} \sigma_z & (-\sigma_z) \end{matrix} & \begin{matrix} \sigma_x & (-\sigma_x) \end{matrix} \\ \hline
\overline{\mathrm{D}}_2 & \begin{matrix} e^{i\pi (t_2+t_3)} \sigma_0 & (-\sigma_0) \end{matrix} & \begin{matrix} -i \sigma_y & (i \sigma_y) \end{matrix} & \begin{matrix} \sigma_z & (-\sigma_z) \end{matrix} & \begin{matrix} \sigma_x & (-\sigma_x) \end{matrix} \\ \hline
\overline{\mathrm{E}}_2 & \begin{matrix} e^{i\pi (t_1+t_2+t_3)} \sigma_0 & (-\sigma_0) \end{matrix} & \begin{matrix} -i \sigma_y & (i \sigma_y) \end{matrix} & \begin{matrix} \sigma_z & (-\sigma_z) \end{matrix} & \begin{matrix} \sigma_x & (-\sigma_x) \end{matrix}
\end{array}
$$
\caption{Summary of irreducible representations and correspondences to characters for double-valued SG \#13.}
\label{table:total_irreps13_dsp}
\end{table}

\clearpage

\begin{table}[htb]
$$
\begin{array}{c | cccc}
\mathrm{Seitz \ symbols} & 
\begin{matrix}
\{ 1 | t_1, t_2, t_3 \} \\ \left( \{ ^\mathrm{d} 1 | 0, 0, 0 \} \right)
\end{matrix} & 
\begin{matrix} 
\{ 2_{010} | 0, 1/2, 1/2 \} \\ \left( \{ ^\mathrm{d} 2_{010} | 0, 1/2, 1/2 \} \right)
\end{matrix} & 
\begin{matrix}
\{ \bar{1} | 0, 0, 0 \} \\ \left( \{ ^\mathrm{d} \bar{1} | 0, 0, 0 \} \right)
\end{matrix} & 
\begin{matrix}
\{ m_{010} | 0, 1/2, 1/2 \} \\ \left( \{ ^\mathrm{d} m_{010} | 0, 1/2, 1/2 \} \right)
\end{matrix} \\ \hline
\mathrm{spin \ half \ integer \ rotation} & \begin{matrix} +\sigma_0 & (-\sigma_0) \end{matrix} & \begin{matrix} -i \sigma_y & (i\sigma_y) \end{matrix} & \begin{matrix} \sigma_0 & (-\sigma_0) \end{matrix} & \begin{matrix} -i \sigma_y & (i\sigma_y) \end{matrix} \\ \hline
\overline{\Gamma}_3 & \begin{matrix} +1 & (-1) \end{matrix} & \begin{matrix} -i & (+i) \end{matrix} & \begin{matrix} +1 & (-1) \end{matrix} & \begin{matrix} -i & (+i) \end{matrix} \\
\overline{\Gamma}_4 & \begin{matrix} +1 & (-1) \end{matrix} & \begin{matrix} +i & (-i) \end{matrix} & \begin{matrix} +1 & (-1) \end{matrix} & \begin{matrix} +i & (-i) \end{matrix} \\
\overline{\Gamma}_5 & \begin{matrix} +1 & (-1) \end{matrix} & \begin{matrix} -i & (+i) \end{matrix} & \begin{matrix} -1 & (+1) \end{matrix} & \begin{matrix} +i & (-i) \end{matrix} \\
\overline{\Gamma}_6 & \begin{matrix} +1 & (-1) \end{matrix} & \begin{matrix} +i & (-i) \end{matrix} & \begin{matrix} -1 & (+1) \end{matrix} & \begin{matrix} -i & (+i) \end{matrix} \\ \hline
\overline{\mathrm{Y}}_3 & \begin{matrix} e^{i\pi t_1} & (-1) \end{matrix} & \begin{matrix} -i & (+i) \end{matrix} & \begin{matrix} +1 & (-1) \end{matrix} & \begin{matrix} -i & (+i) \end{matrix} \\
\overline{\mathrm{Y}}_4 & \begin{matrix} e^{i\pi t_1} & (-1) \end{matrix} & \begin{matrix} +i & (-i) \end{matrix} & \begin{matrix} +1 & (-1) \end{matrix} & \begin{matrix} +i & (-i) \end{matrix} \\
\overline{\mathrm{Y}}_5 & \begin{matrix} e^{i\pi t_1} & (-1) \end{matrix} & \begin{matrix} -i & (+i) \end{matrix} & \begin{matrix} -1 & (+1) \end{matrix} & \begin{matrix} +i & (-i) \end{matrix} \\
\overline{\mathrm{Y}}_6 & \begin{matrix} e^{i\pi t_1} & (-1) \end{matrix} & \begin{matrix} +i & (-i) \end{matrix} & \begin{matrix} -1 & (+1) \end{matrix} & \begin{matrix} -i & (+i) \end{matrix} \\ \hline
\overline{\mathrm{Z}}_2 & \begin{matrix} e^{i\pi t_2} \sigma_0 & (-\sigma_0) \end{matrix} & \begin{matrix} \sigma_x & (-\sigma_x) \end{matrix} & \begin{matrix} \sigma_z & (-\sigma_z) \end{matrix} & \begin{matrix} -i \sigma_y & (i\sigma_y) \end{matrix} \\ \hline
\overline{\mathrm{C}}_2 & \begin{matrix} e^{i\pi (t_1+t_2)} \sigma_0 & (-\sigma_0) \end{matrix} & \begin{matrix} \sigma_x & (-\sigma_x) \end{matrix} & \begin{matrix} \sigma_z & (-\sigma_z) \end{matrix} & \begin{matrix} -i \sigma_y & (i\sigma_y) \end{matrix} \\ \hline
\overline{\mathrm{B}}_2 & \begin{matrix} e^{i\pi t_3} \sigma_0 & (-\sigma_0) \end{matrix} & \begin{matrix} -i \sigma_y & (i \sigma_y) \end{matrix} & \begin{matrix} \sigma_z & (-\sigma_z) \end{matrix} & \begin{matrix} \sigma_x & (-\sigma_x) \end{matrix} \\ \hline
\overline{\mathrm{A}}_2 &\begin{matrix}  e^{i\pi (t_1+t_3)} \sigma_0 & (-\sigma_0) \end{matrix} & \begin{matrix} -i \sigma_y & (i \sigma_y) \end{matrix} & \begin{matrix} \sigma_z & (-\sigma_z) \end{matrix} & \begin{matrix} \sigma_x & (-\sigma_x) \end{matrix} \\ \hline
\overline{\mathrm{D}}_3 & \begin{matrix} e^{i\pi (t_2+t_3)} & (-1) \end{matrix} & \begin{matrix} +1 & (-1) \end{matrix} & \begin{matrix} +1 & (-1) \end{matrix} & \begin{matrix} +1 & (-1) \end{matrix} \\
\overline{\mathrm{D}}_4 & \begin{matrix} e^{i\pi (t_2+t_3)} & (-1) \end{matrix} & \begin{matrix} -1 & (+1) \end{matrix} & \begin{matrix} +1 & (-1) \end{matrix} & \begin{matrix} -1 & (+1) \end{matrix} \\
\overline{\mathrm{D}}_5 & \begin{matrix} e^{i\pi (t_2+t_3)} & (-1) \end{matrix} & \begin{matrix} +1 & (-1) \end{matrix} & \begin{matrix} -1 & (+1) \end{matrix} & \begin{matrix} -1 & (+1) \end{matrix} \\
\overline{\mathrm{D}}_6 & \begin{matrix} e^{i\pi (t_2+t_3)} & (-1) \end{matrix} & \begin{matrix} -1 & (+1) \end{matrix} & \begin{matrix} -1 & (+1) \end{matrix} & \begin{matrix} +1 & (-1) \end{matrix} \\ \hline
\overline{\mathrm{E}}_3 & \begin{matrix} e^{i\pi (t_1+t_2+t_3)} & (-1) \end{matrix} & \begin{matrix} +1 & (-1) \end{matrix} & \begin{matrix} +1 & (-1) \end{matrix} & \begin{matrix} +1 & (-1) \end{matrix} \\
\overline{\mathrm{E}}_4 & \begin{matrix} e^{i\pi (t_1+t_2+t_3)} & (-1) \end{matrix} & \begin{matrix} -1 & (+1) \end{matrix} & \begin{matrix} +1 & (-1) \end{matrix} & \begin{matrix} -1 & (+1) \end{matrix} \\
\overline{\mathrm{E}}_5 & \begin{matrix} e^{i\pi (t_1+t_2+t_3)} & (-1) \end{matrix} & \begin{matrix} +1 & (-1) \end{matrix} & \begin{matrix} -1 & (+1) \end{matrix} & \begin{matrix} -1 & (+1) \end{matrix} \\
\overline{\mathrm{E}}_6 & \begin{matrix} e^{i\pi (t_1+t_2+t_3)} & (-1) \end{matrix} & \begin{matrix} -1 & (+1) \end{matrix} & \begin{matrix} -1 & (+1) \end{matrix} & \begin{matrix} +1 & (-1) \end{matrix}
\end{array}
$$
\caption{Summary of irreducible representations and correspondences to characters for double-valued SG \#14.}
\label{table:total_irreps14_dsp}
\end{table}

\end{widetext}

\bibliographystyle{apsrev4-1}
\bibliography{glide_inversion_biblio2}

\begin{thebibliography}{67}%
\makeatletter
\providecommand \@ifxundefined [1]{%
 \@ifx{#1\undefined}
}%
\providecommand \@ifnum [1]{%
 \ifnum #1\expandafter \@firstoftwo
 \else \expandafter \@secondoftwo
 \fi
}%
\providecommand \@ifx [1]{%
 \ifx #1\expandafter \@firstoftwo
 \else \expandafter \@secondoftwo
 \fi
}%
\providecommand \natexlab [1]{#1}%
\providecommand \enquote  [1]{``#1''}%
\providecommand \bibnamefont  [1]{#1}%
\providecommand \bibfnamefont [1]{#1}%
\providecommand \citenamefont [1]{#1}%
\providecommand \href@noop [0]{\@secondoftwo}%
\providecommand \href [0]{\begingroup \@sanitize@url \@href}%
\providecommand \@href[1]{\@@startlink{#1}\@@href}%
\providecommand \@@href[1]{\endgroup#1\@@endlink}%
\providecommand \@sanitize@url [0]{\catcode `\\12\catcode `\$12\catcode
  `\&12\catcode `\#12\catcode `\^12\catcode `\_12\catcode `\%12\relax}%
\providecommand \@@startlink[1]{}%
\providecommand \@@endlink[0]{}%
\providecommand \url  [0]{\begingroup\@sanitize@url \@url }%
\providecommand \@url [1]{\endgroup\@href {#1}{\urlprefix }}%
\providecommand \urlprefix  [0]{URL }%
\providecommand \Eprint [0]{\href }%
\providecommand \doibase [0]{http://dx.doi.org/}%
\providecommand \selectlanguage [0]{\@gobble}%
\providecommand \bibinfo  [0]{\@secondoftwo}%
\providecommand \bibfield  [0]{\@secondoftwo}%
\providecommand \translation [1]{[#1]}%
\providecommand \BibitemOpen [0]{}%
\providecommand \bibitemStop [0]{}%
\providecommand \bibitemNoStop [0]{.\EOS\space}%
\providecommand \EOS [0]{\spacefactor3000\relax}%
\providecommand \BibitemShut  [1]{\csname bibitem#1\endcsname}%
\let\auto@bib@innerbib\@empty
\bibitem [{\citenamefont {Hasan}\ and\ \citenamefont
  {Kane}(2010)}]{HasanKane2010RevModPhys}%
  \BibitemOpen
  \bibfield  {author} {\bibinfo {author} {\bibfnamefont {M.~Z.}\ \bibnamefont
  {Hasan}}\ and\ \bibinfo {author} {\bibfnamefont {C.~L.}\ \bibnamefont
  {Kane}},\ }\href {\doibase 10.1103/RevModPhys.82.3045} {\bibfield  {journal}
  {\bibinfo  {journal} {Rev. Mod. Phys.}\ }\textbf {\bibinfo {volume} {82}},\
  \bibinfo {pages} {3045} (\bibinfo {year} {2010})}\BibitemShut {NoStop}%
\bibitem [{\citenamefont {Qi}\ and\ \citenamefont
  {Zhang}(2011)}]{QiZhang2011RevModPhys}%
  \BibitemOpen
  \bibfield  {author} {\bibinfo {author} {\bibfnamefont {X.-L.}\ \bibnamefont
  {Qi}}\ and\ \bibinfo {author} {\bibfnamefont {S.-C.}\ \bibnamefont {Zhang}},\
  }\href {\doibase 10.1103/RevModPhys.83.1057} {\bibfield  {journal} {\bibinfo
  {journal} {Rev. Mod. Phys.}\ }\textbf {\bibinfo {volume} {83}},\ \bibinfo
  {pages} {1057} (\bibinfo {year} {2011})}\BibitemShut {NoStop}%
\bibitem [{\citenamefont {Schnyder}\ \emph {et~al.}(2008)\citenamefont
  {Schnyder}, \citenamefont {Ryu}, \citenamefont {Furusaki},\ and\
  \citenamefont {Ludwig}}]{Schnyder2008prb78}%
  \BibitemOpen
  \bibfield  {author} {\bibinfo {author} {\bibfnamefont {A.~P.}\ \bibnamefont
  {Schnyder}}, \bibinfo {author} {\bibfnamefont {S.}~\bibnamefont {Ryu}},
  \bibinfo {author} {\bibfnamefont {A.}~\bibnamefont {Furusaki}}, \ and\
  \bibinfo {author} {\bibfnamefont {A.~W.~W.}\ \bibnamefont {Ludwig}},\ }\href
  {\doibase 10.1103/PhysRevB.78.195125} {\bibfield  {journal} {\bibinfo
  {journal} {Phys. Rev. B}\ }\textbf {\bibinfo {volume} {78}},\ \bibinfo
  {pages} {195125} (\bibinfo {year} {2008})}\BibitemShut {NoStop}%
\bibitem [{\citenamefont {Kitaev}(2009)}]{Kitaev2009aip1134}%
  \BibitemOpen
  \bibfield  {author} {\bibinfo {author} {\bibfnamefont {A.}~\bibnamefont
  {Kitaev}},\ }in\ \href@noop {} {\emph {\bibinfo {booktitle} {AIP Conference
  Proceedings}}},\ Vol.\ \bibinfo {volume} {1134}\ (\bibinfo {organization}
  {AIP},\ \bibinfo {year} {2009})\ pp.\ \bibinfo {pages} {22--30}\BibitemShut
  {NoStop}%
\bibitem [{\citenamefont {Ryu}\ \emph {et~al.}(2010)\citenamefont {Ryu},
  \citenamefont {Schnyder}, \citenamefont {Furusaki},\ and\ \citenamefont
  {Ludwig}}]{Ryu2010njp12}%
  \BibitemOpen
  \bibfield  {author} {\bibinfo {author} {\bibfnamefont {S.}~\bibnamefont
  {Ryu}}, \bibinfo {author} {\bibfnamefont {A.~P.}\ \bibnamefont {Schnyder}},
  \bibinfo {author} {\bibfnamefont {A.}~\bibnamefont {Furusaki}}, \ and\
  \bibinfo {author} {\bibfnamefont {A.~W.}\ \bibnamefont {Ludwig}},\
  }\href@noop {} {\bibfield  {journal} {\bibinfo  {journal} {New Journal of
  Physics}\ }\textbf {\bibinfo {volume} {12}},\ \bibinfo {pages} {065010}
  (\bibinfo {year} {2010})}\BibitemShut {NoStop}%
\bibitem [{\citenamefont {Fu}\ and\ \citenamefont {Kane}(2007)}]{Fu2007prb76}%
  \BibitemOpen
  \bibfield  {author} {\bibinfo {author} {\bibfnamefont {L.}~\bibnamefont
  {Fu}}\ and\ \bibinfo {author} {\bibfnamefont {C.~L.}\ \bibnamefont {Kane}},\
  }\href {\doibase 10.1103/PhysRevB.76.045302} {\bibfield  {journal} {\bibinfo
  {journal} {Phys. Rev. B}\ }\textbf {\bibinfo {volume} {76}},\ \bibinfo
  {pages} {045302} (\bibinfo {year} {2007})}\BibitemShut {NoStop}%
\bibitem [{\citenamefont {Teo}\ \emph {et~al.}(2008)\citenamefont {Teo},
  \citenamefont {Fu},\ and\ \citenamefont {Kane}}]{Teo2008prb78}%
  \BibitemOpen
  \bibfield  {author} {\bibinfo {author} {\bibfnamefont {J.~C.~Y.}\
  \bibnamefont {Teo}}, \bibinfo {author} {\bibfnamefont {L.}~\bibnamefont
  {Fu}}, \ and\ \bibinfo {author} {\bibfnamefont {C.~L.}\ \bibnamefont
  {Kane}},\ }\href {\doibase 10.1103/PhysRevB.78.045426} {\bibfield  {journal}
  {\bibinfo  {journal} {Phys. Rev. B}\ }\textbf {\bibinfo {volume} {78}},\
  \bibinfo {pages} {045426} (\bibinfo {year} {2008})}\BibitemShut {NoStop}%
\bibitem [{\citenamefont {Fu}(2011)}]{Fu2011prl106}%
  \BibitemOpen
  \bibfield  {author} {\bibinfo {author} {\bibfnamefont {L.}~\bibnamefont
  {Fu}},\ }\href {\doibase 10.1103/PhysRevLett.106.106802} {\bibfield
  {journal} {\bibinfo  {journal} {Phys. Rev. Lett.}\ }\textbf {\bibinfo
  {volume} {106}},\ \bibinfo {pages} {106802} (\bibinfo {year}
  {2011})}\BibitemShut {NoStop}%
\bibitem [{\citenamefont {Mong}\ \emph {et~al.}(2010)\citenamefont {Mong},
  \citenamefont {Essin},\ and\ \citenamefont {Moore}}]{Mong2010prb81}%
  \BibitemOpen
  \bibfield  {author} {\bibinfo {author} {\bibfnamefont {R.~S.~K.}\
  \bibnamefont {Mong}}, \bibinfo {author} {\bibfnamefont {A.~M.}\ \bibnamefont
  {Essin}}, \ and\ \bibinfo {author} {\bibfnamefont {J.~E.}\ \bibnamefont
  {Moore}},\ }\href {\doibase 10.1103/PhysRevB.81.245209} {\bibfield  {journal}
  {\bibinfo  {journal} {Phys. Rev. B}\ }\textbf {\bibinfo {volume} {81}},\
  \bibinfo {pages} {245209} (\bibinfo {year} {2010})}\BibitemShut {NoStop}%
\bibitem [{\citenamefont {Hughes}\ \emph {et~al.}(2011)\citenamefont {Hughes},
  \citenamefont {Prodan},\ and\ \citenamefont {Bernevig}}]{Hughes2011prb83}%
  \BibitemOpen
  \bibfield  {author} {\bibinfo {author} {\bibfnamefont {T.~L.}\ \bibnamefont
  {Hughes}}, \bibinfo {author} {\bibfnamefont {E.}~\bibnamefont {Prodan}}, \
  and\ \bibinfo {author} {\bibfnamefont {B.~A.}\ \bibnamefont {Bernevig}},\
  }\href {\doibase 10.1103/PhysRevB.83.245132} {\bibfield  {journal} {\bibinfo
  {journal} {Phys. Rev. B}\ }\textbf {\bibinfo {volume} {83}},\ \bibinfo
  {pages} {245132} (\bibinfo {year} {2011})}\BibitemShut {NoStop}%
\bibitem [{\citenamefont {Fang}\ \emph {et~al.}(2012)\citenamefont {Fang},
  \citenamefont {Gilbert},\ and\ \citenamefont {Bernevig}}]{Fang2012prb86}%
  \BibitemOpen
  \bibfield  {author} {\bibinfo {author} {\bibfnamefont {C.}~\bibnamefont
  {Fang}}, \bibinfo {author} {\bibfnamefont {M.~J.}\ \bibnamefont {Gilbert}}, \
  and\ \bibinfo {author} {\bibfnamefont {B.~A.}\ \bibnamefont {Bernevig}},\
  }\href {\doibase 10.1103/PhysRevB.86.115112} {\bibfield  {journal} {\bibinfo
  {journal} {Phys. Rev. B}\ }\textbf {\bibinfo {volume} {86}},\ \bibinfo
  {pages} {115112} (\bibinfo {year} {2012})}\BibitemShut {NoStop}%
\bibitem [{\citenamefont {Kargarian}\ and\ \citenamefont
  {Fiete}(2013)}]{Kargarian2013prl110}%
  \BibitemOpen
  \bibfield  {author} {\bibinfo {author} {\bibfnamefont {M.}~\bibnamefont
  {Kargarian}}\ and\ \bibinfo {author} {\bibfnamefont {G.~A.}\ \bibnamefont
  {Fiete}},\ }\href {\doibase 10.1103/PhysRevLett.110.156403} {\bibfield
  {journal} {\bibinfo  {journal} {Phys. Rev. Lett.}\ }\textbf {\bibinfo
  {volume} {110}},\ \bibinfo {pages} {156403} (\bibinfo {year}
  {2013})}\BibitemShut {NoStop}%
\bibitem [{\citenamefont {Zhang}\ \emph {et~al.}(2013)\citenamefont {Zhang},
  \citenamefont {Kane},\ and\ \citenamefont {Mele}}]{Zhang2013prl111}%
  \BibitemOpen
  \bibfield  {author} {\bibinfo {author} {\bibfnamefont {F.}~\bibnamefont
  {Zhang}}, \bibinfo {author} {\bibfnamefont {C.~L.}\ \bibnamefont {Kane}}, \
  and\ \bibinfo {author} {\bibfnamefont {E.~J.}\ \bibnamefont {Mele}},\ }\href
  {\doibase 10.1103/PhysRevLett.111.056403} {\bibfield  {journal} {\bibinfo
  {journal} {Phys. Rev. Lett.}\ }\textbf {\bibinfo {volume} {111}},\ \bibinfo
  {pages} {056403} (\bibinfo {year} {2013})}\BibitemShut {NoStop}%
\bibitem [{\citenamefont {Ueno}\ \emph {et~al.}(2013)\citenamefont {Ueno},
  \citenamefont {Yamakage}, \citenamefont {Tanaka},\ and\ \citenamefont
  {Sato}}]{Ueno2013prl111}%
  \BibitemOpen
  \bibfield  {author} {\bibinfo {author} {\bibfnamefont {Y.}~\bibnamefont
  {Ueno}}, \bibinfo {author} {\bibfnamefont {A.}~\bibnamefont {Yamakage}},
  \bibinfo {author} {\bibfnamefont {Y.}~\bibnamefont {Tanaka}}, \ and\ \bibinfo
  {author} {\bibfnamefont {M.}~\bibnamefont {Sato}},\ }\href {\doibase
  10.1103/PhysRevLett.111.087002} {\bibfield  {journal} {\bibinfo  {journal}
  {Phys. Rev. Lett.}\ }\textbf {\bibinfo {volume} {111}},\ \bibinfo {pages}
  {087002} (\bibinfo {year} {2013})}\BibitemShut {NoStop}%
\bibitem [{\citenamefont {Chiu}\ \emph {et~al.}(2013)\citenamefont {Chiu},
  \citenamefont {Yao},\ and\ \citenamefont {Ryu}}]{Chiu2013prb88}%
  \BibitemOpen
  \bibfield  {author} {\bibinfo {author} {\bibfnamefont {C.-K.}\ \bibnamefont
  {Chiu}}, \bibinfo {author} {\bibfnamefont {H.}~\bibnamefont {Yao}}, \ and\
  \bibinfo {author} {\bibfnamefont {S.}~\bibnamefont {Ryu}},\ }\href {\doibase
  10.1103/PhysRevB.88.075142} {\bibfield  {journal} {\bibinfo  {journal} {Phys.
  Rev. B}\ }\textbf {\bibinfo {volume} {88}},\ \bibinfo {pages} {075142}
  (\bibinfo {year} {2013})}\BibitemShut {NoStop}%
\bibitem [{\citenamefont {Morimoto}\ and\ \citenamefont
  {Furusaki}(2013)}]{Morimoto2013prb88}%
  \BibitemOpen
  \bibfield  {author} {\bibinfo {author} {\bibfnamefont {T.}~\bibnamefont
  {Morimoto}}\ and\ \bibinfo {author} {\bibfnamefont {A.}~\bibnamefont
  {Furusaki}},\ }\href {\doibase 10.1103/PhysRevB.88.125129} {\bibfield
  {journal} {\bibinfo  {journal} {Phys. Rev. B}\ }\textbf {\bibinfo {volume}
  {88}},\ \bibinfo {pages} {125129} (\bibinfo {year} {2013})}\BibitemShut
  {NoStop}%
\bibitem [{\citenamefont {Slager}\ \emph {et~al.}(2013)\citenamefont {Slager},
  \citenamefont {Mesaros}, \citenamefont {Juricic},\ and\ \citenamefont
  {Zaanen}}]{Slager2013nphys9}%
  \BibitemOpen
  \bibfield  {author} {\bibinfo {author} {\bibfnamefont {R.-J.}\ \bibnamefont
  {Slager}}, \bibinfo {author} {\bibfnamefont {A.}~\bibnamefont {Mesaros}},
  \bibinfo {author} {\bibfnamefont {V.}~\bibnamefont {Juricic}}, \ and\
  \bibinfo {author} {\bibfnamefont {J.}~\bibnamefont {Zaanen}},\ }\href
  {\doibase 10.1038/nphys2513} {\bibfield  {journal} {\bibinfo  {journal} {Nat.
  Phys.}\ }\textbf {\bibinfo {volume} {9}},\ \bibinfo {pages} {98} (\bibinfo
  {year} {2013})}\BibitemShut {NoStop}%
\bibitem [{\citenamefont {Jadaun}\ \emph {et~al.}(2013)\citenamefont {Jadaun},
  \citenamefont {Xiao}, \citenamefont {Niu},\ and\ \citenamefont
  {Banerjee}}]{Jadaun2013prb88}%
  \BibitemOpen
  \bibfield  {author} {\bibinfo {author} {\bibfnamefont {P.}~\bibnamefont
  {Jadaun}}, \bibinfo {author} {\bibfnamefont {D.}~\bibnamefont {Xiao}},
  \bibinfo {author} {\bibfnamefont {Q.}~\bibnamefont {Niu}}, \ and\ \bibinfo
  {author} {\bibfnamefont {S.~K.}\ \bibnamefont {Banerjee}},\ }\href {\doibase
  10.1103/PhysRevB.88.085110} {\bibfield  {journal} {\bibinfo  {journal} {Phys.
  Rev. B}\ }\textbf {\bibinfo {volume} {88}},\ \bibinfo {pages} {085110}
  (\bibinfo {year} {2013})}\BibitemShut {NoStop}%
\bibitem [{\citenamefont {Alexandradinata}\ \emph {et~al.}(2014)\citenamefont
  {Alexandradinata}, \citenamefont {Fang}, \citenamefont {Gilbert},\ and\
  \citenamefont {Bernevig}}]{Alexandradinata2014prl113}%
  \BibitemOpen
  \bibfield  {author} {\bibinfo {author} {\bibfnamefont {A.}~\bibnamefont
  {Alexandradinata}}, \bibinfo {author} {\bibfnamefont {C.}~\bibnamefont
  {Fang}}, \bibinfo {author} {\bibfnamefont {M.~J.}\ \bibnamefont {Gilbert}}, \
  and\ \bibinfo {author} {\bibfnamefont {B.~A.}\ \bibnamefont {Bernevig}},\
  }\href {\doibase 10.1103/PhysRevLett.113.116403} {\bibfield  {journal}
  {\bibinfo  {journal} {Phys. Rev. Lett.}\ }\textbf {\bibinfo {volume} {113}},\
  \bibinfo {pages} {116403} (\bibinfo {year} {2014})}\BibitemShut {NoStop}%
\bibitem [{\citenamefont {Fulga}\ \emph {et~al.}(2014)\citenamefont {Fulga},
  \citenamefont {van Heck}, \citenamefont {Edge},\ and\ \citenamefont
  {Akhmerov}}]{Fulga2014prb89}%
  \BibitemOpen
  \bibfield  {author} {\bibinfo {author} {\bibfnamefont {I.~C.}\ \bibnamefont
  {Fulga}}, \bibinfo {author} {\bibfnamefont {B.}~\bibnamefont {van Heck}},
  \bibinfo {author} {\bibfnamefont {J.~M.}\ \bibnamefont {Edge}}, \ and\
  \bibinfo {author} {\bibfnamefont {A.~R.}\ \bibnamefont {Akhmerov}},\ }\href
  {\doibase 10.1103/PhysRevB.89.155424} {\bibfield  {journal} {\bibinfo
  {journal} {Phys. Rev. B}\ }\textbf {\bibinfo {volume} {89}},\ \bibinfo
  {pages} {155424} (\bibinfo {year} {2014})}\BibitemShut {NoStop}%
\bibitem [{\citenamefont {Benalcazar}\ \emph {et~al.}(2014)\citenamefont
  {Benalcazar}, \citenamefont {Teo},\ and\ \citenamefont
  {Hughes}}]{Benalcazar2014prb89}%
  \BibitemOpen
  \bibfield  {author} {\bibinfo {author} {\bibfnamefont {W.~A.}\ \bibnamefont
  {Benalcazar}}, \bibinfo {author} {\bibfnamefont {J.~C.~Y.}\ \bibnamefont
  {Teo}}, \ and\ \bibinfo {author} {\bibfnamefont {T.~L.}\ \bibnamefont
  {Hughes}},\ }\href {\doibase 10.1103/PhysRevB.89.224503} {\bibfield
  {journal} {\bibinfo  {journal} {Phys. Rev. B}\ }\textbf {\bibinfo {volume}
  {89}},\ \bibinfo {pages} {224503} (\bibinfo {year} {2014})}\BibitemShut
  {NoStop}%
\bibitem [{\citenamefont {Shiozaki}\ and\ \citenamefont
  {Sato}(2014)}]{Shiozaki2014prb90}%
  \BibitemOpen
  \bibfield  {author} {\bibinfo {author} {\bibfnamefont {K.}~\bibnamefont
  {Shiozaki}}\ and\ \bibinfo {author} {\bibfnamefont {M.}~\bibnamefont
  {Sato}},\ }\href {\doibase 10.1103/PhysRevB.90.165114} {\bibfield  {journal}
  {\bibinfo  {journal} {Phys. Rev. B}\ }\textbf {\bibinfo {volume} {90}},\
  \bibinfo {pages} {165114} (\bibinfo {year} {2014})}\BibitemShut {NoStop}%
\bibitem [{\citenamefont {Parameswaran}\ \emph {et~al.}(2013)\citenamefont
  {Parameswaran}, \citenamefont {Turner}, \citenamefont {Arovas},\ and\
  \citenamefont {Vishwanath}}]{Parameswaran2013nphys9}%
  \BibitemOpen
  \bibfield  {author} {\bibinfo {author} {\bibfnamefont {S.~A.}\ \bibnamefont
  {Parameswaran}}, \bibinfo {author} {\bibfnamefont {A.~M.}\ \bibnamefont
  {Turner}}, \bibinfo {author} {\bibfnamefont {D.~P.}\ \bibnamefont {Arovas}},
  \ and\ \bibinfo {author} {\bibfnamefont {A.}~\bibnamefont {Vishwanath}},\
  }\href@noop {} {\bibfield  {journal} {\bibinfo  {journal} {Nat. Phys.}\
  }\textbf {\bibinfo {volume} {9}},\ \bibinfo {pages} {299} (\bibinfo {year}
  {2013})}\BibitemShut {NoStop}%
\bibitem [{\citenamefont {Liu}\ \emph {et~al.}(2014)\citenamefont {Liu},
  \citenamefont {Zhang},\ and\ \citenamefont {VanLeeuwen}}]{Liu2014prb90}%
  \BibitemOpen
  \bibfield  {author} {\bibinfo {author} {\bibfnamefont {C.-X.}\ \bibnamefont
  {Liu}}, \bibinfo {author} {\bibfnamefont {R.-X.}\ \bibnamefont {Zhang}}, \
  and\ \bibinfo {author} {\bibfnamefont {B.~K.}\ \bibnamefont {VanLeeuwen}},\
  }\href {\doibase 10.1103/PhysRevB.90.085304} {\bibfield  {journal} {\bibinfo
  {journal} {Phys. Rev. B}\ }\textbf {\bibinfo {volume} {90}},\ \bibinfo
  {pages} {085304} (\bibinfo {year} {2014})}\BibitemShut {NoStop}%
\bibitem [{\citenamefont {Young}\ and\ \citenamefont
  {Kane}(2015)}]{Young2015prl115}%
  \BibitemOpen
  \bibfield  {author} {\bibinfo {author} {\bibfnamefont {S.~M.}\ \bibnamefont
  {Young}}\ and\ \bibinfo {author} {\bibfnamefont {C.~L.}\ \bibnamefont
  {Kane}},\ }\href {\doibase 10.1103/PhysRevLett.115.126803} {\bibfield
  {journal} {\bibinfo  {journal} {Phys. Rev. Lett.}\ }\textbf {\bibinfo
  {volume} {115}},\ \bibinfo {pages} {126803} (\bibinfo {year}
  {2015})}\BibitemShut {NoStop}%
\bibitem [{\citenamefont {Fang}\ and\ \citenamefont
  {Fu}(2015)}]{Fang2015prb91}%
  \BibitemOpen
  \bibfield  {author} {\bibinfo {author} {\bibfnamefont {C.}~\bibnamefont
  {Fang}}\ and\ \bibinfo {author} {\bibfnamefont {L.}~\bibnamefont {Fu}},\
  }\href {\doibase 10.1103/PhysRevB.91.161105} {\bibfield  {journal} {\bibinfo
  {journal} {Phys. Rev. B}\ }\textbf {\bibinfo {volume} {91}},\ \bibinfo
  {pages} {161105(R)} (\bibinfo {year} {2015})}\BibitemShut {NoStop}%
\bibitem [{\citenamefont {Shiozaki}\ \emph {et~al.}(2015)\citenamefont
  {Shiozaki}, \citenamefont {Sato},\ and\ \citenamefont
  {Gomi}}]{Shiozaki2015prb91}%
  \BibitemOpen
  \bibfield  {author} {\bibinfo {author} {\bibfnamefont {K.}~\bibnamefont
  {Shiozaki}}, \bibinfo {author} {\bibfnamefont {M.}~\bibnamefont {Sato}}, \
  and\ \bibinfo {author} {\bibfnamefont {K.}~\bibnamefont {Gomi}},\ }\href
  {\doibase 10.1103/PhysRevB.91.155120} {\bibfield  {journal} {\bibinfo
  {journal} {Phys. Rev. B}\ }\textbf {\bibinfo {volume} {91}},\ \bibinfo
  {pages} {155120} (\bibinfo {year} {2015})}\BibitemShut {NoStop}%
\bibitem [{\citenamefont {Varjas}\ \emph {et~al.}(2015)\citenamefont {Varjas},
  \citenamefont {de~Juan},\ and\ \citenamefont {Lu}}]{Varjas2015prb92}%
  \BibitemOpen
  \bibfield  {author} {\bibinfo {author} {\bibfnamefont {D.}~\bibnamefont
  {Varjas}}, \bibinfo {author} {\bibfnamefont {F.}~\bibnamefont {de~Juan}}, \
  and\ \bibinfo {author} {\bibfnamefont {Y.-M.}\ \bibnamefont {Lu}},\ }\href
  {\doibase 10.1103/PhysRevB.92.195116} {\bibfield  {journal} {\bibinfo
  {journal} {Phys. Rev. B}\ }\textbf {\bibinfo {volume} {92}},\ \bibinfo
  {pages} {195116} (\bibinfo {year} {2015})}\BibitemShut {NoStop}%
\bibitem [{\citenamefont {Watanabe}\ \emph {et~al.}(2015)\citenamefont
  {Watanabe}, \citenamefont {Po}, \citenamefont {Vishwanath},\ and\
  \citenamefont {Zaletel}}]{Watanabe2015proc112}%
  \BibitemOpen
  \bibfield  {author} {\bibinfo {author} {\bibfnamefont {H.}~\bibnamefont
  {Watanabe}}, \bibinfo {author} {\bibfnamefont {H.~C.}\ \bibnamefont {Po}},
  \bibinfo {author} {\bibfnamefont {A.}~\bibnamefont {Vishwanath}}, \ and\
  \bibinfo {author} {\bibfnamefont {M.}~\bibnamefont {Zaletel}},\ }\href@noop
  {} {\bibfield  {journal} {\bibinfo  {journal} {Proc. Natl. Acad. Sci. USA}\
  }\textbf {\bibinfo {volume} {112}},\ \bibinfo {pages} {14551} (\bibinfo
  {year} {2015})}\BibitemShut {NoStop}%
\bibitem [{\citenamefont {Po}\ \emph {et~al.}(2016)\citenamefont {Po},
  \citenamefont {Watanabe}, \citenamefont {Zaletel},\ and\ \citenamefont
  {Vishwanath}}]{Po2016sciadv2}%
  \BibitemOpen
  \bibfield  {author} {\bibinfo {author} {\bibfnamefont {H.~C.}\ \bibnamefont
  {Po}}, \bibinfo {author} {\bibfnamefont {H.}~\bibnamefont {Watanabe}},
  \bibinfo {author} {\bibfnamefont {M.~P.}\ \bibnamefont {Zaletel}}, \ and\
  \bibinfo {author} {\bibfnamefont {A.}~\bibnamefont {Vishwanath}},\
  }\href@noop {} {\bibfield  {journal} {\bibinfo  {journal} {Sci. Adv.}\
  }\textbf {\bibinfo {volume} {2}},\ \bibinfo {pages} {e1501782} (\bibinfo
  {year} {2016})}\BibitemShut {NoStop}%
\bibitem [{\citenamefont {Lu}\ \emph {et~al.}(2016)\citenamefont {Lu},
  \citenamefont {Fang}, \citenamefont {Fu}, \citenamefont {Johnson},
  \citenamefont {Joannopoulos},\ and\ \citenamefont
  {Solja{\v{c}}i{\'c}}}]{Lu2016nphys12}%
  \BibitemOpen
  \bibfield  {author} {\bibinfo {author} {\bibfnamefont {L.}~\bibnamefont
  {Lu}}, \bibinfo {author} {\bibfnamefont {C.}~\bibnamefont {Fang}}, \bibinfo
  {author} {\bibfnamefont {L.}~\bibnamefont {Fu}}, \bibinfo {author}
  {\bibfnamefont {S.~G.}\ \bibnamefont {Johnson}}, \bibinfo {author}
  {\bibfnamefont {J.~D.}\ \bibnamefont {Joannopoulos}}, \ and\ \bibinfo
  {author} {\bibfnamefont {M.}~\bibnamefont {Solja{\v{c}}i{\'c}}},\ }\href@noop
  {} {\bibfield  {journal} {\bibinfo  {journal} {Nat. Phys.}\ }\textbf
  {\bibinfo {volume} {12}},\ \bibinfo {pages} {337} (\bibinfo {year}
  {2016})}\BibitemShut {NoStop}%
\bibitem [{\citenamefont {Dong}\ and\ \citenamefont
  {Liu}(2016)}]{Dong2016prb93}%
  \BibitemOpen
  \bibfield  {author} {\bibinfo {author} {\bibfnamefont {X.-Y.}\ \bibnamefont
  {Dong}}\ and\ \bibinfo {author} {\bibfnamefont {C.-X.}\ \bibnamefont {Liu}},\
  }\href {\doibase 10.1103/PhysRevB.93.045429} {\bibfield  {journal} {\bibinfo
  {journal} {Phys. Rev. B}\ }\textbf {\bibinfo {volume} {93}},\ \bibinfo
  {pages} {045429} (\bibinfo {year} {2016})}\BibitemShut {NoStop}%
\bibitem [{\citenamefont {Chen}\ \emph {et~al.}(2016)\citenamefont {Chen},
  \citenamefont {Kim},\ and\ \citenamefont {Kee}}]{Chen2016prb93}%
  \BibitemOpen
  \bibfield  {author} {\bibinfo {author} {\bibfnamefont {Y.}~\bibnamefont
  {Chen}}, \bibinfo {author} {\bibfnamefont {H.-S.}\ \bibnamefont {Kim}}, \
  and\ \bibinfo {author} {\bibfnamefont {H.-Y.}\ \bibnamefont {Kee}},\ }\href
  {\doibase 10.1103/PhysRevB.93.155140} {\bibfield  {journal} {\bibinfo
  {journal} {Phys. Rev. B}\ }\textbf {\bibinfo {volume} {93}},\ \bibinfo
  {pages} {155140} (\bibinfo {year} {2016})}\BibitemShut {NoStop}%
\bibitem [{\citenamefont {Kim}\ and\ \citenamefont
  {Murakami}(2016)}]{Kim2016prb93}%
  \BibitemOpen
  \bibfield  {author} {\bibinfo {author} {\bibfnamefont {H.}~\bibnamefont
  {Kim}}\ and\ \bibinfo {author} {\bibfnamefont {S.}~\bibnamefont {Murakami}},\
  }\href {\doibase 10.1103/PhysRevB.93.195138} {\bibfield  {journal} {\bibinfo
  {journal} {Phys. Rev. B}\ }\textbf {\bibinfo {volume} {93}},\ \bibinfo
  {pages} {195138} (\bibinfo {year} {2016})}\BibitemShut {NoStop}%
\bibitem [{\citenamefont {Shiozaki}\ \emph {et~al.}(2016)\citenamefont
  {Shiozaki}, \citenamefont {Sato},\ and\ \citenamefont
  {Gomi}}]{Shiozaki2016prb93}%
  \BibitemOpen
  \bibfield  {author} {\bibinfo {author} {\bibfnamefont {K.}~\bibnamefont
  {Shiozaki}}, \bibinfo {author} {\bibfnamefont {M.}~\bibnamefont {Sato}}, \
  and\ \bibinfo {author} {\bibfnamefont {K.}~\bibnamefont {Gomi}},\ }\href
  {\doibase 10.1103/PhysRevB.93.195413} {\bibfield  {journal} {\bibinfo
  {journal} {Phys. Rev. B}\ }\textbf {\bibinfo {volume} {93}},\ \bibinfo
  {pages} {195413} (\bibinfo {year} {2016})}\BibitemShut {NoStop}%
\bibitem [{\citenamefont {Wieder}\ and\ \citenamefont
  {Kane}(2016)}]{Wieder2016prb94}%
  \BibitemOpen
  \bibfield  {author} {\bibinfo {author} {\bibfnamefont {B.~J.}\ \bibnamefont
  {Wieder}}\ and\ \bibinfo {author} {\bibfnamefont {C.~L.}\ \bibnamefont
  {Kane}},\ }\href {\doibase 10.1103/PhysRevB.94.155108} {\bibfield  {journal}
  {\bibinfo  {journal} {Phys. Rev. B}\ }\textbf {\bibinfo {volume} {94}},\
  \bibinfo {pages} {155108} (\bibinfo {year} {2016})}\BibitemShut {NoStop}%
\bibitem [{\citenamefont {Zhao}\ and\ \citenamefont
  {Schnyder}(2016)}]{Zhao2016prb94}%
  \BibitemOpen
  \bibfield  {author} {\bibinfo {author} {\bibfnamefont {Y.~X.}\ \bibnamefont
  {Zhao}}\ and\ \bibinfo {author} {\bibfnamefont {A.~P.}\ \bibnamefont
  {Schnyder}},\ }\href {\doibase 10.1103/PhysRevB.94.195109} {\bibfield
  {journal} {\bibinfo  {journal} {Phys. Rev. B}\ }\textbf {\bibinfo {volume}
  {94}},\ \bibinfo {pages} {195109} (\bibinfo {year} {2016})}\BibitemShut
  {NoStop}%
\bibitem [{\citenamefont {Wang}\ \emph {et~al.}(2016)\citenamefont {Wang},
  \citenamefont {Alexandradinata}, \citenamefont {Cava},\ and\ \citenamefont
  {Bernevig}}]{Wang2016nature532}%
  \BibitemOpen
  \bibfield  {author} {\bibinfo {author} {\bibfnamefont {Z.}~\bibnamefont
  {Wang}}, \bibinfo {author} {\bibfnamefont {A.}~\bibnamefont
  {Alexandradinata}}, \bibinfo {author} {\bibfnamefont {R.~J.}\ \bibnamefont
  {Cava}}, \ and\ \bibinfo {author} {\bibfnamefont {B.~A.}\ \bibnamefont
  {Bernevig}},\ }\href@noop {} {\bibfield  {journal} {\bibinfo  {journal}
  {Nature}\ }\textbf {\bibinfo {volume} {532}},\ \bibinfo {pages} {189}
  (\bibinfo {year} {2016})}\BibitemShut {NoStop}%
\bibitem [{\citenamefont {Bzdu{\v{s}}ek}\ \emph {et~al.}(2016)\citenamefont
  {Bzdu{\v{s}}ek}, \citenamefont {Wu}, \citenamefont {R{\"u}egg}, \citenamefont
  {Sigrist},\ and\ \citenamefont {Soluyanov}}]{Bzduvsek2016nature538}%
  \BibitemOpen
  \bibfield  {author} {\bibinfo {author} {\bibfnamefont {T.}~\bibnamefont
  {Bzdu{\v{s}}ek}}, \bibinfo {author} {\bibfnamefont {Q.}~\bibnamefont {Wu}},
  \bibinfo {author} {\bibfnamefont {A.}~\bibnamefont {R{\"u}egg}}, \bibinfo
  {author} {\bibfnamefont {M.}~\bibnamefont {Sigrist}}, \ and\ \bibinfo
  {author} {\bibfnamefont {A.~A.}\ \bibnamefont {Soluyanov}},\ }\href@noop {}
  {\bibfield  {journal} {\bibinfo  {journal} {Nature}\ }\textbf {\bibinfo
  {volume} {538}},\ \bibinfo {pages} {75} (\bibinfo {year} {2016})}\BibitemShut
  {NoStop}%
\bibitem [{\citenamefont {Yang}\ \emph {et~al.}(2017)\citenamefont {Yang},
  \citenamefont {Bojesen}, \citenamefont {Morimoto},\ and\ \citenamefont
  {Furusaki}}]{Yang2017prb95}%
  \BibitemOpen
  \bibfield  {author} {\bibinfo {author} {\bibfnamefont {B.-J.}\ \bibnamefont
  {Yang}}, \bibinfo {author} {\bibfnamefont {T.~A.}\ \bibnamefont {Bojesen}},
  \bibinfo {author} {\bibfnamefont {T.}~\bibnamefont {Morimoto}}, \ and\
  \bibinfo {author} {\bibfnamefont {A.}~\bibnamefont {Furusaki}},\ }\href
  {\doibase 10.1103/PhysRevB.95.075135} {\bibfield  {journal} {\bibinfo
  {journal} {Phys. Rev. B}\ }\textbf {\bibinfo {volume} {95}},\ \bibinfo
  {pages} {075135} (\bibinfo {year} {2017})}\BibitemShut {NoStop}%
\bibitem [{\citenamefont {Takahashi}\ \emph {et~al.}(2017)\citenamefont
  {Takahashi}, \citenamefont {Hirayama},\ and\ \citenamefont
  {Murakami}}]{Takahashi2017prb96}%
  \BibitemOpen
  \bibfield  {author} {\bibinfo {author} {\bibfnamefont {R.}~\bibnamefont
  {Takahashi}}, \bibinfo {author} {\bibfnamefont {M.}~\bibnamefont {Hirayama}},
  \ and\ \bibinfo {author} {\bibfnamefont {S.}~\bibnamefont {Murakami}},\
  }\href {\doibase 10.1103/PhysRevB.96.155206} {\bibfield  {journal} {\bibinfo
  {journal} {Phys. Rev. B}\ }\textbf {\bibinfo {volume} {96}},\ \bibinfo
  {pages} {155206} (\bibinfo {year} {2017})}\BibitemShut {NoStop}%
\bibitem [{\citenamefont {Furusaki}(2017)}]{Furusaki2017scibul62}%
  \BibitemOpen
  \bibfield  {author} {\bibinfo {author} {\bibfnamefont {A.}~\bibnamefont
  {Furusaki}},\ }\href@noop {} {\bibfield  {journal} {\bibinfo  {journal}
  {Science Bulletin}\ }\textbf {\bibinfo {volume} {62}},\ \bibinfo {pages}
  {788} (\bibinfo {year} {2017})}\BibitemShut {NoStop}%
\bibitem [{\citenamefont {Chen}\ \emph {et~al.}(2018)\citenamefont {Chen},
  \citenamefont {Po}, \citenamefont {Neaton},\ and\ \citenamefont
  {Vishwanath}}]{Chen2018nphys14}%
  \BibitemOpen
  \bibfield  {author} {\bibinfo {author} {\bibfnamefont {R.}~\bibnamefont
  {Chen}}, \bibinfo {author} {\bibfnamefont {H.~C.}\ \bibnamefont {Po}},
  \bibinfo {author} {\bibfnamefont {J.~B.}\ \bibnamefont {Neaton}}, \ and\
  \bibinfo {author} {\bibfnamefont {A.}~\bibnamefont {Vishwanath}},\
  }\href@noop {} {\bibfield  {journal} {\bibinfo  {journal} {Nat. Phys.}\
  }\textbf {\bibinfo {volume} {14}},\ \bibinfo {pages} {55} (\bibinfo {year}
  {2018})}\BibitemShut {NoStop}%
\bibitem [{\citenamefont {Benalcazar}\ \emph
  {et~al.}(2017{\natexlab{a}})\citenamefont {Benalcazar}, \citenamefont
  {Bernevig},\ and\ \citenamefont {Hughes}}]{Benalcazar2017sci357}%
  \BibitemOpen
  \bibfield  {author} {\bibinfo {author} {\bibfnamefont {W.~A.}\ \bibnamefont
  {Benalcazar}}, \bibinfo {author} {\bibfnamefont {B.~A.}\ \bibnamefont
  {Bernevig}}, \ and\ \bibinfo {author} {\bibfnamefont {T.~L.}\ \bibnamefont
  {Hughes}},\ }\href@noop {} {\bibfield  {journal} {\bibinfo  {journal}
  {Science}\ }\textbf {\bibinfo {volume} {357}},\ \bibinfo {pages} {61}
  (\bibinfo {year} {2017}{\natexlab{a}})}\BibitemShut {NoStop}%
\bibitem [{\citenamefont {Benalcazar}\ \emph
  {et~al.}(2017{\natexlab{b}})\citenamefont {Benalcazar}, \citenamefont
  {Bernevig},\ and\ \citenamefont {Hughes}}]{Benalcazar2017prb96}%
  \BibitemOpen
  \bibfield  {author} {\bibinfo {author} {\bibfnamefont {W.~A.}\ \bibnamefont
  {Benalcazar}}, \bibinfo {author} {\bibfnamefont {B.~A.}\ \bibnamefont
  {Bernevig}}, \ and\ \bibinfo {author} {\bibfnamefont {T.~L.}\ \bibnamefont
  {Hughes}},\ }\href {\doibase 10.1103/PhysRevB.96.245115} {\bibfield
  {journal} {\bibinfo  {journal} {Phys. Rev. B}\ }\textbf {\bibinfo {volume}
  {96}},\ \bibinfo {pages} {245115} (\bibinfo {year}
  {2017}{\natexlab{b}})}\BibitemShut {NoStop}%
\bibitem [{\citenamefont {Langbehn}\ \emph {et~al.}(2017)\citenamefont
  {Langbehn}, \citenamefont {Peng}, \citenamefont {Trifunovic}, \citenamefont
  {von Oppen},\ and\ \citenamefont {Brouwer}}]{Langbehn2017prl119}%
  \BibitemOpen
  \bibfield  {author} {\bibinfo {author} {\bibfnamefont {J.}~\bibnamefont
  {Langbehn}}, \bibinfo {author} {\bibfnamefont {Y.}~\bibnamefont {Peng}},
  \bibinfo {author} {\bibfnamefont {L.}~\bibnamefont {Trifunovic}}, \bibinfo
  {author} {\bibfnamefont {F.}~\bibnamefont {von Oppen}}, \ and\ \bibinfo
  {author} {\bibfnamefont {P.~W.}\ \bibnamefont {Brouwer}},\ }\href {\doibase
  10.1103/PhysRevLett.119.246401} {\bibfield  {journal} {\bibinfo  {journal}
  {Phys. Rev. Lett.}\ }\textbf {\bibinfo {volume} {119}},\ \bibinfo {pages}
  {246401} (\bibinfo {year} {2017})}\BibitemShut {NoStop}%
\bibitem [{\citenamefont {Song}\ \emph {et~al.}(2017)\citenamefont {Song},
  \citenamefont {Fang},\ and\ \citenamefont {Fang}}]{Song2017prl119}%
  \BibitemOpen
  \bibfield  {author} {\bibinfo {author} {\bibfnamefont {Z.}~\bibnamefont
  {Song}}, \bibinfo {author} {\bibfnamefont {Z.}~\bibnamefont {Fang}}, \ and\
  \bibinfo {author} {\bibfnamefont {C.}~\bibnamefont {Fang}},\ }\href {\doibase
  10.1103/PhysRevLett.119.246402} {\bibfield  {journal} {\bibinfo  {journal}
  {Phys. Rev. Lett.}\ }\textbf {\bibinfo {volume} {119}},\ \bibinfo {pages}
  {246402} (\bibinfo {year} {2017})}\BibitemShut {NoStop}%
\bibitem [{\citenamefont {Schindler}\ \emph
  {et~al.}(2018{\natexlab{a}})\citenamefont {Schindler}, \citenamefont {Cook},
  \citenamefont {Vergniory}, \citenamefont {Wang}, \citenamefont {Parkin},
  \citenamefont {Bernevig},\ and\ \citenamefont
  {Neupert}}]{Schindler2018sciadv4}%
  \BibitemOpen
  \bibfield  {author} {\bibinfo {author} {\bibfnamefont {F.}~\bibnamefont
  {Schindler}}, \bibinfo {author} {\bibfnamefont {A.~M.}\ \bibnamefont {Cook}},
  \bibinfo {author} {\bibfnamefont {M.~G.}\ \bibnamefont {Vergniory}}, \bibinfo
  {author} {\bibfnamefont {Z.}~\bibnamefont {Wang}}, \bibinfo {author}
  {\bibfnamefont {S.~S.}\ \bibnamefont {Parkin}}, \bibinfo {author}
  {\bibfnamefont {B.~A.}\ \bibnamefont {Bernevig}}, \ and\ \bibinfo {author}
  {\bibfnamefont {T.}~\bibnamefont {Neupert}},\ }\href@noop {} {\bibfield
  {journal} {\bibinfo  {journal} {Sci. Adv.}\ }\textbf {\bibinfo {volume}
  {4}},\ \bibinfo {pages} {eaat0346} (\bibinfo {year}
  {2018}{\natexlab{a}})}\BibitemShut {NoStop}%
\bibitem [{\citenamefont {Khalaf}(2018)}]{Khalaf2018prb97}%
  \BibitemOpen
  \bibfield  {author} {\bibinfo {author} {\bibfnamefont {E.}~\bibnamefont
  {Khalaf}},\ }\href {\doibase 10.1103/PhysRevB.97.205136} {\bibfield
  {journal} {\bibinfo  {journal} {Phys. Rev. B}\ }\textbf {\bibinfo {volume}
  {97}},\ \bibinfo {pages} {205136} (\bibinfo {year} {2018})}\BibitemShut
  {NoStop}%
\bibitem [{\citenamefont {Schindler}\ \emph
  {et~al.}(2018{\natexlab{b}})\citenamefont {Schindler}, \citenamefont {Wang},
  \citenamefont {Vergniory}, \citenamefont {Cook}, \citenamefont {Murani},
  \citenamefont {Sengupta}, \citenamefont {Kasumov}, \citenamefont {Deblock},
  \citenamefont {Jeon}, \citenamefont {Drozdov} \emph
  {et~al.}}]{Schindler2018nphys14}%
  \BibitemOpen
  \bibfield  {author} {\bibinfo {author} {\bibfnamefont {F.}~\bibnamefont
  {Schindler}}, \bibinfo {author} {\bibfnamefont {Z.}~\bibnamefont {Wang}},
  \bibinfo {author} {\bibfnamefont {M.~G.}\ \bibnamefont {Vergniory}}, \bibinfo
  {author} {\bibfnamefont {A.~M.}\ \bibnamefont {Cook}}, \bibinfo {author}
  {\bibfnamefont {A.}~\bibnamefont {Murani}}, \bibinfo {author} {\bibfnamefont
  {S.}~\bibnamefont {Sengupta}}, \bibinfo {author} {\bibfnamefont {A.~Y.}\
  \bibnamefont {Kasumov}}, \bibinfo {author} {\bibfnamefont {R.}~\bibnamefont
  {Deblock}}, \bibinfo {author} {\bibfnamefont {S.}~\bibnamefont {Jeon}},
  \bibinfo {author} {\bibfnamefont {I.}~\bibnamefont {Drozdov}},  \emph
  {et~al.},\ }\href@noop {} {\bibfield  {journal} {\bibinfo  {journal} {Nat.
  Phys.}\ }\textbf {\bibinfo {volume} {14}},\ \bibinfo {pages} {918} (\bibinfo
  {year} {2018}{\natexlab{b}})}\BibitemShut {NoStop}%
\bibitem [{\citenamefont {Song}\ \emph
  {et~al.}(2018{\natexlab{a}})\citenamefont {Song}, \citenamefont {Zhang},\
  and\ \citenamefont {Fang}}]{Song2018prx8}%
  \BibitemOpen
  \bibfield  {author} {\bibinfo {author} {\bibfnamefont {Z.}~\bibnamefont
  {Song}}, \bibinfo {author} {\bibfnamefont {T.}~\bibnamefont {Zhang}}, \ and\
  \bibinfo {author} {\bibfnamefont {C.}~\bibnamefont {Fang}},\ }\href {\doibase
  10.1103/PhysRevX.8.031069} {\bibfield  {journal} {\bibinfo  {journal} {Phys.
  Rev. X}\ }\textbf {\bibinfo {volume} {8}},\ \bibinfo {pages} {031069}
  (\bibinfo {year} {2018}{\natexlab{a}})}\BibitemShut {NoStop}%
\bibitem [{\citenamefont {Khalaf}\ \emph {et~al.}(2018)\citenamefont {Khalaf},
  \citenamefont {Po}, \citenamefont {Vishwanath},\ and\ \citenamefont
  {Watanabe}}]{Khalaf2018prx8}%
  \BibitemOpen
  \bibfield  {author} {\bibinfo {author} {\bibfnamefont {E.}~\bibnamefont
  {Khalaf}}, \bibinfo {author} {\bibfnamefont {H.~C.}\ \bibnamefont {Po}},
  \bibinfo {author} {\bibfnamefont {A.}~\bibnamefont {Vishwanath}}, \ and\
  \bibinfo {author} {\bibfnamefont {H.}~\bibnamefont {Watanabe}},\ }\href
  {\doibase 10.1103/PhysRevX.8.031070} {\bibfield  {journal} {\bibinfo
  {journal} {Phys. Rev. X}\ }\textbf {\bibinfo {volume} {8}},\ \bibinfo {pages}
  {031070} (\bibinfo {year} {2018})}\BibitemShut {NoStop}%
\bibitem [{\citenamefont {Freed}\ and\ \citenamefont
  {Moore}(2013)}]{Freed2013springer}%
  \BibitemOpen
  \bibfield  {author} {\bibinfo {author} {\bibfnamefont {D.~S.}\ \bibnamefont
  {Freed}}\ and\ \bibinfo {author} {\bibfnamefont {G.~W.}\ \bibnamefont
  {Moore}},\ }in\ \href@noop {} {\emph {\bibinfo {booktitle} {Annales Henri
  Poincar{\'e}}}},\ Vol.~\bibinfo {volume} {14}\ (\bibinfo {organization}
  {Springer},\ \bibinfo {year} {2013})\ pp.\ \bibinfo {pages}
  {1927--2023}\BibitemShut {NoStop}%
\bibitem [{\citenamefont {Read}(2017)}]{Read2017prb95}%
  \BibitemOpen
  \bibfield  {author} {\bibinfo {author} {\bibfnamefont {N.}~\bibnamefont
  {Read}},\ }\href {\doibase 10.1103/PhysRevB.95.115309} {\bibfield  {journal}
  {\bibinfo  {journal} {Phys. Rev. B}\ }\textbf {\bibinfo {volume} {95}},\
  \bibinfo {pages} {115309} (\bibinfo {year} {2017})}\BibitemShut {NoStop}%
\bibitem [{\citenamefont {Shiozaki}\ \emph {et~al.}(2017)\citenamefont
  {Shiozaki}, \citenamefont {Sato},\ and\ \citenamefont
  {Gomi}}]{Shiozaki2017prb95}%
  \BibitemOpen
  \bibfield  {author} {\bibinfo {author} {\bibfnamefont {K.}~\bibnamefont
  {Shiozaki}}, \bibinfo {author} {\bibfnamefont {M.}~\bibnamefont {Sato}}, \
  and\ \bibinfo {author} {\bibfnamefont {K.}~\bibnamefont {Gomi}},\ }\href
  {\doibase 10.1103/PhysRevB.95.235425} {\bibfield  {journal} {\bibinfo
  {journal} {Phys. Rev. B}\ }\textbf {\bibinfo {volume} {95}},\ \bibinfo
  {pages} {235425} (\bibinfo {year} {2017})}\BibitemShut {NoStop}%
\bibitem [{\citenamefont {Shiozaki}\ \emph
  {et~al.}(2018{\natexlab{a}})\citenamefont {Shiozaki}, \citenamefont {Sato},\
  and\ \citenamefont {Gomi}}]{Shiozaki2018arXiv180206694}%
  \BibitemOpen
  \bibfield  {author} {\bibinfo {author} {\bibfnamefont {K.}~\bibnamefont
  {Shiozaki}}, \bibinfo {author} {\bibfnamefont {M.}~\bibnamefont {Sato}}, \
  and\ \bibinfo {author} {\bibfnamefont {K.}~\bibnamefont {Gomi}},\ }\href@noop
  {} {\bibfield  {journal} {\bibinfo  {journal} {arXiv preprint
  arXiv:1802.06694}\ } (\bibinfo {year} {2018}{\natexlab{a}})}\BibitemShut
  {NoStop}%
\bibitem [{\citenamefont {Shiozaki}\ \emph
  {et~al.}(2018{\natexlab{b}})\citenamefont {Shiozaki}, \citenamefont {Xiong},\
  and\ \citenamefont {Gomi}}]{Shiozaki2018arXiv181000801}%
  \BibitemOpen
  \bibfield  {author} {\bibinfo {author} {\bibfnamefont {K.}~\bibnamefont
  {Shiozaki}}, \bibinfo {author} {\bibfnamefont {C.~Z.}\ \bibnamefont {Xiong}},
  \ and\ \bibinfo {author} {\bibfnamefont {K.}~\bibnamefont {Gomi}},\
  }\href@noop {} {\bibfield  {journal} {\bibinfo  {journal} {arXiv preprint
  arXiv:1810.00801}\ } (\bibinfo {year} {2018}{\natexlab{b}})}\BibitemShut
  {NoStop}%
\bibitem [{\citenamefont {Kruthoff}\ \emph {et~al.}(2017)\citenamefont
  {Kruthoff}, \citenamefont {de~Boer}, \citenamefont {van Wezel}, \citenamefont
  {Kane},\ and\ \citenamefont {Slager}}]{Kruthoff2017prx7}%
  \BibitemOpen
  \bibfield  {author} {\bibinfo {author} {\bibfnamefont {J.}~\bibnamefont
  {Kruthoff}}, \bibinfo {author} {\bibfnamefont {J.}~\bibnamefont {de~Boer}},
  \bibinfo {author} {\bibfnamefont {J.}~\bibnamefont {van Wezel}}, \bibinfo
  {author} {\bibfnamefont {C.~L.}\ \bibnamefont {Kane}}, \ and\ \bibinfo
  {author} {\bibfnamefont {R.-J.}\ \bibnamefont {Slager}},\ }\href {\doibase
  10.1103/PhysRevX.7.041069} {\bibfield  {journal} {\bibinfo  {journal} {Phys.
  Rev. X}\ }\textbf {\bibinfo {volume} {7}},\ \bibinfo {pages} {041069}
  (\bibinfo {year} {2017})}\BibitemShut {NoStop}%
\bibitem [{\citenamefont {Po}\ \emph {et~al.}(2017)\citenamefont {Po},
  \citenamefont {Vishwanath},\ and\ \citenamefont {Watanabe}}]{Po2017ncommun}%
  \BibitemOpen
  \bibfield  {author} {\bibinfo {author} {\bibfnamefont {H.~C.}\ \bibnamefont
  {Po}}, \bibinfo {author} {\bibfnamefont {A.}~\bibnamefont {Vishwanath}}, \
  and\ \bibinfo {author} {\bibfnamefont {H.}~\bibnamefont {Watanabe}},\
  }\href@noop {} {\bibfield  {journal} {\bibinfo  {journal} {Nat. Commun.}\
  }\textbf {\bibinfo {volume} {8}},\ \bibinfo {pages} {50} (\bibinfo {year}
  {2017})}\BibitemShut {NoStop}%
\bibitem [{\citenamefont {Bradlyn}\ \emph {et~al.}(2017)\citenamefont
  {Bradlyn}, \citenamefont {Elcoro}, \citenamefont {Cano}, \citenamefont
  {Vergniory}, \citenamefont {Wang}, \citenamefont {Felser}, \citenamefont
  {Aroyo},\ and\ \citenamefont {Bernevig}}]{Bradlyn2017nature547}%
  \BibitemOpen
  \bibfield  {author} {\bibinfo {author} {\bibfnamefont {B.}~\bibnamefont
  {Bradlyn}}, \bibinfo {author} {\bibfnamefont {L.}~\bibnamefont {Elcoro}},
  \bibinfo {author} {\bibfnamefont {J.}~\bibnamefont {Cano}}, \bibinfo {author}
  {\bibfnamefont {M.}~\bibnamefont {Vergniory}}, \bibinfo {author}
  {\bibfnamefont {Z.}~\bibnamefont {Wang}}, \bibinfo {author} {\bibfnamefont
  {C.}~\bibnamefont {Felser}}, \bibinfo {author} {\bibfnamefont
  {M.}~\bibnamefont {Aroyo}}, \ and\ \bibinfo {author} {\bibfnamefont {B.~A.}\
  \bibnamefont {Bernevig}},\ }\href@noop {} {\bibfield  {journal} {\bibinfo
  {journal} {Nature}\ }\textbf {\bibinfo {volume} {547}},\ \bibinfo {pages}
  {298} (\bibinfo {year} {2017})}\BibitemShut {NoStop}%
\bibitem [{\citenamefont {Watanabe}\ \emph {et~al.}(2018)\citenamefont
  {Watanabe}, \citenamefont {Po},\ and\ \citenamefont
  {Vishwanath}}]{Watanabe2018sciadv4}%
  \BibitemOpen
  \bibfield  {author} {\bibinfo {author} {\bibfnamefont {H.}~\bibnamefont
  {Watanabe}}, \bibinfo {author} {\bibfnamefont {H.~C.}\ \bibnamefont {Po}}, \
  and\ \bibinfo {author} {\bibfnamefont {A.}~\bibnamefont {Vishwanath}},\
  }\href@noop {} {\bibfield  {journal} {\bibinfo  {journal} {Sci. Adv.}\
  }\textbf {\bibinfo {volume} {4}},\ \bibinfo {pages} {eaat8685} (\bibinfo
  {year} {2018})}\BibitemShut {NoStop}%
\bibitem [{\citenamefont {Altland}\ and\ \citenamefont
  {Zirnbauer}(1997)}]{AZ1997prb55}%
  \BibitemOpen
  \bibfield  {author} {\bibinfo {author} {\bibfnamefont {A.}~\bibnamefont
  {Altland}}\ and\ \bibinfo {author} {\bibfnamefont {M.~R.}\ \bibnamefont
  {Zirnbauer}},\ }\href {\doibase 10.1103/PhysRevB.55.1142} {\bibfield
  {journal} {\bibinfo  {journal} {Phys. Rev. B}\ }\textbf {\bibinfo {volume}
  {55}},\ \bibinfo {pages} {1142} (\bibinfo {year} {1997})}\BibitemShut
  {NoStop}%
\bibitem [{\citenamefont {Aroyo}\ \emph {et~al.}(2006)\citenamefont {Aroyo},
  \citenamefont {Perez-Mato}, \citenamefont {Capillas}, \citenamefont
  {Kroumova}, \citenamefont {Ivantchev}, \citenamefont {Madariaga},
  \citenamefont {Kirov},\ and\ \citenamefont {Wondratschek}}]{Aroyo2006bilbao}%
  \BibitemOpen
  \bibfield  {author} {\bibinfo {author} {\bibfnamefont {M.~I.}\ \bibnamefont
  {Aroyo}}, \bibinfo {author} {\bibfnamefont {J.~M.}\ \bibnamefont
  {Perez-Mato}}, \bibinfo {author} {\bibfnamefont {C.}~\bibnamefont
  {Capillas}}, \bibinfo {author} {\bibfnamefont {E.}~\bibnamefont {Kroumova}},
  \bibinfo {author} {\bibfnamefont {S.}~\bibnamefont {Ivantchev}}, \bibinfo
  {author} {\bibfnamefont {G.}~\bibnamefont {Madariaga}}, \bibinfo {author}
  {\bibfnamefont {A.}~\bibnamefont {Kirov}}, \ and\ \bibinfo {author}
  {\bibfnamefont {H.}~\bibnamefont {Wondratschek}},\ }\href@noop {} {\bibfield
  {journal} {\bibinfo  {journal} {Zeitschrift f{\"u}r
  Kristallographie-Crystalline Materials}\ }\textbf {\bibinfo {volume} {221}},\
  \bibinfo {pages} {15} (\bibinfo {year} {2006})}\BibitemShut {NoStop}%
\bibitem [{\citenamefont {Song}\ \emph
  {et~al.}(2018{\natexlab{b}})\citenamefont {Song}, \citenamefont {Zhang},
  \citenamefont {Fang},\ and\ \citenamefont {Fang}}]{Song2018ncommun9}%
  \BibitemOpen
  \bibfield  {author} {\bibinfo {author} {\bibfnamefont {Z.}~\bibnamefont
  {Song}}, \bibinfo {author} {\bibfnamefont {T.}~\bibnamefont {Zhang}},
  \bibinfo {author} {\bibfnamefont {Z.}~\bibnamefont {Fang}}, \ and\ \bibinfo
  {author} {\bibfnamefont {C.}~\bibnamefont {Fang}},\ }\href@noop {} {\bibfield
   {journal} {\bibinfo  {journal} {Nat. Commun.}\ }\textbf {\bibinfo {volume}
  {9}},\ \bibinfo {pages} {3530} (\bibinfo {year}
  {2018}{\natexlab{b}})}\BibitemShut {NoStop}%
\bibitem [{\citenamefont {Murakami}(2007)}]{Murakami2007njp9}%
  \BibitemOpen
  \bibfield  {author} {\bibinfo {author} {\bibfnamefont {S.}~\bibnamefont
  {Murakami}},\ }\href@noop {} {\bibfield  {journal} {\bibinfo  {journal} {New
  Journal of Physics}\ }\textbf {\bibinfo {volume} {9}},\ \bibinfo {pages}
  {356} (\bibinfo {year} {2007})}\BibitemShut {NoStop}%
\bibitem [{\citenamefont {Wan}\ \emph {et~al.}(2011)\citenamefont {Wan},
  \citenamefont {Turner}, \citenamefont {Vishwanath},\ and\ \citenamefont
  {Savrasov}}]{Wan2011prb83}%
  \BibitemOpen
  \bibfield  {author} {\bibinfo {author} {\bibfnamefont {X.}~\bibnamefont
  {Wan}}, \bibinfo {author} {\bibfnamefont {A.~M.}\ \bibnamefont {Turner}},
  \bibinfo {author} {\bibfnamefont {A.}~\bibnamefont {Vishwanath}}, \ and\
  \bibinfo {author} {\bibfnamefont {S.~Y.}\ \bibnamefont {Savrasov}},\ }\href
  {\doibase 10.1103/PhysRevB.83.205101} {\bibfield  {journal} {\bibinfo
  {journal} {Phys. Rev. B}\ }\textbf {\bibinfo {volume} {83}},\ \bibinfo
  {pages} {205101} (\bibinfo {year} {2011})}\BibitemShut {NoStop}%
\bibitem [{\citenamefont {Ono}\ and\ \citenamefont
  {Watanabe}(2018)}]{Ono2018prb98}%
  \BibitemOpen
  \bibfield  {author} {\bibinfo {author} {\bibfnamefont {S.}~\bibnamefont
  {Ono}}\ and\ \bibinfo {author} {\bibfnamefont {H.}~\bibnamefont {Watanabe}},\
  }\href {\doibase 10.1103/PhysRevB.98.115150} {\bibfield  {journal} {\bibinfo
  {journal} {Phys. Rev. B}\ }\textbf {\bibinfo {volume} {98}},\ \bibinfo
  {pages} {115150} (\bibinfo {year} {2018})}\BibitemShut {NoStop}%
\end{thebibliography}%

\end{document}